\documentclass[12pt]{article}

\usepackage[margin=0.8in]{geometry}
\usepackage{setspace}
\usepackage{amsmath,amsthm}
\usepackage[charter]{mathdesign}

\makeatletter
\@ifundefined{circledS}{\usepackage{amssymb}}{}
\makeatother

\usepackage{graphicx}
\graphicspath{{Figures/}}
\usepackage{caption,subcaption}
\usepackage{booktabs}
\usepackage{rotating}
\usepackage{pdflscape}
\usepackage{enumerate}
\DeclareCaptionFormat{cont}{#1 (cont.)#2#3\par}

\usepackage{verbatim}   
\usepackage{color}
\usepackage{url}
\usepackage[title]{appendix}

\usepackage[colon]{natbib}
\usepackage{xr-hyper}
\usepackage[pdftex,colorlinks=true]{hyperref}
\hypersetup{linkcolor=blue,citecolor=blue,urlcolor=blue,filecolor=red,
  pdfauthor={Xibin Zhang}}

\theoremstyle{definition}
\newtheorem{assumption}{Assumption}
\newtheorem{theorem}{Theorem}

\newtheorem{proposition}{Proposition}

\begin{document}

\begin{center}
{\Large An entropy-adjusted invariant likelihood for incidental parameters}

\medskip 

\medskip 

{\large Xibin Zhang}\footnote{Address: 900 Dandenong Road, Caulfield East, Victoria 3145, Australia. Telephone: +61 3 99032130. \\ Email: {\em\color{blue}xibin.zhang@monash.edu}.} 

\medskip 
Department of Econometrics and Business Statistics, Monash University, Australia

\medskip
\medskip 
\today
\end{center}

\vspace{0.3cm}
\baselineskip 0.78cm

\noindent{\bf Abstract}:
Incidental parameters can make maximum likelihood inconsistent in short
panels.  When they act through a known translation group, integration with
respect to Haar measure eliminates them exactly and gives an invariant
likelihood, whose score is shown to be the conditional expectation, given
the maximal invariant, of the Neyman-orthogonalised score of every order.
Under interactive fixed effects the orbit is an unknown factor space, and
the invariant score for it has a drift equal to minus the derivative of the
entropy of the retained contrasts.  An entropy adjustment removes this drift
and, at the true structural parameter, identifies the factor space uniquely
when the loadings span it.  The adjusted estimating equations are the
gradient of a single criterion if and only if the entropy separates into an
orbit term and an error-law term, as in scale and elliptical families;
consistency then follows from a design condition, and in general from an
equilibrium-identification condition, with asymptotic normality for fixed
panel length.  A boundary is derived for recovering a skewness sign
discarded by a two-period invariant reduction.  An application to
Vietnamese rice farming rejects equal household-effect weights across
survey waves and finds lower returns to scale.

\medskip

\noindent\textbf{Keywords:} entropy adjustment; Haar integration; interactive fixed effects; Neyman orthogonality; translation orbit.

\medskip
\noindent\textbf{JEL Classification:} C13; C23; C33; Q12


\baselineskip 0.78cm


\section{Introduction}
\label{sec:introduction}

Incidental parameters are common in models for panel and network data.  Each
individual, firm or node may have its own parameter, such as a fixed effect
or a factor loading, to account for unobserved heterogeneity
\citep{JochmansWeidner2019}.  The number of such parameters grows with the
sample, but the number of observations on each unit does not.
\citet{NeymanScott1948} showed that maximum likelihood can then be
inconsistent for the parameters of interest.  The problem is serious in
practice because many panels are short.  Household surveys and firm panels
often follow thousands of units over only a few periods, and in such data
the bias does not go away when more units are added.

The literature offers three kinds of remedy, and each has a limit
\citep{Lancaster2000,ArellanoHahn2007,FernandezValWeidner2018}.  The first
keeps the incidental parameters and corrects the bias from estimating them,
analytically or by the jackknife
\citep{HahnNewey2004,HahnKuersteiner2011,DhaeneJochmans2015,FernandezValWeidner2016}.
Modified profile likelihoods
\citep{BarndorffNielsen1983,CoxReid1987,Sartori2003} and integrated
likelihoods with suitable weights
\citep{KalbfleischSprott1970,BergerLiseoWolpert1999,ArellanoBonhomme2009}
also address the distortion introduced by nuisance parameters.  Their
general bias-reduction guarantees typically require the number of periods
to grow, although special models admit exact corrections.  The second
remedy replaces the score by an estimating equation that is insensitive to the incidental parameters
up to a chosen order.  The idea goes back to the $C(\alpha)$ tests of
\citet{Neyman1959} and to projected scores
\citep{SmallMcLeish1989,WatermanLindsay1996}.  It also underlies double
machine learning \citep{ChernozhukovEtAl2018}, and its higher-order version
\citep{BonhommeJochmansNeweyWeidner2026,BonhommeJochmansWeidner2026} allows
shorter panels.  In general, finite-order orthogonalisation leaves a
remainder and requires preliminary estimates of the incidental parameters.
The third remedy removes the incidental parameters completely, by conditioning on a sufficient statistic
\citep{Andersen1970,Chamberlain1980}, by functional differencing
\citep{Bonhomme2012}, or by using a maximal invariant
\citep{ChamberlainMoreira2009,Moreira2009}.  It is exact for any panel
length, but it needs a model with special structure.

This paper studies the third remedy in the form that rests on group
invariance.  When each incidental parameter moves the data along the orbits
of a known group, the likelihood can be integrated along the orbits with
respect to Haar measure.  The result is the likelihood of a maximal
invariant, which does not contain the incidental parameters
\citep{Wijsman1967,Eaton1989}.  Differencing and the within transformation
are simple examples.

Three questions are addressed.  First, what is the exact relation between
Haar integration and higher-order orthogonalisation?  Recent work
distinguishes finite-order and full orthogonality in likelihood models
\citep{BonhommeJochmansWeidnerOrthogonal2026}; the comparison here
identifies the conditional-expectation relation induced by a translation
quotient.  Second, what information does this reduction discard, and when
can that information be recovered?  Third, and most important for applied
work, how can likelihood-based inference proceed when the group is unknown?
In models with interactive fixed effects
\citep{Bai2009,MoonWeidner2015,Bai2024} the loadings move the data along an
unknown factor space.  With short panels these models are usually estimated
by quasi-differencing and the generalised method of moments
\citep{HoltzEakinNeweyRosen1988,AhnLeeSchmidt2001,AhnLeeSchmidt2013}, which
use second moments only.  A likelihood method is needed when the error law
itself is of interest, as in a stochastic frontier model, where the
asymmetry of the error measures inefficiency.  \citet{Zhang2026IFE} showed
that maximising the invariant likelihood over the factor space
can be inconsistent with non-Gaussian errors, and gave a repair that
requires errors that are independent and identically distributed over time.
That repair does not cover period-specific error variances, which are common
in survey data.

The first contribution is to identify the orbit, namely the set along which
an incidental parameter moves the data, as the object that decides what Haar
integration can deliver.  When the orbit is known and free of the structural
parameter, as with additive unit effects, Haar integration amounts to
rotating the data orthogonally and deleting the coordinates along the orbit.
The Haar score is the conditional expectation, given the retained
coordinates, of the orthogonalised score of every order.  It is therefore
insensitive to the incidental parameters at all orders, and it needs no
preliminary estimate, no sample splitting and no condition on the panel
length.  Exactness has a price: the covariance matrix of the Haar score is
bounded above, in the positive semidefinite order, by that of the
orthogonalised score evaluated at the true effects, and the two scores
coincide from the second order onwards under spherical Gaussian errors.
When the orbit moves with the structural parameter, the Haar score drifts.
In the dynamic panel model the orbit depends on the autoregressive
coefficient, the bias of \citet{Nickell1981} is exactly this drift, and
removing it gives the adjusted likelihood of \citet{Lancaster2002} and
\citet{DhaeneJochmans2016}, whose equations are shown to be the
orthogonalised score equations of every order from the second, evaluated at
the unit-level maximum likelihood estimates of the effects.  An orbit that
is itself unknown is beyond both constructions.

The second and main contribution is an entropy adjustment that repairs the
invariant likelihood when the orbit is itself an unknown parameter, as the
factor space is under interactive fixed effects.  Each candidate orbit then
defines a different set of retained contrasts, and the invariant likelihood
can favour orbits whose contrasts have lower entropy.  Its drift in the
factor direction is shown to be exactly minus the derivative of this
entropy, whenever the entropy is finite and the required derivatives and
expectations exist.  The entropy depends on the error law and on the
candidate factor space, but not on the data or the loadings, so it can be
added to the invariant likelihood.  At the true structural parameter, the
expected adjusted criterion is minus a Kullback--Leibler divergence, and it
is uniquely maximised at the true factor space when the loadings span that
space.  The errors may be non-Gaussian, heteroskedastic or dependent over
time.  For Gaussian errors the adjusted criterion reduces to principal
components or to generalised least squares, so the adjustment extends these
methods to non-Gaussian error laws.  The estimator is proved to be
consistent and asymptotically normal for fixed panel length under a global
equilibrium-identification condition.  The estimating equations are the
gradient of a single criterion if and only if the entropy separates into a
term in the orbit and a term in the error parameters, as it does for a known
error law and for scale and elliptical families; consistency then follows
from a condition on the design alone.

The third contribution is to locate the limits of the method, and to
separate the losses that can be repaired from those that cannot.  The first
kind is caused by the reduction itself.  With two periods and an asymmetric
error law, the invariant experiment loses the sign of the asymmetry, so a
production frontier cannot be told from a cost frontier.  This loss is
recoverable: a third period restores the sign, and so do moments whose
expectation does not involve the incidental parameters.  Recovery has its
own limit.  In the two-period problem it is uniform when the cross-sectional
variance of the incidental parameters is $o(N)$, a rate that cannot be
improved under an additional tail condition, and the minimax testing error
equals $1/2$ for every sample size when the effects are unrestricted.  The
second kind belongs to the data.  When the variation that identifies the parameter of
interest lies inside the orbit, the parameter is not identified in the
original experiment, and no method can recover it.

Simulations and an application show how much these results matter in
practice.  In a stochastic frontier model with fixed effects, the
second-order orthogonalised score retains a drift that is not negligible in
short panels with many units, whereas the Haar score is exactly centred and
keeps between two fifths and nine tenths of the information about the
parameter that separates noise from inefficiency.  When error variances
differ across periods, the unadjusted invariant likelihood and the
second-moment estimator show a factor-space error that does not shrink as
the cross-section grows in the reported simulations, whereas the error of
the entropy-adjusted estimator falls at about the rate $N^{-1/2}$.  In an
application to rice production by 800 Vietnamese households over five survey
waves, a Wald test based on the entropy-adjusted estimator rejects the usual
assumption that a household effect shifts output equally in every wave, and
relaxing this assumption lowers the estimated returns to scale from one to
about 0.96.  When measurement error is added to the first two waves, the
unadjusted and second-moment estimates of the returns to scale rise above
one, while the entropy-adjusted estimate stays at 0.96.

The remainder of the paper is organised as follows.
Section~\ref{sec:methodology} presents Haar integration for known orbits and
its relation to orthogonalisation, including the dynamic panel model, in
which the orbit moves with the structural parameter.
Section~\ref{sec:entropy} treats unknown orbits: the entropy adjustment, the
estimator and its limit distribution.  Section~\ref{sec:limits} measures the
information that the reduction loses and locates its limits, with the stochastic frontier model
as an illustration.  Section~\ref{sec:application} applies the estimator to
rice farming in Vietnam.  Section~\ref{sec:conclusion} concludes.  Proofs
and additional simulation results are in
Appendices~\ref{app:proofs}--\ref{app:proofs_application}\csname appendixlocation\endcsname.


\section{Known orbits: Haar integration and orthogonality}
\label{sec:methodology}

Consider a model with a structural parameter $\theta$ of fixed dimension and
unit-specific parameters $\xi_1,\ldots,\xi_N$ whose number grows with the
sample while the information about each one stays fixed.  When each
incidental parameter moves the data along the orbits of a known group, the
likelihood can be integrated along those orbits with respect to Haar
measure.  This section treats orbits that are known, or known up to the
structural parameter. Section~\ref{sec:entropy} treats an unknown orbit,
and Section~\ref{sec:limits} examines the limits of the reduction.

\subsection{The Neyman--Scott problem}
\label{subsec:neyman_scott}

Consider the Gaussian location panel
\begin{equation}
    Y_{it}=\alpha_i+\varepsilon_{it},
    \qquad
    \varepsilon_{it}\overset{\mathrm{iid}}{\sim}N(0,\sigma^2),
    \qquad
    i=1,2,\ldots,N,\quad t=1,2,\ldots,T,
\label{eq:NS_model}
\end{equation}
with fixed $T\ge2$.  The structural parameter is $\sigma^2$ and the locations
$\alpha_1,\alpha_2,\ldots,\alpha_N$ are unrestricted.  Write
$Y_i=(Y_{i1},Y_{i2},\ldots,Y_{iT})'$,
$\overline{Y}_i=T^{-1}\sum_tY_{it}$ and $W_i=\sum_t(Y_{it}-\overline{Y}_i)^2$.

Maximising the likelihood over $\alpha_i$ gives $\widehat\alpha_i=\overline{Y}_i$,
and the profile score of unit $i$ for $\sigma^2$,
$-T/(2\sigma^2)+W_i/(2\sigma^4)$, has mean $-1/(2\sigma^2)$ at the true
value, for every unit and every $N$.  The maximum likelihood estimator
$\sum_iW_i/(NT)$ therefore converges to $(T-1)\sigma^2/T$: one dimension per
unit has been spent on $\alpha_i$, but all $T$ observations are used as if
they were available for $\sigma^2$.  A bias correction multiplies the
estimator by $T/(T-1)$, and the second-order orthogonalised score of
\citet{BonhommeJochmansWeidner2026} removes the bias and no longer involves
$\alpha_i$.  The third remedy of Section~\ref{sec:introduction} changes the
experiment.  The group
$(\mathbb R,+)$ acts by
\begin{equation}
    Y_i\longmapsto Y_i+a\mathbf 1_T,
    \qquad
    \alpha_i\longmapsto\alpha_i+a,
    \qquad a\in\mathbb R,
\label{eq:NS_action}
\end{equation}
and this action leaves $\sigma^2$ unchanged.  The Haar measure of the
translation group is Lebesgue measure, and integrating the unit likelihood
along the orbit gives
\begin{equation}
L_{H,i}(\sigma^2)
=
\int_{\mathbb R}
(2\pi\sigma^2)^{-T/2}
\exp\left\{-\frac{1}{2\sigma^2}
\sum_{t=1}^T(Y_{it}-\alpha)^2\right\}d\alpha
=
\frac{1}{\sqrt T}
(2\pi\sigma^2)^{-(T-1)/2}
\exp\left\{-\frac{W_i}{2\sigma^2}\right\}.
\label{eq:NS_haar}
\end{equation}

\begin{proposition}[Neyman--Scott]
\label{prop:NS}
Let $B_T$ be a $T\times(T-1)$ matrix with $B_T'B_T=I_{T-1}$ and
$B_T'\mathbf 1_T=0$.  In model \eqref{eq:NS_model}, for every $N\ge1$,
$T\ge2$ and $(\alpha_1,\alpha_2,\ldots,\alpha_N)\in\mathbb R^N$:
\begin{enumerate}
\item[(a)] $B_T'Y_i$ is a maximal invariant under \eqref{eq:NS_action}, and
$L_{H,i}(\sigma^2)$ in \eqref{eq:NS_haar} equals $T^{-1/2}$ times the density
of $B_T'Y_i$ evaluated at the data.
\item[(b)] The Haar score
\begin{equation}
    s_{H,i}(\sigma^2)=\frac{W_i-(T-1)\sigma^2}{2\sigma^4}
\label{eq:NS_score}
\end{equation}
has mean zero at the true value, and it coincides with the second-order
orthogonalised score of \citet{BonhommeJochmansWeidner2026}.
\item[(c)] The Haar estimator
$\widehat\sigma_H^2=\sum_iW_i/\{N(T-1)\}$ satisfies
$N(T-1)\widehat\sigma_H^2/\sigma^2\sim\chi^2_{N(T-1)}$.  It is therefore
unbiased, with variance $2\sigma^4/\{N(T-1)\}$.
\end{enumerate}
\end{proposition}

\noindent{\em Proof.} See Appendix~\ref{app:proof_NS}.

The action \eqref{eq:NS_action} is known and preserves the structural
parameter. The quotient $B_T'Y_i$ retains variation informative about
$\sigma^2$. The Haar integral is thus
an integrated likelihood
\citep{KalbfleischSprott1970,BergerLiseoWolpert1999} whose weight is fixed
by the group rather than chosen, and it is the likelihood of a smaller
experiment in which the incidental parameter does not exist.

\subsection{Translation orbits and the Haar likelihood}
\label{subsec:haar_reduction}

Eliminating a translation effect does not require normality or independence
over time.  Let unit $i$ contribute
\begin{equation}
    Y_i=\mu_i(\beta)+A\xi_i+\varepsilon_i,
    \qquad i=1,\ldots,N,
\label{eq:orbit_model}
\end{equation}
where $Y_i\in\mathbb R^T$, $A$ is a known $T\times r$ matrix of rank $r<T$,
$\xi_i\in\mathbb R^r$ is an unrestricted incidental parameter, $\mu_i(\cdot)$
is a known function that may depend on covariates $X_i$ and on common
parameters, and $\varepsilon_i$ has density $g(\cdot;\eta)$ on $\mathbb R^T$
conditional on $X_i$.  The structural parameter is
$\theta=(\beta',\eta')'$, with true value $\theta_0$.  The group
$(\mathbb R^r,+)$ acts by
\[
    Y_i\longmapsto Y_i+Aa,
    \qquad
    \xi_i\longmapsto\xi_i+a,
    \qquad a\in\mathbb R^r,
\]
and preserves $\theta$.  Define the Haar likelihood
\begin{equation}
L_{H,i}(\theta)
=
|A'A|^{1/2}
\int_{\mathbb R^r}
g\{Y_i-\mu_i(\beta)-A\xi;\eta\}\,d\xi.
\label{eq:haar_lik}
\end{equation}
Let $B$ be a $T\times(T-r)$ matrix with $B'B=I_{T-r}$ and $B'A=0$, and let
\[
    h(w;\eta)
    =\int_{\mathbb R^r}
    g\{Bw+A(A'A)^{-1/2}c;\eta\}\,dc,
    \qquad w\in\mathbb R^{T-r},
\]
which is the density of $B'\varepsilon_i$.

\begin{assumption}[Regular quotient family]
\label{ass:regular}
The functions $\mu_i$ are twice continuously differentiable.  The density
$h(w;\eta)$ is positive, twice continuously differentiable in $(w,\eta)$, and
vanishes as $\|w\|\to\infty$.  Differentiation with respect to $(w,\eta)$ can
be interchanged twice with integration over $w$, and the scores of the family
$\{h(\cdot-B'\mu_i(\beta);\eta)\}$ are square integrable.
\end{assumption}

\begin{proposition}[Haar likelihood of a translation orbit]
\label{prop:haar_translation}
In model \eqref{eq:orbit_model}, for every $T>r$ and every
$\xi_i\in\mathbb R^r$:
\begin{enumerate}
\item[(a)] $B'Y_i$ is a maximal invariant, and its distribution depends on
$(\theta,\xi_i)$ only through $\theta$.
\item[(b)] The Haar likelihood is the density of the maximal invariant,
\begin{equation}
    L_{H,i}(\theta)=h\{B'(Y_i-\mu_i(\beta));\eta\}.
\label{eq:haar_equals_quotient}
\end{equation}
It is finite for almost every $Y_i$ and is unchanged when $A$ is replaced by
$AH$ for any nonsingular $r\times r$ matrix $H$.
\item[(c)] Under Assumption~\ref{ass:regular}, the Haar score
$s_{H,i}(\theta)=\partial\log L_{H,i}(\theta)/\partial\theta$ satisfies
$\mathbb E_{\theta_0,\xi_i}s_{H,i}(\theta_0)=0$ and the information identity
$\mathbb E_{\theta_0,\xi_i}\{s_{H,i}s_{H,i}'\}
=-\mathbb E_{\theta_0,\xi_i}\{\partial s_{H,i}/\partial\theta'\}$
at $\theta_0$.
\item[(d)] For every $\psi$ with
$\mathbb E|\psi(B'Y_i)|<\infty$, the mean
$\mathbb E_{\theta,\xi_i}\psi(B'Y_i)$ does not depend on $\xi_i$.  If, in
addition, $g$ is $p$ times differentiable in its argument, differentiation
can be interchanged with integration, and the resulting products are
integrable, then $\psi(B'Y_i)$ has zero covariance with the $p$-th Bhattacharyya function
$v_p=\nabla^{(p)}_{\xi}g/g$ of the incidental parameter
\citep{Bhattacharyya1946}.
\end{enumerate}
\end{proposition}

\noindent{\em Proof.} See Appendix~\ref{app:proof_haar_translation}.

Part (b) is the general form of \eqref{eq:NS_haar}, and by part (c) the
profile-score bias of Section~\ref{subsec:neyman_scott} cannot arise at any
$T>r$.  Parts (a)--(c) are classical consequences of invariance
\citep{Wijsman1967,Eaton1989}; for interactive effects with a given factor
space they are Propositions~1--3 of \citet{Zhang2026IFE}.  Part (d) is the
link to orthogonalisation that is used next.

\medskip 
\noindent{\bf An orthogonal transformation.}
The proof of Proposition~\ref{prop:haar_translation} shows what the Haar
integral does to the data.  Let $Q=A(A'A)^{-1/2}$.  The matrix $(Q,B)$ is
orthogonal, and it rotates the observation into two blocks,
\begin{equation}
    Q'Y_i=Q'\mu_i(\beta)+(A'A)^{1/2}\xi_i+Q'\varepsilon_i,
    \qquad
    B'Y_i=B'\mu_i(\beta)+B'\varepsilon_i.
\label{eq:rotation}
\end{equation}
The incidental parameter is confined to the first block, where it is a free
location.  Haar integration integrates the first block out and keeps the
second.  In the translation model, Haar integration is therefore the same
operation as an orthogonal transformation of the data followed by the
deletion of $r$ coordinates.

This description invites a comparison with the second remedy of
Section~\ref{sec:introduction}, which is also called orthogonalisation.  There,
orthogonality is a property of the estimating equation and not of the data.
Let
$u_i(\theta,\xi_i)
=\partial\log g\{Y_i-\mu_i(\beta)-A\xi_i;\eta\}/\partial\theta$ be the score
of the original likelihood, let $\mathcal V_q$ be the linear span of the
components of the Bhattacharyya functions $v_1,\ldots,v_q$ of
Proposition~\ref{prop:haar_translation}(d), and let
\begin{equation}
    u^*_{q,i}=u_i-\Pi(u_i\mid\mathcal V_q)
\label{eq:projected_score}
\end{equation}
be the residual from the mean-square projection of $u_i$ on $\mathcal V_q$
under the distribution indexed by $(\theta,\xi_i)$.  This is the
orthogonalised score of order $q$ of \citet{BonhommeJochmansWeidner2026},
obtained from their main theorem when the estimating function is the score.
It is Neyman-orthogonal to order $q$, and in general it still depends on
$\xi_i$.

\begin{theorem}[Haar integration and orthogonalisation]
\label{thm:haar_orthogonal}
Consider model \eqref{eq:orbit_model} under Assumption~\ref{ass:regular}, and
suppose that $u_i$ and $v_1,\ldots,v_q$ are square integrable and that
differentiation can be interchanged with integration.  For every $T>r$ and
every $\xi_i\in\mathbb R^r$:
\begin{enumerate}
\item[(a)] The Haar score is the projection of the original score on the
retained block,
\[
    s_{H,i}(\theta)
    =\mathbb E_{\theta,\xi_i}\{u_i(\theta,\xi_i)\mid B'Y_i\}.
\]
\item[(b)] For every order $q\ge1$,
\[
    s_{H,i}(\theta)=\mathbb E_{\theta,\xi_i}\{u^*_{q,i}\mid B'Y_i\}.
\]
Consequently $s_{H,i}$ is uncorrelated with $v_p$ for every $p$ satisfying
the same integrability and differentiation conditions, and
$\operatorname{Var}(u^*_{q,i})-\operatorname{Var}(s_{H,i})$ is positive
semidefinite.
\item[(c)] If $\varepsilon_i\sim N(0,\sigma^2I_T)$ and
$\theta=(\beta',\sigma^2)'$, then $u^*_{q,i}=s_{H,i}$ for every $q\ge2$.  For
$q=1$ the equality holds for the components of $\beta$ but not for
$\sigma^2$.
\end{enumerate}
\end{theorem}

\noindent{\em Proof.} See Appendix~\ref{app:proof_haar_orthogonal}.

The word orthogonal thus applies twice to the Haar reduction: the data are
rotated by an orthogonal matrix, and by part (a) the score is projected
orthogonally on functions of the retained block.  By part (b), the
orthogonalised score of any order, averaged over the discarded block, is the
Haar score.  The Haar score is therefore orthogonal to the incidental
parameter at every order and not at a chosen order $q$, and since it does
not involve $\xi_i$ it needs no preliminary estimate of the incidental
parameters, no sample splitting, and no condition on the relative size of
$N$ and $T$.  Part (c) shows that with spherical Gaussian errors the two
constructions coincide from the second order onwards;
Proposition~\ref{prop:NS}(b) is the simplest instance.

Two qualifications follow.  Outside the Gaussian case the inequality in
part (b) can be strict: the orthogonalised score may use the discarded block
$Q'Y_i$, so the advantage of the Haar reduction is exactness at fixed $T$,
not additional information (Sections~\ref{subsec:information}
and~\ref{sec:frontier}).  The reduction also requires a known nuisance
subspace, so that the rotation \eqref{eq:rotation} can be applied to the
data, which finite-order orthogonalisation does not.  The price of exactness
is the group.

\medskip 
\noindent{\bf Two-way effects and other groups.}
With unit and time effects, fixed $T$ and large $N$, only the unit effects
are incidental, and model \eqref{eq:orbit_model} applies with
$A=\mathbf 1_T$ and the time effects included in $\mu_i$
\citep{HongKingZhang2026}.  Groups other than translations
\citep{Wijsman1967,Eaton1989,Davis1954,Moreira2009,BondarMilnes1981,ChamberlainMoreira2009}
are discussed in Appendix~\ref{app:additional_groups}.

\subsection{Conditions for an exact solution}
\label{subsec:principle}

The reduction solves an incidental-parameter problem when five requirements
hold.

\begin{enumerate}
\item[H1.] \emph{Known action.}  The group acts on the sample space through
transformations that do not depend on any unknown parameter, and the induced
action on the parameter space leaves $\theta$ fixed.
\item[H2.] \emph{Proper quotient likelihood.}  The Haar integral is finite and
is the density of a maximal invariant with respect to a measure that does not
depend on the parameters.
\item[H3.] \emph{Elimination or compression.}  The maximal invariant of the
incidental parameter is empty or has a dimension that does not grow with the
sample.
\item[H4.] \emph{Identification in the quotient.}  The distribution of the
maximal invariant determines $\theta$.
\item[H5.] \emph{Positive information.}  The Fisher information for $\theta$
in the quotient, after any remaining common nuisance has been partialled out,
is nonsingular.
\end{enumerate}

In the translation model \eqref{eq:orbit_model}, H1 holds by construction,
H2 holds by Proposition~\ref{prop:haar_translation}(b) when $T>r$, and H3
holds with complete elimination.  The remaining requirements are properties of
the quotient family.  Let
$Q_N(\theta)=N^{-1}\sum_{i=1}^N\log L_{H,i}(\theta)$,
let $\widehat\theta_H$ maximise $Q_N$ over a compact set $\Theta$ with
$\theta_0$ in its interior, and let
$\mathrm{KL}_i(\theta)$ denote the Kullback--Leibler divergence
\citep{KullbackLeibler1951}
$\operatorname{KL}(p_{i,\theta_0}\,\|\,p_{i,\theta})$, where
$p_{i,\theta}$ is the density of $B'Y_i$ under $\theta$.

Assumption~\ref{ass:quotient} in
Appendix~\ref{app:proof_haar_reduction_principle} collects the conditions on
this family: a uniform law of large numbers for $Q_N$, separation of the
average divergence $N^{-1}\sum_i\mathrm{KL}_i(\theta)$ from zero away from
$\theta_0$ (H4), a positive definite limit $I_H$ of the average information
of the Haar scores (H5), and Lindeberg's condition.

\begin{proposition}[Haar reduction principle]
\label{prop:haar_reduction_principle}
Consider model \eqref{eq:orbit_model} with $T>r$ fixed, and suppose
Assumptions~\ref{ass:regular} and~\ref{ass:quotient} hold.
\begin{enumerate}
\item[(a)] For every $N$ and every $(\xi_1,\xi_2,\ldots,\xi_N)$,
$\mathbb EQ_N(\theta_0)-\mathbb EQ_N(\theta)
=N^{-1}\sum_i\mathrm{KL}_i(\theta)\ge0$, and neither side depends on the
incidental parameters.
\item[(b)] $\widehat\theta_H\overset p\to\theta_0$ as $N\to\infty$.
\item[(c)] $\sqrt N(\widehat\theta_H-\theta_0)\overset d\to N(0,I_H^{-1})$.
\end{enumerate}
\end{proposition}

\noindent{\em Proof.} See Appendix~\ref{app:proof_haar_reduction_principle}.

No condition restricts the incidental parameters, their dispersion, or the
relative size of $N$ and $T$, and no preliminary estimate or sample split is
used: the incidental-parameter problem has been replaced by an ordinary one.

\subsection{Parameter-dependent orbits: the dynamic panel model}
\label{sec:dynamic}

Requirement H1 can fail even when the incidental parameter is an additive
unit effect.  The linear dynamic panel with fixed effects, the model in
which the incidental-parameter problem is met most often in applied work
\citep{Nickell1981,Lancaster2002,ChamberlainMoreira2009,Moreira2009,DhaeneJochmans2016},
is the leading case.  Let
\begin{equation}
    y_{it}=\rho\,y_{i,t-1}+x_{it}'\beta+\alpha_i+\varepsilon_{it},
    \qquad i=1,2,\ldots,N,\quad t=1,2,\ldots,T,
\label{eq:dyn_model}
\end{equation}
with $T\ge2$ fixed, initial observations $y_{i0}$, strictly exogenous
covariates $x_{it}$, and unrestricted effects $\alpha_i$.  Conditional on
$(y_{i0},X_i,\alpha_i)$ the errors $\varepsilon_{i1},\ldots,\varepsilon_{iT}$
are uncorrelated with mean zero and common variance $\sigma^2$; normality is
not assumed unless stated.  The structural parameter is
$\theta=(\rho,\beta',\sigma^2)'$.  Write $y_i=(y_{i1},\ldots,y_{iT})'$,
$y_{i,-}=(y_{i0},\ldots,y_{i,T-1})'$, $X_i=(x_{i1},\ldots,x_{iT})'$,
$e_i(\rho,\beta)=y_i-\rho y_{i,-}-X_i\beta$ and
$M=I_T-T^{-1}\mathbf 1_T\mathbf 1_T'$.  Integrating the conditional Gaussian
likelihood of the observed path over $\alpha_i$, with the normalisation of
\eqref{eq:haar_lik}, gives the criterion
\begin{equation}
    \ell_{H,i}(\theta)
    =-\frac{T-1}{2}\log(2\pi\sigma^2)
     -\frac{1}{2\sigma^2}\,e_i(\rho,\beta)'M\,e_i(\rho,\beta),
\label{eq:dyn_naive}
\end{equation}
whose maximiser in $(\rho,\beta)$ is the within-group estimator.  Unlike the
construction in Proposition~\ref{prop:haar_translation}, this is not the
likelihood of a parameter-independent quotient.  The translation
$y_i\mapsto y_i+a\mathbf 1_T$ also shifts the lagged values in $y_{i,-}$, and
solving the recursion shows that the effect moves the data along
$c(\rho)=\mathcal L(\rho)\mathbf 1_T$, where $\mathcal L(\rho)$ is the
lower-triangular matrix with $(t,s)$ entry $\rho^{\,t-s}$ for $t\ge s$.
This direction is $\mathbf 1_T$ only when $\rho=0$. The orbit moves with the structural
parameter, and the score for $\rho$ drifts.  At $\theta_0$, for every value
of $(\alpha_i,y_{i0},X_i)$, the scores of \eqref{eq:dyn_naive} for $\beta$
and $\sigma^2$ have mean zero, and
\begin{equation}
    \mathbb E\,\frac{\partial\ell_{H,i}}{\partial\rho}=-b_T(\rho_0),
    \qquad
    b_T(\rho)=\frac1T\sum_{t=1}^{T-1}(T-t)\rho^{\,t-1}
\label{eq:dyn_bT}
\end{equation}
(Proposition~\ref{prop:dyn_drift} in Appendix~\ref{app:proof_dyn_drift}).
This drift is the bias of \citet{Nickell1981}.  It is a known function of the
structural parameter, so it can be removed, and since $\rho$ is a scalar the
correction integrates to an adjustment of the criterion itself.  Let
\begin{equation}
    \ell_{A,i}(\theta)=\ell_{H,i}(\theta)+a_T(\rho),
    \qquad
    a_T(\rho)=\frac1T\sum_{t=1}^{T-1}\frac{T-t}{t}\,\rho^{\,t},
\label{eq:dyn_adjusted}
\end{equation}
so that $a_T'(\rho)=b_T(\rho)$ and all scores of \eqref{eq:dyn_adjusted}
have mean zero at $\theta_0$, for every $T\ge2$, for every configuration of
effects and initial observations, and without normality.  This is the
adjusted likelihood of \citet{Lancaster2002} and \citet{DhaeneJochmans2016}.

How does the recentred score compare with the orthogonalised score of
\citet{BonhommeJochmansWeidner2026}?  Their construction does not need a
group, so it applies directly to the Gaussian likelihood of $y_i$ given
$(y_{i0},X_i)$, with $\alpha_i$ as the incidental parameter.  Let
$u^*_{q,i}(\theta,\alpha)$ be the orthogonalised score of order $q$, defined
as in \eqref{eq:projected_score}, and let
$\overline{e}_i(\theta)=T^{-1}\mathbf 1_T'e_i(\rho,\beta)$, which is the maximum
likelihood estimator of $\alpha_i$ for given $\theta$.  Let $K(\rho)$ be the
$T\times T$ matrix with $(t,j)$ entry $\rho^{\,t-1-j}$ for $j<t$ and zero
otherwise, and let $g(\rho)=K(\rho)'\mathbf 1_T$.

\begin{theorem}[Recentring is orthogonalisation]
\label{thm:dyn_ortho}
In model \eqref{eq:dyn_model} with the Gaussian likelihood, for every
$q\ge2$, every $T\ge2$, and all values of $(\theta,\alpha)$:
\begin{enumerate}
\item[(a)] The components of $u^*_{q,i}(\theta,\alpha)$ for $\beta$ and
$\sigma^2$ do not depend on $\alpha$ and equal
$\partial\ell_{H,i}/\partial\beta$ and
$\partial\ell_{H,i}/\partial\sigma^2$.
\item[(b)] The component for $\rho$ is
\begin{equation}
    u^*_{\rho,q,i}(\theta,\alpha)
    =\frac{\partial\ell_{A,i}(\theta)}{\partial\rho}
     +\frac{\{\overline{e}_i(\theta)-\alpha\}\,
            g(\rho)'M\,e_i(\rho,\beta)}{\sigma^2}.
\label{eq:dyn_ortho}
\end{equation}
\item[(c)] Consequently
$u^*_{q,i}\{\theta,\overline{e}_i(\theta)\}
=\partial\ell_{A,i}(\theta)/\partial\theta$ identically in $\theta$ and in
the data.  The likelihood equations of the adjusted criterion
\eqref{eq:dyn_adjusted} are the orthogonalised score equations of every order
$q\ge2$, evaluated at the unit-level maximum likelihood estimates of the
effects.
\end{enumerate}
\end{theorem}

\noindent{\em Proof.} See Appendix~\ref{app:proof_dyn_ortho}.

The orthogonalised score \eqref{eq:dyn_ortho} still contains $\alpha_i$, but
with a coefficient that has mean zero at the true values, so neither
independent preliminary estimates nor a condition on the relative size of
$N$ and $T$ is needed.  Part (c) has a precedent in
\citet{LiLindsayWaterman2003}, who show that a second-order projected score
evaluated at the unit-level estimate is unbiased to first order in general
models; in the Gaussian autoregression this holds exactly at every order
from the second.  Haar integration thus contributes an explanation of the
Nickell bias and not a new estimator.  Centred scores do not settle
requirements H4 and H5: persistent short panels are weakly identified, and
with two periods the estimator must be defined as an interior local
maximiser (Appendices~\ref{subsec:dyn_T2}, \ref{subsec:dyn_numerical}
and~\ref{subsec:dyn_nongaussian}).  The model also admits two groups that
satisfy H1, at the price of an assumption on the initial condition or of
normality (Appendix~\ref{subsec:dyn_groups}).  A free common orbit requires
a different repair, which is the subject of the next section.


\section{Unknown orbits: the entropy-adjusted likelihood}
\label{sec:entropy}

This section studies the case in which the orbit is a free common
parameter: why the invariant likelihood then fails, the entropy adjustment
that repairs it, and the properties of the resulting estimator.

\subsection{Orbit-score drift and the entropy identity}
\label{subsec:unknown_orbit}

When the direction in which the incidental parameter moves the data is
itself unknown, requirement H1 fails in a way that recentring cannot repair.  The interactive fixed-effects model is the
leading case:
\begin{equation}
    Y_i=X_i\beta+F\lambda_i+\varepsilon_i,
    \qquad i=1,2,\ldots,N,
\label{eq:IFE_model}
\end{equation}
where $F$ is an unknown $T\times r$ matrix of common factors with rank
$r<T$, $\lambda_i\in\mathbb R^r$ is an incidental loading, and
$\varepsilon_i$ has density $g(\cdot;\eta)$.  Let
$Q_0=F(F'F)^{-1/2}$ and let $B_0$ be an orthonormal basis of the orthogonal
complement of $\operatorname{span}(F)$.  If $F$ were known,
\eqref{eq:IFE_model} would be model \eqref{eq:orbit_model} with $A=F$, and
Proposition~\ref{prop:haar_translation} would apply: the loadings are
eliminated exactly, the factor $|F'F|^{1/2}$ in \eqref{eq:haar_lik} makes the
result depend on $F$ only through its column space, and the scores for
$\theta$ are centred.  The common nuisance that remains is a point on the
Grassmann manifold $\operatorname{Gr}(r,T)$, of dimension $r(T-r)$.
\citet{HallinMoreiraOnatski2013} use the same invariance for tests in a
Gaussian factor model.

It is tempting to conclude that the incidental-parameter problem has been
reduced to a parametric one, and to maximise
\begin{equation}
    \ell_{H,i}(\theta,U)
    =\log\int_{\mathbb R^r}g(Y_i-X_i\beta-Uc;\eta)\,dc,
    \qquad U'U=I_r,
\label{eq:IFE_criterion}
\end{equation}
jointly over $\theta$ and the candidate basis $U$.  This step is not
justified.  The transformation $Y_i\mapsto Y_i+Fa$ cannot be applied without
knowing $F$, so each $U$ defines a different maximal invariant, and
\eqref{eq:IFE_criterion} compares densities of different statistics.

\citet[Proposition~4]{Zhang2026IFE} makes the failure precise.  Let
$W_i=B_0'\varepsilon_i$, $C_i=Q_0'\varepsilon_i$ and
$s_w(w)=\nabla_w\log h(w;\eta_0)$, where $h$ is the density of $W_i$.  Under
smoothness conditions on $g$, the derivative of \eqref{eq:IFE_criterion}
with respect to the orbit at the true values $(\theta_0,Q_0)$, in a
direction $\Delta=B_0H$ with $H\in\mathbb R^{(T-r)\times r}$, has mean
\begin{equation}
    \mathbb E\,D_U\ell_{H,i}(\theta_0,Q_0)[\Delta]
    =-\mathbb E\{s_w(W_i)'HC_i\}.
\label{eq:orbit_drift}
\end{equation}
The loading does not appear on the right-hand side, which is zero when $W_i$
and $C_i$ are independent, as with spherical Gaussian errors, but not in
general.  A criterion whose score has nonzero mean at the truth does not, in
general, have its maximiser at the truth, and adding units does not help
because the drift is the same for every unit.

That analysis leaves two questions open: what the drift is for a general
panel length, rank and error law, and whether it can be removed within the
likelihood, without restricting the error law.  The entropy of the retained
contrasts answers both.  For a candidate
basis $U$, let $B_U$ be an orthonormal basis of the orthogonal complement of
$\operatorname{span}(U)$, let $h_U(\cdot;\eta)$ be the density of
$B_U'\varepsilon_i$, so that
$\ell_{H,i}(\theta,U)=\log h_U\{B_U'(Y_i-X_i\beta);\eta\}$, and let
\begin{equation}
    \mathcal E(U;\eta)=-\int h_U(w;\eta)\log h_U(w;\eta)\,dw
\label{eq:contrast_entropy}
\end{equation}
be the differential entropy of the retained contrasts.  The entropy depends
on the error law and on the candidate orbit, and on nothing else: no data and
no loadings enter it.  Define the entropy-adjusted criterion
\begin{equation}
    \ell_{E,i}(\theta,U)=\ell_{H,i}(\theta,U)+\mathcal E(U;\eta).
\label{eq:entropy_adjusted}
\end{equation}

\begin{theorem}[Entropy adjustment]
\label{thm:entropy}
Consider model \eqref{eq:IFE_model} with an arbitrary error density
$g(\cdot;\eta_0)$ for which $\mathcal E(U;\eta_0)$ is finite for every $U$,
and let $a_i=(F'F)^{1/2}\lambda_i$ and
$\mathcal A_N=N^{-1}\sum_ia_ia_i'$.
\begin{enumerate}
\item[(a)] For every $U$ and every $\lambda_i$,
\begin{equation}
    \mathbb E\,\ell_{E,i}(\theta_0,U)
    =-\operatorname{KL}\!\big\{h_U\,\big\|\,h_U(\cdot+B_U'Q_0a_i)\big\}
    \le0,
\label{eq:entropy_KL}
\end{equation}
where the divergence is between the density of $B_U'\varepsilon_i$ and its
translate, both at $\eta_0$.  Equality holds if and only if
$B_U'Q_0a_i=0$.
\item[(b)] If $\mathcal A_N$ is positive definite, then
$N^{-1}\sum_i\mathbb E\,\ell_{E,i}(\theta_0,U)$ is maximised over
$\operatorname{Gr}(r,T)$ only at $\operatorname{span}(U)=\operatorname{span}(F)$.
\item[(c)] Suppose that $\mathcal E(U;\eta_0)$ is differentiable in $U$ at
$Q_0$ and that $\mathbb E\,\ell_{H,i}(\theta_0,U)$ may be differentiated
under the expectation there.  Then, for every loading,
\[
    \mathbb E\,D_U\ell_{H,i}(\theta_0,Q_0)=-D_U\mathcal E(Q_0;\eta_0),
    \qquad
    \mathbb E\,D_U\ell_{E,i}(\theta_0,Q_0)=0.
\]
The drift \eqref{eq:orbit_drift} is minus the derivative of the entropy, for
every $T>r$ and every such error law.
\item[(d)] If $\varepsilon_i\sim N(m_\eta,\Sigma_\eta)$ with $\Sigma_\eta$
nonsingular, then, with $z_i=Y_i-X_i\beta-m_\eta$,
\[
    \ell_{E,i}(\theta,U)
    =\frac{T-r}2-\frac12\min_{c\in\mathbb R^r}
     (z_i-Uc)'\Sigma_\eta^{-1}(z_i-Uc).
\]
\end{enumerate}
\end{theorem}

\noindent{\em Proof.} See Appendix~\ref{app:proof_entropy}.

\begin{figure}[t]
\centering
\includegraphics[width=0.78\textwidth]{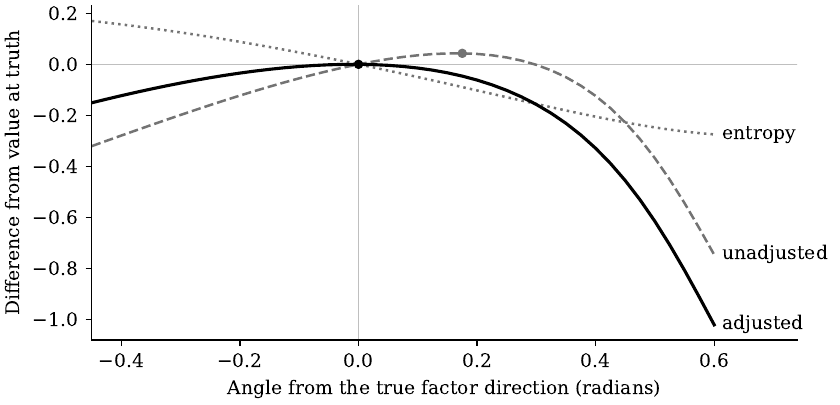}
\caption{Expected Haar criterion (unadjusted), entropy of the retained
contrasts, and their sum (adjusted) along a path of candidate factor
directions through the true one, at the true structural parameter, in the
heteroskedastic design of Section~\ref{subsec:entropy_numerical}.  Each curve
is shown as a difference from its value at the true direction.}
\label{fig:entropy}
\end{figure}

Part (c) extends a two-period observation of \citet{Zhang2026IFE} to
every panel length, rank and error law, and it does so without computing the
drift: the expected adjusted criterion has a maximum at the true orbit, so
its derivative there is zero, and the entropy is not random.  The expected
Haar criterion contains a negative entropy term that can favour orbits with
lower-entropy contrasts, and the adjustment cancels this term
(Figure~\ref{fig:entropy}).  By the
entropy power inequality \citep{Stam1959}, with two periods, one factor and
independent errors from a common density with finite variance and entropy,
this drift vanishes at every factor direction only for Gaussian errors.

Parts (a) and (b) say more than centring.  At the true structural parameter,
the expected adjusted criterion is minus a Kullback--Leibler divergence
between a density and its translate, so it is maximised at the true orbit
globally, for every $T>r$, for every configuration of loadings with
$\mathcal A_N$ positive definite, and for every error law, Gaussian or not,
independent over time or not.  The negative Hessian of the expected average criterion at the
true orbit is $\mathcal A_N\otimes J_W$, where $J_W$ is the Fisher
information for location of the contrast density
(Theorem~\ref{thm:entropy_limit}(d)): the information about the orbit, with
the structural parameter fixed, comes from the second moments of the
loadings, weighted by how sharply the error law locates a shift.

The adjustment applies to the orbit and not to the structural parameter.
The entropy depends on $\eta$, so adding it would displace the scores for
$\eta$, which Proposition~\ref{prop:haar_translation}(c) shows are already
centred at the true orbit.  The repaired system is therefore
\begin{equation}
    \frac1N\sum_{i=1}^N
    \frac{\partial\ell_{H,i}(\theta,U)}{\partial\theta}=0,
    \qquad
    U\in\arg\max_{U}\frac1N\sum_{i=1}^N\ell_{E,i}(\theta,U),
\label{eq:entropy_system}
\end{equation}
with $\dim(\theta)+r(T-r)$ equations, all of which have mean zero at
$(\theta_0,Q_0)$.  The rest of this section studies the resulting
estimator.  Requirement H1 fails in yet another way when the group does not
act on the observed data at all, as in binary panels, where translations act
on a latent index (Appendix~\ref{app:additional_latent}).

\subsection{The entropy-adjusted estimator}
\label{subsec:entropy_estimator}

The estimator $(\widehat\theta_E,\widehat U_E)$ solves
\eqref{eq:entropy_system}, and the two steps can be alternated until neither
changes.  The model, the structural step and its computation by Gaussian
quadrature are those of \citet{Zhang2026IFE}; the orbit step is new.

Three remarks place the estimator.  First, it requires the same input as the
Haar likelihood itself, namely the error law up to a finite-dimensional
parameter, and one further computation, the entropy
\eqref{eq:contrast_entropy}, which is an expectation under the error law
alone and can be simulated once for each candidate $(U,\eta)$, with common
random numbers across candidates.  Second, by
Theorem~\ref{thm:entropy}(d) the estimator reduces to known procedures in
Gaussian models: with errors $N(0,\sigma^2I_T)$ the orbit step is principal
components, and with errors $N(0,\Sigma_\eta)$ it is principal components in
the metric $\Sigma_\eta^{-1}$.  Third, the estimator does not use the moment structure
of the errors.  When error variances differ across periods, or the errors
are serially dependent, the leading eigenspace used by the second-moment
equation of \citet{Zhang2026IFE} need not coincide with the factor space,
while Theorem~\ref{thm:entropy}(a)--(b) continues to hold.

\subsection{Consistency and asymptotic normality}
\label{subsec:entropy_limit}

The structural step maximises the Haar likelihood and the orbit step
maximises the adjusted likelihood, each with the other parameter held
fixed. Their scores are the gradient of a single criterion only under the
condition of Theorem~\ref{thm:separable}(a). In general, solutions that
maximise each criterion in its own argument are equilibria of two criteria, and consistency is
established through the gap by which a candidate fails to be one, a device
from the theory of noncooperative equilibria \citep{NikaidoIsoda1955}.  Let
$Q_N(\theta,U)=N^{-1}\sum_i\ell_{H,i}(\theta,U)$ and
$L_N(\theta,U)=N^{-1}\sum_i\ell_{E,i}(\theta,U)$, where in $L_N$ the entropy
may be replaced by a simulated version, let $\overline{Q}_N$ and $\overline{L}_N$ be
their expectations with the exact entropy, and let
\begin{equation}
    \Delta_N(\theta,U)
    =\Big\{\sup_{\theta'\in\Theta}\overline{Q}_N(\theta',U)-\overline{Q}_N(\theta,U)\Big\}
    +\Big\{\sup_{U'}\overline{L}_N(\theta,U')-\overline{L}_N(\theta,U)\Big\},
\label{eq:equilibrium_gap}
\end{equation}
with the second supremum over $\operatorname{Gr}(r,T)$.  The gap is
nonnegative and is zero exactly at the equilibria of the population system.
Distances on $\Theta\times\operatorname{Gr}(r,T)$ are measured by
\[
    d\{(\theta,U),(\theta_0,Q_0)\}
    =\|\theta-\theta_0\|+\|UU'-Q_0Q_0'\|.
\]

\begin{assumption}[Equilibrium identification]
\label{ass:entropy_consistency}
The units are independent, $\Theta$ is compact with $\theta_0$ in its
interior, and Assumption~\ref{ass:regular} holds at the true orbit.  In addition:
\begin{enumerate}
\item[(i)] $\sup|Q_N-\overline{Q}_N|\overset p\to0$ and
$\sup|L_N-\overline{L}_N|\overset p\to0$, the suprema being over
$\Theta\times\operatorname{Gr}(r,T)$;
\item[(ii)] for every $\epsilon>0$,
$\liminf_{N\to\infty}\inf_{d\{(\theta,U),(\theta_0,Q_0)\}\ge\epsilon}
\Delta_N(\theta,U)>0$.
\end{enumerate}
\end{assumption}

For the limit distribution, parametrise orbits near the truth by
$U(H)=(Q_0+B_0H)(I_r+H'H)^{-1/2}$ with
$H\in\mathbb R^{(T-r)\times r}$, write
$\varphi=(\theta',\operatorname{vec}(H)')'$ with true value $\varphi_0$, and
let
\[
    g_i(\varphi)
    =\begin{pmatrix}
        \partial\ell_{H,i}\{\theta,U(H)\}/\partial\theta\\[1mm]
        \partial\ell_{E,i}\{\theta,U(H)\}/\partial\operatorname{vec}(H)
     \end{pmatrix},
    \qquad
    G_N(\varphi)=\frac1N\sum_{i=1}^Ng_i(\varphi).
\]

Assumption~\ref{ass:entropy} in Appendix~\ref{app:proof_entropy_limit}
states the regularity conditions on these estimating equations.  Under it,
$\Gamma$ denotes the nonsingular limit of
$\partial G_N(\varphi_0)/\partial\varphi'$ and $\Omega$ the limit of the
average of the covariance matrices of the $g_i(\varphi_0)$.

\begin{theorem}[Entropy-adjusted estimator]
\label{thm:entropy_limit}
Consider model \eqref{eq:IFE_model} with $T>r$ fixed.  For part (a), assume
the score regularity in Assumption~\ref{ass:regular} at the true orbit and
the differentiability conditions of Theorem~\ref{thm:entropy}(c).
\begin{enumerate}
\item[(a)] For every $N$ and every configuration of loadings,
$\mathbb E\,G_N(\varphi_0)=0$ and $\Delta_N(\theta_0,Q_0)=0$.
\item[(b)] Under Assumption~\ref{ass:entropy_consistency}, let
$(\widehat\theta_E,\widehat U_E)$ satisfy
\[
\begin{split}
    Q_N(\widehat\theta_E,\widehat U_E)
    &\ge\sup_\theta Q_N(\theta,\widehat U_E)-o_p(1),\\
    L_N(\widehat\theta_E,\widehat U_E)
    &\ge\sup_U L_N(\widehat\theta_E,U)-o_p(1).
\end{split}
\]
Then $\widehat\theta_E\overset p\to\theta_0$ and
$\widehat U_E\widehat U_E'\overset p\to Q_0Q_0'$.
\item[(c)] If, in addition, Assumption~\ref{ass:entropy} holds, the
entropy is exact and the two maximisations in (b) are exact, then
$\sqrt N(\widehat\varphi-\varphi_0)\overset d\to
N\{0,\Gamma^{-1}\Omega(\Gamma^{-1})'\}$.
\item[(d)] If the second-order expansion of the divergence in
\eqref{eq:entropy_KL} may be differentiated and the expected orbit Hessians
converge to the corresponding block of $\Gamma$, then that block for the
orbit equation and the orbit parameter is
\[
    -\lim_{N\to\infty}(\mathcal A_N\otimes J_W).
\]
\end{enumerate}
\end{theorem}

\noindent{\em Proof.} See Appendix~\ref{app:proof_entropy_limit}.

Part (b) is the consistency of the joint system, and a converged alternating
algorithm of Section~\ref{subsec:entropy_estimator} has this form when both
steps attain their respective global maxima.  The argument does not need
$\overline{Q}_N$ or $\overline{L}_N$ to converge, which matters because both depend on
the loadings away from the true orbit.  Condition (i) is a uniform law of
large numbers that includes the simulation error of the entropy.
Condition (ii) is the identification condition of the system.  The gap is
zero at the truth by part (a). On the slice $U=Q_0$ it is at least the
quotient divergence of Proposition~\ref{prop:haar_reduction_principle}(a).
Condition (ii) requires uniform separation away from the truth as $N$
increases. Section~\ref{subsec:separable} derives such separation for a
single criterion under conditions on the design and error family.

\citet[Theorem~1]{Zhang2026IFE} gives the corresponding limit for the
second-moment system under errors that are independent and identically
distributed over time and a local identification condition; the present
result changes the orbit equation, drops that restriction on the errors, and
replaces local by global identification through the gap
\eqref{eq:equilibrium_gap}.  The limit is that of a just-identified system
and not that of a maximum likelihood estimator, because
\eqref{eq:entropy_system} is in general not the gradient of a single
criterion: the
covariance has the sandwich form, and $\Omega$ contains, besides the term
$\mathcal A_N\otimes J_W$ that mirrors $\Gamma$, a contribution from the
errors alone. By part (d), holding the structural parameter fixed, the
orbit criterion has negative definite limiting curvature if the limiting
second-moment matrix of the loadings and the location information of the
contrasts are both positive definite. Under an additional uniform
expansion for the simulated orbit score, simulation from $M_N$ error vectors independent of the data leaves
the limit in part (c) unchanged if $M_N/N\to\infty$; with $M_N=mN$ it
generally adds a covariance term of order $1/m$.

\subsection{Separable entropy and identification from the design}
\label{subsec:separable}

Assumption~\ref{ass:entropy_consistency}(ii) imposes a separation condition
on the population system. When the entropy separates, conditions on the
design and the error family give identification through a single criterion. Say that the
error family has \emph{separable contrast entropy} if there are functions
$a$ and $b$ such that
\begin{equation}
    \mathcal E(U;\eta)=a(U)+b(\eta)
    \qquad\text{for every }U\in\operatorname{Gr}(r,T)\text{ and every }\eta,
\label{eq:separable}
\end{equation}
and in that case let $P_N(\theta,U)=Q_N(\theta,U)+a(U)$, with expectation
$\overline{P}_N$. In this subsection $L_N$ uses the exact entropy
$\mathcal E(U;\eta)$. Let
\[
    s_i(\beta,U)=B_U'\{F\lambda_i+X_i(\beta_0-\beta)\}
\]
be the part of the systematic component of unit $i$ that the contrasts of a
candidate orbit retain at a candidate slope.

\begin{assumption}[Design and error family]
\label{ass:design}
\begin{enumerate}
\item[(i)] For every $\epsilon>0$ there are $\delta>0$ and $\pi>0$ such
that, for all large $N$ and every $(\beta,U)$ with
$\|\beta-\beta_0\|+\|UU'-Q_0Q_0'\|\ge\epsilon$, at least $\pi N$ units have
$\|s_i(\beta,U)\|\ge\delta$.
\item[(ii)] The set of values of $\eta$ is compact and
$\eta\mapsto g(\cdot;\eta)$ is continuous in $L^1$.  For every $U$, every
$s\in\mathbb R^{T-r}$ and every $\eta$, the densities
$h_U(\cdot-s;\eta_0)$ and $h_U(\cdot;\eta)$ coincide almost everywhere only
if $s=0$ and $\eta=\eta_0$.
\end{enumerate}
\end{assumption}

\begin{theorem}[Separable entropy]
\label{thm:separable}
Consider model \eqref{eq:IFE_model} with $T>r$ fixed.  Suppose that
$\mathcal E(U;\eta)$ is finite and continuously differentiable in $U$, that
$\ell_{H,i}$ is continuously differentiable in $(\theta,U)$, and that any
two points of $\Theta$ can be joined by a piecewise continuously
differentiable path in $\Theta$.
\begin{enumerate}
\item[(a)] The estimating functions of system \eqref{eq:entropy_system},
namely $\partial Q_N/\partial\theta$ and $D_UL_N$, are the gradient of a
single function on $\Theta\times\operatorname{Gr}(r,T)$ if and only if the
entropy is separable.  The function is then $P_N$, up to a constant.
\item[(b)] If the entropy is separable, exact maximisation at either step
of the alternating algorithm weakly increases $P_N$. For every $N$, every
$(\theta,U)$ and every configuration of loadings,
\begin{equation}
    \overline{P}_N(\theta,U)
    =-b(\eta_0)-\frac1N\sum_{i=1}^N
      \operatorname{KL}\big\{h_U(\cdot-s_i(\beta,U);\eta_0)\,\big\|\,
      h_U(\cdot;\eta)\big\}
    \le\overline{P}_N(\theta_0,Q_0).
\label{eq:potential_KL}
\end{equation}
\item[(c)] If the entropy is separable, the uniform law for $Q_N$ in
Assumption~\ref{ass:entropy_consistency}(i) and
Assumption~\ref{ass:design} hold, then every $(\widehat\theta,\widehat U)$
with
$P_N(\widehat\theta,\widehat U)\ge\sup_{\theta,U}P_N(\theta,U)-o_p(1)$
satisfies $\widehat\theta\overset p\to\theta_0$ and
$\widehat U\widehat U'\overset p\to Q_0Q_0'$.
\item[(d)] The entropy is separable, and Assumption~\ref{ass:design}(ii)
holds, in each of the following families:
(1) a known error law, with $b=0$;
(2) a scale family $\varepsilon_i=\sigma\varepsilon_i^\circ$, where
$\varepsilon_i^\circ$ has a known density on $\mathbb R^T$ and
$\sigma\in[\underline\sigma,\overline\sigma]$ with $\underline\sigma>0$, for
which $a(U)$ is the entropy of $B_U'\varepsilon_i^\circ$ and
$b(\sigma)=(T-r)\log\sigma$;
(3) an elliptical family $\varepsilon_i=\Sigma_0^{1/2}\zeta_i$, where
$\Sigma_0$ is a known positive definite matrix and $\zeta_i$ is spherically
symmetric with a density that belongs to an identified family indexed
continuously in $L^1$ by $\eta$ in a compact set. Here
$a(U)=\tfrac12\log\det(B_U'\Sigma_0B_U)$ and $b(\eta)$ is the entropy of
the first $T-r$ coordinates of $\zeta_i$.
\end{enumerate}
\end{theorem}

\noindent{\em Proof.} See Appendix~\ref{app:proof_separable}.

Part (b) extends the Kullback--Leibler identity in
Theorem~\ref{thm:entropy}(a) to joint variation in the structural
parameter and the orbit. Part (c) uses uniform separation of the expected
criterion to establish consistency for approximate global maximisers.
Assumption~\ref{ass:design}(i) requires a nonvanishing share of units to
retain systematic variation in the contrasts away from the truth. It
fails when a regressor lies in the factor space, in agreement with
Proposition~\ref{prop:geometric}. Assumption~\ref{ass:design}(ii)
identifies the error law from its contrasts even in the presence of an
unknown translation. The scale family leaves the shape of the error law and its
dependence over time unrestricted, provided they are known, and the
elliptical family covers Gaussian errors with covariance $\sigma^2\Sigma_0$
and multivariate Student errors with unknown scale and degrees of freedom.
When $\Sigma_0=I_T$ the function $a$ is constant, and the unadjusted
invariant likelihood is itself the single criterion.

For nonseparable entropy, identification requires a separate argument.
With no regressors and Gaussian errors of unrestricted covariance, an
adjustment to the error covariance absorbs the second moments of the factor
component, and the gap \eqref{eq:equilibrium_gap} vanishes at every orbit
(Proposition~\ref{prop:gaussian_gap} in Appendix~\ref{app:gaussian_gap}).
The frontier family used below lies between the two cases.  Its entropy is
separable in the common scale $\sigma$ but not in the ratio $\lambda$ or in
period-specific noise variances, so the condition remains a restriction on
the error family \citep[see also][]{BaiMones2026}.  In the designs used below, including one
calibrated to the application, a numerical check finds a positive gap at
every candidate orbit examined (Appendix~\ref{app:gap_check}).

\subsection{Simulation evidence}
\label{subsec:entropy_numerical}

Table~\ref{tab:ife_sim} reports an experiment for the one-factor frontier
model of \citet{Zhang2026IFE},
$y_{it}=f_t\lambda_i+v_{it}-u_{it}$ with $T=3$, factor vector
$F=(1,0.6,1.4)'$, loadings $\lambda_i\sim N(0.5,0.5^2)$, half-normal
$u_{it}$ with $\sigma_u=1$, and normal $v_{it}$.  In the first design the
noise is homoskedastic, with $\sigma_{v}=0.5$ in every period.  In the
second the noise standard deviations are $(0.3,0.5,0.9)$ across the three
periods, and the correct period-specific error law is used by every
estimator.  Three estimators are compared.  The raw estimator maximises the
Haar likelihood jointly over the structural parameter and the orbit.  The
second-moment estimator combines the Haar score for the structural parameter
with the leading eigenvector of the centred second-moment matrix.  The
entropy-adjusted estimator solves \eqref{eq:entropy_system}, with the
entropy computed from $8N$ simulated error vectors.  The orbit error is the
sine of the angle between the estimated and the true factor directions.

\begin{table}[t]
\centering
\caption{Estimators in the one-factor frontier model with $T=3$: mean and,
in parentheses, standard deviation over replications (200 for $N=500$, 100
for $N=2000$).  True values are $\sigma_{v1}=0.5$ (homoskedastic design) or
$0.3$ (heteroskedastic design) and $\sigma_u=1$.  The last column is the sine
of the angle between the estimated and true factor directions. The
integral over the loading is computed by the piecewise rule of
Appendix~\ref{app:quadrature}.}
\label{tab:ife_sim}
\begin{tabular}{llrccc}
\hline
Design & Estimator & $N$ & $\widehat\sigma_{v1}$ & $\widehat\sigma_u$ & Orbit error\\
\hline
Homoskedastic   & Raw Haar         & 500  & 0.500 (0.049) & 0.988 (0.076) & 0.041 (0.024)\\
                & Second moment    & 500  & 0.494 (0.050) & 0.998 (0.078) & 0.043 (0.023)\\
                & Entropy adjusted & 500  & 0.487 (0.052) & 1.009 (0.079) & 0.044 (0.023)\\
                & Raw Haar         & 2000 & 0.508 (0.026) & 0.986 (0.038) & 0.025 (0.014)\\
                & Second moment    & 2000 & 0.500 (0.026) & 0.998 (0.038) & 0.023 (0.012)\\
                & Entropy adjusted & 2000 & 0.495 (0.027) & 1.007 (0.039) & 0.025 (0.012)\\[1mm]
Heteroskedastic & Raw Haar         & 500  & 0.476 (0.143) & 0.743 (0.073) & 0.277 (0.043)\\
                & Second moment    & 500  & 0.468 (0.135) & 0.756 (0.077) & 0.262 (0.046)\\
                & Entropy adjusted & 500  & 0.281 (0.184) & 0.967 (0.108) & 0.091 (0.063)\\
                & Raw Haar         & 2000 & 0.502 (0.063) & 0.740 (0.037) & 0.277 (0.024)\\
                & Second moment    & 2000 & 0.499 (0.041) & 0.750 (0.040) & 0.264 (0.027)\\
                & Entropy adjusted & 2000 & 0.285 (0.109) & 0.998 (0.048) & 0.041 (0.022)\\
\hline
\end{tabular}
\end{table}

In the homoskedastic design the three estimators are close.  The drift of
the raw orbit score is small relative to the information from the loadings,
so the raw estimator shows only a slight bias in $\sigma_u$, and the entropy
adjustment brings no gain in precision over the second-moment equation when
the conditions of that equation hold.  The heteroskedastic design separates
them.  The results for the raw and second-moment estimators are consistent
with convergence to an incorrect orbit: the orbit error stays near $0.27$ as
$N$ increases from 500 to 2000 while its standard deviation falls, and the
wrong orbit carries the structural estimates with it, with $\sigma_u$
underestimated by a quarter.  The second-moment equation fails because the
noise covariance is not a multiple of the identity, and the raw likelihood
fails because the entropy of the contrasts now varies strongly with the
orbit.  The entropy-adjusted estimates are closer to the true values, and
its orbit error falls from $0.091$ to $0.041$, close to the rate $N^{-1/2}$.
The ranking is the same with a longer panel and with two factors
(Appendix~\ref{app:ife_rk}). The smallest noise standard deviation is weakly
identified at this panel length.  The entropy-adjusted estimate of
$\sigma_{v1}$ is at its lower bound in 41 of the 200 samples with $N=500$
and in 6 of the 100 with $N=2000$, which explains its large standard
deviation; its median is $0.30$ at both sample sizes, and $\sigma_u$ and the
orbit are not affected.  A Gauss--Hermite rule understates the likelihood
near that bound and hides these solutions
(Appendix~\ref{app:quadrature}).

\medskip
\noindent{\bf Scope.}
The adjustment uses a parametric error law. In the misspecification
experiments of Appendix~\ref{app:misspec}, the largest orbit error in the
large-sample entropy-adjusted fits is $0.08$ after rounding. The adjustment
applies to an orbit that is a free common parameter. An orbit tied to the
structural parameter, as in the dynamic panel, calls for the recentring of
Section~\ref{sec:dynamic}. A nearly singular second-moment matrix of the
loadings leaves some orbit directions with little curvature.


\section{Information loss and the limits of the reduction}
\label{sec:limits}

The invariant reduction is exact, but it is not free.  This section shows
what the quotient discards, when the loss can be recovered, and where
recovery stops.

\subsection{Information loss and recovery}
\label{subsec:information}

When H1--H3 hold, the quotient is a valid experiment, but it is a smaller
one.  Requirement H4 can fail in the quotient while the original data still
identify the target.  The simplest example has two periods and an asymmetric
error:
\begin{equation}
    Y_{it}=\alpha_i+Z_{it},
    \qquad t=1,2,
\label{eq:two_period_asym}
\end{equation}
where the $Z_{it}$ are independent across $i$ and $t$ and have a common
centred law with variance $v$, third moment $\kappa$ and finite sixth moment. Let
$D_i=Y_{i1}-Y_{i2}$, $A_i=D_i^2$ and $S_i=Y_{i1}+Y_{i2}$.

\begin{proposition}[Loss and recovery with two periods]
\label{prop:two_period}
In model \eqref{eq:two_period_asym}:
\begin{enumerate}
\item[(a)] The maximal invariant $D_i$ has the same distribution when the law
of $Z_{it}$ is replaced by the law of $-Z_{it}$.  The sign of $\kappa$ is
therefore not identified in the quotient.
\item[(b)] For every $\alpha_i$,
$\mathbb E(A_i-2v)=0$ and $\mathbb E\{(A_i-2v)S_i\}=2\kappa$.
\item[(c)] With a third period, the invariant statistic
$K_i=\tfrac32\sum_{t=1}^3(Y_{it}-\overline{Y}_i)^3$ has mean $\kappa$ and a
distribution that does not depend on $\alpha_i$.
\end{enumerate}
\end{proposition}

\noindent{\em Proof.} See Appendix~\ref{app:proof_two_period}.

The statistic $(A_i-2v)S_i$ in part (b) is not invariant, but its
expectation is free of $\alpha_i$, so it is a functional-differencing moment
in the sense of \citet{Bonhomme2012}.  Functions of the maximal invariant
are always of this kind, by Proposition~\ref{prop:haar_translation}(d), and
the example shows that the converse fails:
\begin{equation}
\{\text{invariant statistics}\}
\ \subsetneq\
\{\text{statistics whose mean is free of the incidental parameter}\}.
\label{eq:inv_subset_robust}
\end{equation}
The strict inclusion concerns classes of moments, whereas the
conditional-expectation identity of Theorem~\ref{thm:haar_orthogonal}(b)
concerns the projected scores: the orthogonalised score of any order differs
from the Haar score by a term with conditional mean zero given the maximal
invariant, a property that $(A_i-2v)S_i$ need not have.

This suggests an augmentation: combining the Haar score with moments of
the larger class restores local identification when the added moments supply
rank in the directions in which the quotient information is singular,
although a global alias such as the reflection in
Proposition~\ref{prop:two_period}(a) requires moments that separate the
aliased values themselves.  Part (c) shows another possibility: more
replication within the unit enlarges the quotient and restores the
information without leaving the invariant experiment.
Proposition~\ref{prop:sfm_orientation} gives an instance with economic
content, in which the lost sign is the orientation of a stochastic frontier.
The loss is specific to fixed effects: with random effects independent of
the errors, two periods can identify the error law under the conditions of
Kotlarski's identity \citep{ArellanoBonhomme2012,KatoSasakiUra2019}.

\subsection{Limits of recovery}
\label{subsec:boundaries}

Augmentation cannot restore information that the data do not contain.  Two
boundaries of this kind are relevant, one geometric and one asymptotic.

\medskip 
\noindent{\bf Structural variation inside the nuisance orbit.}
\begin{proposition}[Geometric boundary]
\label{prop:geometric}
Consider model \eqref{eq:orbit_model} with $\mu_i(\beta)=X_i\beta$, where
$X_i$ is a $T\times k$ matrix, and unrestricted $\xi_i$.
\begin{enumerate}
\item[(a)] If there is $h\neq0$ with
$X_ih\in\operatorname{col}(A)$ for every $i$, then, provided both values
belong to the parameter space, $\beta$ and $\beta+h$ generate the same
distribution of the full data for suitable incidental parameters, and
$\beta$ is not identified in the original experiment.
\item[(b)] Under Assumption~\ref{ass:regular}, the quotient information for
$\beta$ at given $\eta$ is $\sum_iX_i'BJ_WB'X_i$, where
$J_W=\mathbb E\{s_w(W_i)s_w(W_i)'\}$, $W_i=B'\varepsilon_i$ and
$s_w(w)=\nabla_w\log h(w;\eta)$.  When $J_W$ is positive definite, this
matrix is singular if and only if the condition in (a) holds.
\end{enumerate}
\end{proposition}

\noindent{\em Proof.} See Appendix~\ref{app:proof_geometric}.

For slope coefficients and a known error law with positive definite
location information, the null directions of the quotient information are
exactly the directions that the original experiment cannot identify.  The
familiar cases follow: a time-invariant regressor with unit effects, a
regressor in the span of unit and time dummies, and a regressor lying in the
factor space of \eqref{eq:IFE_model} \citep{Zhang2026IFE}.  What matters is
the codimension of the orbit, not the number of periods as such.  A value
that the quotient cannot distinguish from the truth, but the original
experiment can, is a loss caused by the reduction and may be recovered by
augmentation; a value that the original experiment cannot distinguish from
the truth cannot be recovered by any method.

\medskip 
\noindent{\bf Asymptotic non-separation.}
Identification for each fixed value of the incidental parameters does not
imply that the target can be estimated uniformly over them.  The two-period
model shows that uniform recovery can fail even though a moment free of the
incidental parameters exists.  Let $F_+$ be a centred law with a density,
variance $v>0$, third moment $\kappa>0$ and finite sixth moment, and let
$F_-$ be the law of $-Z$ when $Z\sim F_+$.  Consider testing $F_+$ against
$F_-$ in \eqref{eq:two_period_asym} over the class
\[
\mathcal A_N(H_N)
=\Big\{\alpha\in\mathbb R^N:
\frac1N\sum_{i=1}^N(\alpha_i-\overline{\alpha}_N)^2\le H_N\Big\},
\]
and let
\[
R_N^*
=\inf_\phi\max\Big\{
\sup_{\alpha\in\mathcal A_N(H_N)}\mathbb E_{F_+,\alpha}\phi,\
\sup_{\alpha\in\mathcal A_N(H_N)}\mathbb E_{F_-,\alpha}(1-\phi)\Big\}
\]
be the minimax error over tests $\phi$ that assign probability $\phi$ to
deciding for $F_-$. The estimator of the third moment
suggested by Proposition~\ref{prop:two_period}(b) is
\begin{equation}
\widehat\kappa_N
=\frac{1}{2(N-1)}\sum_{i=1}^N(A_i-\overline{A})(S_i-\overline{S}).
\label{eq:kappa_hat_boundary}
\end{equation}

\begin{theorem}[Uniform recovery and its limit]
\label{thm:minimax}
In the two-period model with error law $F_+$ or $F_-$:
\begin{enumerate}
\item[(a)] For every $\alpha\in\mathbb R^N$ and $N\ge2$,
$\widehat\kappa_N$ is unbiased for the third moment of the error law, and
there is a constant $c_F$ depending only on $F_+$ such that
\[
\sup_{\alpha\in\mathcal A_N(H_N)}
\operatorname{Var}(\widehat\kappa_N)
\le c_F\,\frac{1+H_N}{N}.
\]
\item[(b)] The test that decides for $F_+$ when $\widehat\kappa_N>0$ has
maximal error over $\mathcal A_N(H_N)$ of at most
$c_F(1+H_N)/(N\kappa^2)$.  Hence $R_N^*\to0$ whenever $H_N=o(N)$.
\item[(c)] If the effects are unrestricted, so that the class is
$\mathbb R^N$, then $R_N^*=1/2$ for every $N\ge1$.
\end{enumerate}
\end{theorem}

\noindent{\em Proof.} See Appendix~\ref{app:proof_minimax}.

Part (c) is an impossibility result for the original experiment, not for the
quotient.  The moment in Proposition~\ref{prop:two_period}(b) has mean free of
the effects, but its variance is not, and effects that are sufficiently
dispersed mask the sign of the skewness completely.  Parts (a) and (b) show
that recovery is uniform as long as the cross-sectional dispersion of the
effects grows more slowly than $N$.  Under an additional tail condition this
rate cannot be improved: once the dispersion is of order $N$, every test has
a worst-case error probability bounded away from zero
(Proposition~\ref{prop:minimax_sharp} in Appendix~\ref{app:minimax_sharp}).
With three periods, Proposition~\ref{prop:two_period}(c) supplies an
effect-invariant estimator of $\kappa$, whose variance does not depend on
the effects.

\subsection{Exactness and information in a stochastic frontier model}
\label{sec:frontier}

The stochastic frontier model with fixed effects shows both sides of the
trade between exactness and information.  Let
\begin{equation}
    y_{it}=\alpha_i+x_{it}'\beta+v_{it}-\varsigma\,u_{it},
    \qquad i=1,2,\ldots,N,\quad t=1,2,\ldots,T,
\label{eq:sfm_model}
\end{equation}
with noise $v_{it}\sim N(0,\sigma_v^2)$ and transient inefficiency
$u_{it}=|u^\circ_{it}|$, $u^\circ_{it}\sim N(0,\sigma_u^2)$, all independent
across $i$ and $t$ and of the strictly exogenous covariates.  The sign
$\varsigma$ is $1$ for a production frontier and $-1$ for a cost frontier
\citep{AignerLovellSchmidt1977}.  The effects $\alpha_i$ are unrestricted,
and the incidental-parameter problem they create is a recognised difficulty
in applied work \citep{Greene2005}.  With
$\sigma^2=\sigma_v^2+\sigma_u^2$ and $\lambda=\sigma_u/\sigma_v$, the composed
error $\varepsilon_{it}=v_{it}-\varsigma u_{it}$ has density
\[
    f(\varepsilon)
    =\frac2\sigma\,\phi\!\left(\frac\varepsilon\sigma\right)
     \Phi\!\left(-\frac{\varsigma\lambda\varepsilon}{\sigma}\right),
\]
and the structural parameter is $\theta=(\beta',\sigma,\lambda)'$.

The model is the translation model \eqref{eq:orbit_model} with
$A=\mathbf 1_T$, so requirements H1--H3 hold for every $T\ge2$.  Its Haar
likelihood is known: \citet{HongKingZhang2026} obtain it by integrating the
unit likelihood over the location group and use it as the restricted
likelihood of score tests for inefficiency determinants.  The same
likelihood is the within-transformed likelihood of \citet{ChenSchmidtWang2014} and the integrated likelihood of
\citet{BellioGrassetti2024}; related transformation approaches are those of
\citet{WangHo2010} and \citet{BelottiIlardi2018}.  With $e_i=y_i-X_i\beta$,
$M=I_T-T^{-1}\mathbf 1_T\mathbf 1_T'$, $\widetilde{e}_i=Me_i$ and
$Z\sim N(0,1)$, and with the normalisation of \eqref{eq:haar_lik},
\begin{equation}
    L_{H,i}(\theta)
    =2^T(2\pi\sigma^2)^{-(T-1)/2}
     \exp\!\left\{-\frac{\widetilde{e}_i'\widetilde{e}_i}{2\sigma^2}\right\}
     \mathbb E_Z\!\left[\prod_{t=1}^T
     \Phi\!\left\{-\varsigma\lambda\left(\frac{\widetilde{e}_{it}}\sigma
     +\frac Z{\sqrt T}\right)\right\}\right].
\label{eq:sfm_haar}
\end{equation}
A Gauss--Hermite rule evaluates the one-dimensional expectation.  By
Proposition~\ref{prop:haar_translation}, \eqref{eq:sfm_haar} is the density
of the contrasts $B_T'y_i$, so its scores are centred for every
$T\ge2$, and Theorem~\ref{thm:haar_orthogonal}(a)--(b) and
Proposition~\ref{prop:haar_reduction_principle} apply.  The geometric boundary of
Proposition~\ref{prop:geometric} is also known in this model: persistent
inefficiency, a term that does not vary over time, lies in the orbit of the
unit effect and cannot be separated from it \citep{HongKingZhang2026}.

The information boundary of Section~\ref{subsec:information} has a
consequence for this model that appears not to have been noted.

\begin{proposition}[Orientation of the frontier]
\label{prop:sfm_orientation}
In model \eqref{eq:sfm_model}:
\begin{enumerate}
\item[(a)] For $T=2$, \eqref{eq:sfm_haar} takes the same value for
$\varsigma=1$ and $\varsigma=-1$.
\item[(b)] For $T\ge3$, the statistic $K_i$ of
Proposition~\ref{prop:two_period}(c), computed from three components of
$e_i(\beta_0)$, has mean
$-\varsigma\,\sigma_u^3\sqrt{2/\pi}\,(4/\pi-1)$.
\end{enumerate}
\end{proposition}

\noindent{\em Proof.} See Appendix~\ref{app:proof_sfm_orientation}.

With two periods the quotient cannot tell a production frontier from a cost
frontier, and by Theorem~\ref{thm:minimax}(c) neither can any test based
on the original data when the effects are unrestricted.  A third period
restores the orientation through the skewness of the contrasts.

\medskip
\noindent{\bf Exactness: the residual drift at the second order.}
Because the frontier model is a location model in $\alpha_i$, the
orthogonalised score \eqref{eq:projected_score} depends on the data and on
$\alpha_i$ only through $e_i(\beta)-\alpha_i\mathbf 1_T$, and its projection
coefficients depend on $\theta$ alone.  A feasible version
replaces $\alpha_i$ by its unit-level maximum likelihood estimate.  For
$q=1$ this returns the profile score, because the first
Bhattacharyya function vanishes at the unit-level estimate.  For $q=2$ it
gives a corrected score that has no closed form but is easily simulated.

Table~\ref{tab:sfm_scores} reports, for $\sigma_v=0.5$ and $\sigma_u=1$, the
mean of each estimating function at the true parameter divided by its
standard deviation, from $10^6$ simulated units.  For a component with
standardised mean $\delta$, the mean of the summed estimating function
equals one of its standard deviations in absolute value when
$N=\delta^{-2}$; the table reports this number as $N^*$, computed from the
unrounded means.

\begin{table}[ht]
\centering
\caption{Estimating functions for $(\sigma,\lambda)$ at the true parameter in
the frontier model.  The profile and second-order scores are evaluated at the
unit-level maximum likelihood estimate of the effect.  The Monte Carlo
standard error of each standardised mean is $0.001$.  The last two columns
give the variance of the Haar score as a share of the variance of the first-
and second-order orthogonalised scores evaluated at the true effect.}
\label{tab:sfm_scores}
\begin{tabular}{rlcrrrcrrccc}
\hline
 & && \multicolumn{3}{c}{Standardised mean} && \multicolumn{2}{c}{$N^*$}
 && \multicolumn{2}{c}{Variance retained}\\
\cline{4-6}\cline{8-9}\cline{11-12}
$T$ & && Profile & 2nd order & Haar && Profile & 2nd order && 1st order & 2nd order\\
\hline
3  & $\sigma$  && $-0.515$ & $-0.025$ & $0.000$  && 4  & 1{,}600  && 0.66 & 0.96\\
3  & $\lambda$ && $0.708$  & $0.032$  & $0.001$  && 2  & 980      && 0.40 & 0.60\\
5  & $\sigma$  && $-0.357$ & $-0.008$ & $0.002$  && 8  & 15{,}000 && 0.80 & 0.99\\
5  & $\lambda$ && $0.423$  & $0.009$  & $-0.002$ && 6  & 13{,}000 && 0.59 & 0.75\\
10 & $\sigma$  && $-0.237$ & $-0.003$ & $0.000$  && 18 & 86{,}000 && 0.90 & 1.00\\
10 & $\lambda$ && $0.262$  & $0.004$  & $0.000$  && 15 & 57{,}000 && 0.78 & 0.88\\
\hline
\end{tabular}
\end{table}

The profile score is substantially distorted at these panel lengths: its
drift dominates sampling noise once there are more than a handful of units,
even at $T=10$.  Second-order orthogonalisation removes most of the drift
but not all of it, in contrast to the exact equality in the Gaussian case of
Theorem~\ref{thm:haar_orthogonal}(c).  The residual drift becomes comparable
to sampling variation in the summed scores at about a thousand units when
$T=3$, and at tens of thousands when $T=5$ or $T=10$.  The Haar score shows
no drift, as Proposition~\ref{prop:haar_translation}(c) requires.  This is
the sense in which the Haar reduction does better than orthogonalisation of
a fixed order, and it is a limited one: the gain over the second order
matters for short panels with large cross-sections.

In the table the unit-level estimate comes from the same observations.  The
complete procedure of \citet{BonhommeJochmansWeidner2026} uses preliminary
estimates from an independent sample.  With the periods split into halves
and $N=1000$ (Appendix~\ref{app:sfm_split}), its second-order estimator of
$\lambda=2$ has mean $1.27$ at $T=4$, $1.82$ at $T=10$ and $1.95$ at $T=20$,
against $2.03$, $2.02$ and $2.00$ for the Haar estimator, and at $T=20$ the
two are equally precise.

\medskip
\noindent{\bf The price: information discarded by the quotient.}
The last two columns of Table~\ref{tab:sfm_scores} measure the inequality in
Theorem~\ref{thm:haar_orthogonal}(b).  For the scale $\sigma$ the quotient
gives up little.  For $\lambda$, which governs the split between noise and
inefficiency, the Haar score retains 40 per cent of the first-order variance
and 60 per cent of the second-order variance at $T=3$, and 78 and 88 per
cent at $T=10$.  Both benchmarks are infeasible at fixed $T$, since they are
evaluated at the true effect, but they show where the information lies: part
of what the data say about the asymmetry of the error is carried by the
level of the unit, and the quotient discards the level.  Exactness is
nevertheless what makes estimation possible at these panel lengths.  For
$T=3$ and $T=5$ the criterion of the maximum likelihood estimator that keeps
the effects as parameters increases as $\sigma_v\to0$, so that one of the
two error components disappears from the fitted model, while the Haar
estimator has little bias at each panel length
(Appendix~\ref{app:sfm_estimators}).

%

\section{An application to rice farming in Vietnam}
\label{sec:application}

Does the direction along which a household effect acts matter in practice?
A fixed effect in a production frontier is assumed to shift output by the
same amount in every period, whereas the interactive model
\eqref{eq:IFE_model} lets the data decide.  This section applies the
entropy-adjusted estimator to rice production by households in the Viet Nam
Access to Resources Household Survey \citep{BrandtTarp2017}, using the
2008--2016 panel files that UNU-WIDER distributes for research
\citep{TarpSinghal2020}.

The survey follows rural households in twelve provinces over five biennial
waves, from 2008 to 2016.  From its balanced panel of 2{,}131 households the
sample keeps those that grew rice in every wave, reported positive
expenditure on seed, fertiliser and pesticide, had at least one member
working in agriculture, and had a yield between one and twelve tonnes per
hectare, which leaves $N=800$ households and $T=5$ waves.  Output is the
quantity of rice harvested.  The inputs are the area under rice, the number
of household members working in agriculture, expenditure on seed, on
fertiliser and on pesticide, and expenditure on hired services, which
combines hired labour with the rental of machinery and draught animals.
Expenditures are deflated by the consumer price index.  Hired-services
expenditure is zero in 13 per cent of observations; for these the log
variable is set to zero and an indicator for zero expenditure is included.

The model is
\[
    y_{it}=g(x_{it};\beta)+\gamma_t+\lambda_if_t+v_{it}-u_{it},
\]
where $y_{it}$ is log output, $g$ is a translog function
\citep{ChristensenJorgensonLau1973} of the six inputs with logarithms centred at their sample means, $\gamma_t$ is a wave effect,
$v_{it}\sim N(0,\sigma_{vt}^2)$ with a separate variance in each wave, and
$u_{it}$ is half-normal with scale $\sigma_u$.  The loading $\lambda_i$
collects what is fixed about a household, including land quality, skill and
any persistent inefficiency, and $f_t$ is the weight with which it acts in
wave $t$.  This is the frontier with time-varying individual effects of
\citet{LeeSchmidt1993}, estimated as in \citet{Zhang2026IFE} but with
noise variances that differ across waves, which his second-moment equation
does not allow.  The ordinary fixed-effects frontier is the case of equal
weights.

Four estimators are compared: the Haar likelihood with equal weights, and
the second-moment, raw and entropy-adjusted estimators of
Section~\ref{subsec:entropy_numerical} with the direction 
$F=(f_1,\ldots,f_5)'$ unrestricted.  For the entropy-adjusted
estimator, standard errors use the sandwich formula of
Theorem~\ref{thm:entropy_limit}(c), treating the simulated entropy as
exact.  This approximation is examined in Appendix~\ref{app:rice_coverage}.

\begin{table}[t]
\centering
\caption{Rice frontier for 800 Vietnamese households in five waves.
Standard errors, in parentheses, are reported for the equal-weights and
entropy-adjusted estimators.  The factor direction is normalised to unit
length; equal weights correspond to $0.447$ in every wave.}
\label{tab:rice}
\begin{tabular}{lcccc}
\hline
 & Equal weights & Second moment & Raw Haar & Entropy adjusted\\
\hline
\multicolumn{5}{l}{\emph{Output elasticities at the sample mean}}\\
~~~Land & 0.758 (0.018) & 0.739 & 0.743 & 0.744 (0.017)\\
~~~Labour & 0.057 (0.014) & 0.052 & 0.048 & 0.046 (0.012)\\
~~~Seed & 0.060 (0.008) & 0.053 & 0.050 & 0.049 (0.008)\\
~~~Fertiliser & 0.073 (0.009) & 0.067 & 0.063 & 0.062 (0.009)\\
~~~Pesticide & 0.013 (0.007) & 0.011 & 0.013 & 0.015 (0.006)\\
~~~Hired services & 0.037 (0.008) & 0.037 & 0.037 & 0.039 (0.008)\\
~~~Returns to scale & 0.998 (0.018) & 0.958 & 0.955 & 0.955 (0.016)\\[1mm]
\multicolumn{5}{l}{\emph{Error components}}\\
~~~$\sigma_u$ & 0.278 (0.015) & 0.302 & 0.306 & 0.313 (0.014)\\
~~~$\sigma_v$ (2008) & 0.170 (0.020) & 0.138 & 0.134 & 0.130 (0.015)\\
~~~$\sigma_v$ (2010) & 0.178 (0.012) & 0.161 & 0.159 & 0.158 (0.011)\\
~~~$\sigma_v$ (2012) & 0.109 (0.017) & 0.091 & 0.093 & 0.083 (0.031)\\
~~~$\sigma_v$ (2014) & 0.171 (0.013) & 0.131 & 0.122 & 0.111 (0.032)\\
~~~$\sigma_v$ (2016) & 0.111 (0.019) & 0.101 & 0.098 & 0.088 (0.025)\\[1mm]
\multicolumn{5}{l}{\emph{Factor direction}}\\
~~~2008 & 0.447 & 0.333 & 0.269 & 0.246 (0.038)\\
~~~2010 & 0.447 & 0.352 & 0.314 & 0.247 (0.049)\\
~~~2012 & 0.447 & 0.498 & 0.489 & 0.519 (0.046)\\
~~~2014 & 0.447 & 0.593 & 0.651 & 0.638 (0.033)\\
~~~2016 & 0.447 & 0.406 & 0.408 & 0.448 (0.042)\\[1mm]
Haar log-likelihood & 278.3 & 337.7 & 344.1 & 338.4\\
\hline
\end{tabular}
\end{table}

Table~\ref{tab:rice} gives three findings, and two further exercises
examine them.  First, equal weights are rejected.  The entropy-adjusted direction puts a weight of about $0.25$ on
the household effect in 2008 and 2010 and between $0.45$ and $0.64$ from
2012 onwards, and a Wald test of equal weights gives $13.4$ on four degrees
of freedom, with $p$-value $0.010$.  Under the model, the estimated
contribution of persistent household heterogeneity is roughly twice as large in the later waves as in the
earlier ones.

Second, the restriction is not innocuous for the structural parameters.
With equal weights the returns to scale are $1.00$.  With the direction
estimated they are $0.955$ with standard error $0.016$, significantly below
one.  The estimated scale of transient inefficiency rises from $0.278$ to
$0.313$, the noise standard deviation falls in every wave, and the implied
mean of $\exp(-u_{it})$ falls from $0.81$ to $0.79$.  A Wald test rejects
Cobb--Douglas ($81.1$ on 21 degrees of freedom).

Third, on these data the three estimated directions are broadly similar.
Their components differ by at most $0.11$, compared with marginal standard
errors of about $0.04$ for the entropy-adjusted estimator, although this
comparison is not a formal test of equality across estimators.  Two features
of the data help explain the similarity.  The estimated noise standard
deviations differ across waves by a factor of less than two (a Wald test of
their equality gives $9.5$ on four degrees of freedom, $p$-value $0.051$),
and the household effect is strong: the leading eigenvalue of the
second-moment matrix of the residuals is $0.58$, against $0.04$ to $0.07$
for the others.  Starting the algorithm from the equal-weights estimate and
from the second-moment estimate gave the same entropy-adjusted solution.

How fragile is that agreement?  Appendix~\ref{app:rice_contaminated} adds
independent normal errors with standard deviation $0.25$ to log output in
the first two waves, as a stylised model of additional recall error in the
early survey rounds.  The model stays within the same family, and only the
noise standard deviations of the two waves rise, to about three times those
of the later waves.  The entropy-adjusted direction stays within $0.06$ of
the original estimate in every component, and the returns to scale remain
at $0.96$.  The second-moment and raw estimators now place almost all the
weight of the household effect on the contaminated waves, and both report
returns to scale of about $1.03$ and a smaller inefficiency scale.  This is
a targeted sensitivity exercise, rather than evidence of robustness to
arbitrary measurement error.

In designs calibrated to the estimates of Table~\ref{tab:rice}
(Appendix~\ref{app:rice_coverage}) the reported standard errors of the
direction and of the returns to scale are accurate, that of $\sigma_u$ is
about a fifth too small, and the Wald test of equal weights is somewhat
liberal at the 5 per cent level; the simulated $p$-value of the statistic for the
full sample is $0.01$.  The evidence against equal weights is stronger
among the larger farms (Appendix~\ref{app:rice_contaminated}), and
Appendix~\ref{app:rice_quadrature} checks the quadrature at the reported
estimates.  One limit
of the analysis should be recorded: by Proposition~\ref{prop:geometric}, persistent inefficiency is part of the
loading and cannot be separated from other fixed household characteristics,
so $\sigma_u$ measures transient inefficiency only.


\section{Conclusion}
\label{sec:conclusion}

Haar integration solves an incidental-parameter problem exactly when the
incidental parameters act through a known group that leaves the structural
parameter fixed and when the resulting quotient is proper, identifying and
informative.  For a known translation orbit the invariant score is
the conditional expectation, given the maximal invariant, of the
orthogonalised score of every order.  When the orbit moves
with the structural parameter, as in the dynamic panel model, recentring the
integrated score reproduces the orthogonalised score equations at the
estimated effects.  Exactness has a price: the quotient gives up information that only
the discarded coordinates carry.  Some of it can be recovered from moments whose mean is free of the
effects, but only while the cross-sectional variance of the effects is of smaller
order than the number of units.

The case that the classical construction does not cover is an orbit that is
itself an unknown parameter.  The entropy adjustment handles it within the
likelihood: the orbit-score drift is minus the gradient of the entropy of
the retained contrasts, and adding that entropy gives a criterion whose
expectation, at the true structural parameter, is uniquely maximised at the
true orbit when the loadings span it.  Under an equilibrium-identification
condition the resulting estimator is consistent and asymptotically normal
for fixed panel length, and it does not require the errors to be Gaussian,
homoskedastic or independent over time.  When the entropy separates into an
orbit term and an error-law term, as in scale and elliptical families, a
design condition replaces the equilibrium condition.  In the rice-farming application the estimator
rejects equal weights of the household effect across survey waves, and
relaxing them lowers the estimated returns to scale from one to about $0.96$.

Three extensions would widen the use of the method: primitive conditions for a unique
equilibrium when the entropy is not separable, as in the frontier family
with unknown shape; more efficient entropy calculations for several factors;
and a version that does not require a parametric error law, since under a
misspecified law the estimator converges to a pseudo-true orbit
(Appendix~\ref{app:misspec}).  Whether a given incidental-parameter
problem is easy or hard is decided by the geometry of the orbit and by the
information that lies outside it.

\baselineskip 0.65cm
\setlength{\bibsep}{0pt}
\bibliography{references/Haar}

@article{NeymanScott1948,
  author  = {Neyman, Jerzy and Scott, Elizabeth L.},
  title   = {Consistent Estimates Based on Partially Consistent Observations},
  journal = {Econometrica},
  year    = {1948},
  volume  = {16},
  number  = {1},
  pages   = {1--32}
}

@article{Lancaster2000,
  author  = {Lancaster, Tony},
  title   = {The Incidental Parameter Problem Since 1948},
  journal = {Journal of Econometrics},
  year    = {2000},
  volume  = {95},
  number  = {2},
  pages   = {391--413}
}

@article{Lancaster2002,
  author  = {Lancaster, Tony},
  title   = {Orthogonal Parameters and Panel Data},
  journal = {Review of Economic Studies},
  year    = {2002},
  volume  = {69},
  number  = {3},
  pages   = {647--666}
}

@article{HahnNewey2004,
  author  = {Hahn, Jinyong and Newey, Whitney K.},
  title   = {Jackknife and Analytical Bias Reduction for Nonlinear Panel Models},
  journal = {Econometrica},
  year    = {2004},
  volume  = {72},
  number  = {4},
  pages   = {1295--1319}
}

@article{Bonhomme2012,
  author  = {Bonhomme, St{\'e}phane},
  title   = {Functional Differencing},
  journal = {Econometrica},
  year    = {2012},
  volume  = {80},
  number  = {4},
  pages   = {1337--1385}
}

@article{ArellanoBonhomme2012,
  author  = {Arellano, Manuel and Bonhomme, St{\'e}phane},
  title   = {Identifying Distributional Characteristics in Random Coefficients Panel Data Models},
  journal = {Review of Economic Studies},
  year    = {2012},
  volume  = {79},
  number  = {3},
  pages   = {987--1020}
}

@unpublished{KatoSasakiUra2019,
  author = {Kato, Kengo and Sasaki, Yuya and Ura, Takuya},
  title  = {Inference Based on Kotlarski's Identity},
  note   = {Working paper, arXiv:1808.09375v3},
  year   = {2019}
}

@article{Chamberlain2010,
  author  = {Chamberlain, Gary},
  title   = {Binary Response Models for Panel Data: Identification and Information},
  journal = {Econometrica},
  year    = {2010},
  volume  = {78},
  number  = {1},
  pages   = {159--168}
}

@article{Moreira2009,
  author  = {Moreira, Marcelo J.},
  title   = {A Maximum Likelihood Method for the Incidental Parameter Problem},
  journal = {The Annals of Statistics},
  year    = {2009},
  volume  = {37},
  number  = {6A},
  pages   = {3660--3696}
}

@article{ChamberlainMoreira2009,
  author  = {Chamberlain, Gary and Moreira, Marcelo J.},
  title   = {Decision Theory Applied to a Linear Panel Data Model},
  journal = {Econometrica},
  year    = {2009},
  volume  = {77},
  number  = {1},
  pages   = {107--133}
}

@techreport{HallinMoreiraOnatski2013,
  author      = {Hallin, Marc and Moreira, Marcelo J. and Onatski, Alexei},
  title       = {Group Invariance, Likelihood Ratio Tests, and the Incidental Parameter Problem in a High-Dimensional Linear Model},
  institution = {SFB 823, TU Dortmund},
  type        = {Discussion Paper},
  number      = {01/2013},
  year        = {2013}
}

@unpublished{BonhommeJochmansWeidner2026,
  author = {Bonhomme, St{\'e}phane and Jochmans, Koen and Weidner, Martin},
  title  = {A {Neyman}-Orthogonalization Approach to the Incidental Parameter Problem in Likelihood Models},
  note   = {arXiv:2412.10304v4, August 17, 2026},
  year   = {2026}
}

@unpublished{BonhommeJochmansWeidnerOrthogonal2026,
  author = {Bonhomme, St{\'e}phane and Jochmans, Koen and Weidner, Martin},
  title  = {Orthogonal Moments in Likelihood Models},
  note   = {TSE Working Paper 26-1774, arXiv:2609.31561},
  year   = {2026},
  url    = {https://arxiv.org/abs/2609.31561}
}

@article{WatermanLindsay1996,
  author  = {Waterman, Richard P. and Lindsay, Bruce G.},
  title   = {Projected Score Methods for Approximating Conditional Scores},
  journal = {Biometrika},
  year    = {1996},
  volume  = {83},
  number  = {1},
  pages   = {1--13}
}

@article{LiLindsayWaterman2003,
  author  = {Li, Haihong and Lindsay, Bruce G. and Waterman, Richard P.},
  title   = {Efficiency of Projected Score Methods in Rectangular Array Asymptotics},
  journal = {Journal of the Royal Statistical Society: Series B},
  year    = {2003},
  volume  = {65},
  number  = {1},
  pages   = {191--208}
}

@incollection{ArellanoHahn2007,
  author    = {Arellano, Manuel and Hahn, Jinyong},
  title     = {Understanding Bias in Nonlinear Panel Models: Some Recent Developments},
  booktitle = {Advances in Economics and Econometrics: Ninth World Congress},
  editor    = {Blundell, Richard and Newey, Whitney K. and Persson, Torsten},
  volume    = {III},
  publisher = {Cambridge University Press},
  year      = {2007}
}

@article{FernandezValWeidner2018,
  author  = {Fern{\'a}ndez-Val, Iv{\'a}n and Weidner, Martin},
  title   = {Fixed Effects Estimation of Large-{T} Panel Data Models},
  journal = {Annual Review of Economics},
  year    = {2018},
  volume  = {10},
  pages   = {109--138}
}

@article{HahnKuersteiner2011,
  author  = {Hahn, Jinyong and Kuersteiner, Guido},
  title   = {Bias Reduction for Dynamic Nonlinear Panel Models with Fixed Effects},
  journal = {Econometric Theory},
  year    = {2011},
  volume  = {27},
  number  = {6},
  pages   = {1152--1191}
}

@article{DhaeneJochmans2015,
  author  = {Dhaene, Geert and Jochmans, Koen},
  title   = {Split-Panel Jackknife Estimation of Fixed-Effect Models},
  journal = {The Review of Economic Studies},
  year    = {2015},
  volume  = {82},
  number  = {3},
  pages   = {991--1030}
}

@article{FernandezValWeidner2016,
  author  = {Fern{\'a}ndez-Val, Iv{\'a}n and Weidner, Martin},
  title   = {Individual and Time Effects in Nonlinear Panel Models with Large {N}, {T}},
  journal = {Journal of Econometrics},
  year    = {2016},
  volume  = {192},
  number  = {1},
  pages   = {291--312}
}

@article{BarndorffNielsen1983,
  author  = {Barndorff-Nielsen, Ole E.},
  title   = {On a Formula for the Distribution of the Maximum Likelihood Estimator},
  journal = {Biometrika},
  year    = {1983},
  volume  = {70},
  number  = {2},
  pages   = {343--365}
}

@article{CoxReid1987,
  author  = {Cox, D. R. and Reid, N.},
  title   = {Parameter Orthogonality and Approximate Conditional Inference},
  journal = {Journal of the Royal Statistical Society: Series B},
  year    = {1987},
  volume  = {49},
  number  = {1},
  pages   = {1--39}
}

@article{Sartori2003,
  author  = {Sartori, Nicola},
  title   = {Modified Profile Likelihoods in Models with Stratum Nuisance Parameters},
  journal = {Biometrika},
  year    = {2003},
  volume  = {90},
  number  = {3},
  pages   = {533--549}
}

@article{KalbfleischSprott1970,
  author  = {Kalbfleisch, J. D. and Sprott, D. A.},
  title   = {Application of Likelihood Methods to Models Involving Large Numbers of Parameters},
  journal = {Journal of the Royal Statistical Society: Series B},
  year    = {1970},
  volume  = {32},
  number  = {2},
  pages   = {175--208}
}

@article{BergerLiseoWolpert1999,
  author  = {Berger, James O. and Liseo, Brunero and Wolpert, Robert L.},
  title   = {Integrated Likelihood Methods for Eliminating Nuisance Parameters},
  journal = {Statistical Science},
  year    = {1999},
  volume  = {14},
  number  = {1},
  pages   = {1--28}
}

@article{ArellanoBonhomme2009,
  author  = {Arellano, Manuel and Bonhomme, St{\'e}phane},
  title   = {Robust Priors in Nonlinear Panel Data Models},
  journal = {Econometrica},
  year    = {2009},
  volume  = {77},
  number  = {2},
  pages   = {489--536}
}

@incollection{Neyman1959,
  author    = {Neyman, J.},
  title     = {Optimal Asymptotic Tests of Composite Statistical Hypotheses},
  booktitle = {Probability and Statistics: The Harald Cram{\'e}r Volume},
  editor    = {Grenander, U.},
  publisher = {Almqvist and Wiksell},
  address   = {Stockholm},
  year      = {1959},
  pages     = {213--234}
}

@article{SmallMcLeish1989,
  author  = {Small, Christopher G. and McLeish, D. L.},
  title   = {Projection as a Method for Increasing Sensitivity and Eliminating Nuisance Parameters},
  journal = {Biometrika},
  year    = {1989},
  volume  = {76},
  number  = {4},
  pages   = {693--703}
}

@article{ChernozhukovEtAl2018,
  author  = {Chernozhukov, Victor and Chetverikov, Denis and Demirer, Mert and Duflo, Esther and Hansen, Christian and Newey, Whitney and Robins, James},
  title   = {Double/Debiased Machine Learning for Treatment and Structural Parameters},
  journal = {The Econometrics Journal},
  year    = {2018},
  volume  = {21},
  number  = {1},
  pages   = {C1--C68}
}

@misc{BonhommeJochmansNeweyWeidner2026,
  author       = {Bonhomme, St{\'e}phane and Jochmans, Koen and Newey, Whitney K. and Weidner, Martin},
  title        = {Higher-Order {N}eyman Orthogonality in Moment-Condition Models},
  howpublished = {arXiv:2605.10842},
  year         = {2026}
}

@article{Andersen1970,
  author  = {Andersen, Erling B.},
  title   = {Asymptotic Properties of Conditional Maximum-Likelihood Estimators},
  journal = {Journal of the Royal Statistical Society: Series B},
  year    = {1970},
  volume  = {32},
  number  = {2},
  pages   = {283--301}
}

@article{Chamberlain1980,
  author  = {Chamberlain, Gary},
  title   = {Analysis of Covariance with Qualitative Data},
  journal = {The Review of Economic Studies},
  year    = {1980},
  volume  = {47},
  number  = {1},
  pages   = {225--238}
}

@article{JochmansWeidner2019,
  author  = {Jochmans, Koen and Weidner, Martin},
  title   = {Fixed-Effect Regressions on Network Data},
  journal = {Econometrica},
  year    = {2019},
  volume  = {87},
  number  = {5},
  pages   = {1543--1560}
}

@incollection{Wijsman1967,
  author    = {Wijsman, Robert A.},
  title     = {Cross-Sections of Orbits and Their Application to Densities of Maximal Invariants},
  booktitle = {Proceedings of the Fifth Berkeley Symposium on Mathematical Statistics and Probability},
  volume    = {1},
  pages     = {389--400},
  publisher = {University of California Press},
  address   = {Berkeley},
  year      = {1967}
}

@book{Eaton1989,
  author    = {Eaton, Morris L.},
  title     = {Group Invariance Applications in Statistics},
  series    = {NSF-CBMS Regional Conference Series in Probability and Statistics},
  volume    = {1},
  publisher = {Institute of Mathematical Statistics},
  address   = {Hayward, CA},
  year      = {1989}
}

@phdthesis{Davis1954,
  author = {Davis, Harry F.},
  title  = {Decomposition Theorem for the Haar Integral},
  school = {Massachusetts Institute of Technology},
  year   = {1954}
}

@article{BondarMilnes1981,
  author  = {Bondar, J. V. and Milnes, P.},
  title   = {Amenability: A Survey for Statistical Applications of {H}unt--{S}tein and Related Conditions on Groups},
  journal = {Zeitschrift f{\"u}r Wahrscheinlichkeitstheorie und Verwandte Gebiete},
  year    = {1981},
  volume  = {57},
  number  = {1},
  pages   = {103--128}
}

@article{Bhattacharyya1946,
  author  = {Bhattacharyya, A.},
  title   = {On Some Analogues of the Amount of Information and Their Use in Statistical Estimation},
  journal = {Sankhy{\=a}},
  year    = {1946},
  volume  = {8},
  number  = {1},
  pages   = {1--14}
}

@article{KullbackLeibler1951,
  author  = {Kullback, Solomon and Leibler, Richard A.},
  title   = {On Information and Sufficiency},
  journal = {The Annals of Mathematical Statistics},
  year    = {1951},
  volume  = {22},
  number  = {1},
  pages   = {79--86}
}

@article{Stam1959,
  author  = {Stam, A. J.},
  title   = {Some Inequalities Satisfied by the Quantities of Information of {F}isher and {S}hannon},
  journal = {Information and Control},
  year    = {1959},
  volume  = {2},
  number  = {2},
  pages   = {101--112}
}

@article{Hoeffding1948,
  author  = {Hoeffding, Wassily},
  title   = {A Class of Statistics with Asymptotically Normal Distribution},
  journal = {The Annals of Mathematical Statistics},
  year    = {1948},
  volume  = {19},
  number  = {3},
  pages   = {293--325}
}

@book{Tsybakov2009,
  author    = {Tsybakov, Alexandre B.},
  title     = {Introduction to Nonparametric Estimation},
  publisher = {Springer},
  address   = {New York},
  year      = {2009}
}

@article{NikaidoIsoda1955,
  author  = {Nikaid{\^o}, Hukukane and Isoda, Kazuo},
  title   = {Note on Non-Cooperative Convex Games},
  journal = {Pacific Journal of Mathematics},
  year    = {1955},
  volume  = {5},
  number  = {S1},
  pages   = {807--815}
}

@article{Nickell1981,
  author  = {Nickell, Stephen},
  title   = {Biases in Dynamic Models with Fixed Effects},
  journal = {Econometrica},
  year    = {1981},
  volume  = {49},
  number  = {6},
  pages   = {1417--1426}
}

@article{ArellanoBond1991,
  author  = {Arellano, Manuel and Bond, Stephen},
  title   = {Some Tests of Specification for Panel Data: {M}onte {C}arlo Evidence and an Application to Employment Equations},
  journal = {Review of Economic Studies},
  year    = {1991},
  volume  = {58},
  number  = {2},
  pages   = {277--297}
}

@article{HsiaoPesaranTahmiscioglu2002,
  author  = {Hsiao, Cheng and Pesaran, M. Hashem and Tahmiscioglu, A. Kamil},
  title   = {Maximum Likelihood Estimation of Fixed Effects Dynamic Panel Data Models Covering Short Time Periods},
  journal = {Journal of Econometrics},
  year    = {2002},
  volume  = {109},
  number  = {1},
  pages   = {107--150}
}

@article{DhaeneJochmans2016,
  author  = {Dhaene, Geert and Jochmans, Koen},
  title   = {Likelihood Inference in an Autoregression with Fixed Effects},
  journal = {Econometric Theory},
  year    = {2016},
  volume  = {32},
  number  = {5},
  pages   = {1178--1215}
}

@article{Bai2009,
  author  = {Bai, Jushan},
  title   = {Panel Data Models With Interactive Fixed Effects},
  journal = {Econometrica},
  year    = {2009},
  volume  = {77},
  number  = {4},
  pages   = {1229--1279}
}

@article{Bai2024,
  author  = {Bai, Jushan},
  title   = {Likelihood Approach to Dynamic Panel Models with Interactive Effects},
  journal = {Journal of Econometrics},
  year    = {2024},
  volume  = {240},
  pages   = {105636}
}

@unpublished{BaiMones2026,
  author = {Bai, Jushan and Mones, Pablo},
  title  = {Global Identification of Dynamic Panel Models with Interactive Effects},
  note   = {arXiv:2504.14354v3, July 16, 2026},
  year   = {2026}
}

@misc{Zhang2026IFE,
  author       = {Zhang, Xibin},
  title        = {Invariant Likelihood for Non-{G}aussian Panels with Interactive Fixed Effects},
  howpublished = {SSRN working paper 7649594},
  year         = {2026},
  doi          = {10.2139/ssrn.7649594}
}

@article{HoltzEakinNeweyRosen1988,
  author  = {Holtz-Eakin, Douglas and Newey, Whitney and Rosen, Harvey S.},
  title   = {Estimating Vector Autoregressions with Panel Data},
  journal = {Econometrica},
  year    = {1988},
  volume  = {56},
  number  = {6},
  pages   = {1371--1395}
}

@article{AhnLeeSchmidt2001,
  author  = {Ahn, Seung C. and Lee, Young H. and Schmidt, Peter},
  title   = {{GMM} Estimation of Linear Panel Data Models with Time-Varying Individual Effects},
  journal = {Journal of Econometrics},
  year    = {2001},
  volume  = {101},
  number  = {2},
  pages   = {219--255}
}

@article{AhnLeeSchmidt2013,
  author  = {Ahn, Seung C. and Lee, Young H. and Schmidt, Peter},
  title   = {Panel Data Models with Multiple Time-Varying Individual Effects},
  journal = {Journal of Econometrics},
  year    = {2013},
  volume  = {174},
  number  = {1},
  pages   = {1--14}
}

@article{MoonWeidner2015,
  author  = {Moon, Hyungsik Roger and Weidner, Martin},
  title   = {Linear Regression for Panel with Unknown Number of Factors as Interactive Fixed Effects},
  journal = {Econometrica},
  year    = {2015},
  volume  = {83},
  number  = {4},
  pages   = {1543--1579}
}

@article{AignerLovellSchmidt1977,
  author  = {Aigner, Dennis and Lovell, C. A. Knox and Schmidt, Peter},
  title   = {Formulation and Estimation of Stochastic Frontier Production Function Models},
  journal = {Journal of Econometrics},
  year    = {1977},
  volume  = {6},
  number  = {1},
  pages   = {21--37}
}

@article{Greene2005,
  author  = {Greene, William},
  title   = {Reconsidering Heterogeneity in Panel Data Estimators of the Stochastic Frontier Model},
  journal = {Journal of Econometrics},
  year    = {2005},
  volume  = {126},
  number  = {2},
  pages   = {269--303}
}

@article{WangHo2010,
  author  = {Wang, Hung-Jen and Ho, Chia-Wen},
  title   = {Estimating Fixed-Effect Panel Stochastic Frontier Models by Model Transformation},
  journal = {Journal of Econometrics},
  year    = {2010},
  volume  = {157},
  number  = {2},
  pages   = {286--296}
}

@article{ChenSchmidtWang2014,
  author  = {Chen, Yi-Yi and Schmidt, Peter and Wang, Hung-Jen},
  title   = {Consistent Estimation of the Fixed Effects Stochastic Frontier Model},
  journal = {Journal of Econometrics},
  year    = {2014},
  volume  = {181},
  number  = {2},
  pages   = {65--76}
}

@article{BelottiIlardi2018,
  author  = {Belotti, Federico and Ilardi, Giuseppe},
  title   = {Consistent Inference in Fixed-Effects Stochastic Frontier Models},
  journal = {Journal of Econometrics},
  year    = {2018},
  volume  = {202},
  number  = {2},
  pages   = {161--177}
}

@article{BellioGrassetti2024,
  author  = {Bellio, Ruggero and Grassetti, Luca},
  title   = {Efficient Estimation of True Fixed-Effects Stochastic Frontier Models},
  journal = {Journal of Productivity Analysis},
  year    = {2024},
  volume  = {62},
  pages   = {111--118}
}

@misc{HongKingZhang2026,
  author       = {Hong, Yongmiao and King, Maxwell L. and Zhang, Xibin},
  title        = {Fourier Tests for Inefficiency Determinants in Stochastic Frontier Models with Two-Way Fixed Effects},
  howpublished = {SSRN working paper},
  year         = {2026}
}

@incollection{LeeSchmidt1993,
  author    = {Lee, Young H. and Schmidt, Peter},
  title     = {A Production Frontier Model with Flexible Temporal Variation in Technical Efficiency},
  booktitle = {The Measurement of Productive Efficiency: Techniques and Applications},
  editor    = {Fried, Harold O. and Lovell, C. A. Knox and Schmidt, Shelton S.},
  publisher = {Oxford University Press},
  address   = {New York},
  year      = {1993},
  pages     = {237--255}
}

@article{ChristensenJorgensonLau1973,
  author  = {Christensen, Laurits R. and Jorgenson, Dale W. and Lau, Lawrence J.},
  title   = {Transcendental Logarithmic Production Frontiers},
  journal = {The Review of Economics and Statistics},
  year    = {1973},
  volume  = {55},
  number  = {1},
  pages   = {28--45}
}

@incollection{BrandtTarp2017,
  author    = {Brandt, Kasper and Tarp, Finn},
  title     = {Characteristics of the {VARHS} Data and Other Data Sources},
  booktitle = {Growth, Structural Transformation, and Rural Change in {V}iet {N}am: A Rising Dragon on the Move},
  editor    = {Tarp, Finn},
  series    = {UNU-WIDER Studies in Development Economics},
  publisher = {Oxford University Press},
  address   = {Oxford},
  year      = {2017},
  chapter   = {2}
}

@misc{TarpSinghal2020,
  author       = {Tarp, Finn and Singhal, Saurabh and {the VARHS team}},
  title        = {User Guide to the {VARHS} 2008--2016 Panel Data},
  howpublished = {UNU-WIDER, Helsinki},
  year         = {2020}
}

\pagebreak

\baselineskip 0.78cm

\noindent{\large Supplementary material}

\appendix
\counterwithin{assumption}{section}
\counterwithin{proposition}{section}

\section{Proofs and additional results for Section~\ref{sec:methodology}}
\label{app:proofs}

Throughout, expectations are taken at the true parameter values and
conditional on the covariates.

\subsection{Proof of Proposition~\ref{prop:NS}}
\label{app:proof_NS}

\noindent{\bf Part (a)}

\noindent{\bf Invariant $B_T'Y_i$.}
Since $B_T'\mathbf 1_T=0$, $B_T'(Y_i+a\mathbf 1_T)=B_T'Y_i$ for every real
$a$.

\medskip 
\noindent{\bf The invariant is maximal.}
Suppose $B_T'Y_i=B_T'\widetilde{Y}_i$.  Then $Y_i-\widetilde{Y}_i$ lies in the null
space of $B_T'$.  The $T\times T$ matrix $(T^{-1/2}\mathbf 1_T,B_T)$ is
orthogonal, so this null space is the span of $\mathbf 1_T$.  Hence
$\widetilde{Y}_i=Y_i+a\mathbf 1_T$ for some $a$, and the two points are on the
same orbit.  An invariant that takes different values on different orbits is
a maximal invariant.

\medskip 
\noindent{\bf The density of $B_T'Y_i$.}
The model is $Y_i=\alpha_i\mathbf 1_T+\varepsilon_i$ with
$\varepsilon_i\sim N(0,\sigma^2I_T)$.  Since $B_T'\mathbf 1_T=0$,
$B_T'Y_i=B_T'\varepsilon_i$, which is normal with mean zero and covariance
$\sigma^2B_T'B_T=\sigma^2I_{T-1}$.  The orthogonality of
$(T^{-1/2}\mathbf 1_T,B_T)$ also gives
$B_TB_T'=I_T-T^{-1}\mathbf 1_T\mathbf 1_T'$, so
\[
    \|B_T'Y_i\|^2=Y_i'(I_T-T^{-1}\mathbf 1_T\mathbf 1_T')Y_i
    =\sum_{t=1}^T(Y_{it}-\overline{Y}_i)^2=W_i .
\]
The density of $B_T'Y_i$ at the data is therefore
$(2\pi\sigma^2)^{-(T-1)/2}\exp\{-W_i/(2\sigma^2)\}$.

\medskip 
\noindent{\bf The Haar integral.}
As $\sum_t(Y_{it}-\overline{Y}_i)=0$, the cross term vanishes in
\[
    \sum_{t=1}^T(Y_{it}-\alpha)^2=W_i+T(\overline{Y}_i-\alpha)^2 .
\]
The integrand in \eqref{eq:NS_haar} is therefore
$(2\pi\sigma^2)^{-T/2}\exp\{-W_i/(2\sigma^2)\}
\exp\{-T(\overline{Y}_i-\alpha)^2/(2\sigma^2)\}$, and
\[
    \int_{\mathbb R}\exp\Big\{-\frac{T(\overline{Y}_i-\alpha)^2}{2\sigma^2}\Big\}d\alpha
    =\Big(\frac{2\pi\sigma^2}{T}\Big)^{1/2}.
\]
Multiplying the two factors gives
$T^{-1/2}(2\pi\sigma^2)^{-(T-1)/2}\exp\{-W_i/(2\sigma^2)\}$, which is
$T^{-1/2}$ times the density of $B_T'Y_i$ derived above.

\medskip
\noindent{\bf Part (b)}

\noindent{\bf The score.}
From \eqref{eq:NS_haar},
$\log L_{H,i}(\sigma^2)=-\tfrac12\log T-\tfrac{T-1}2\log(2\pi\sigma^2)
-W_i/(2\sigma^2)$.  Its derivative with respect to $\sigma^2$ is
$-(T-1)/(2\sigma^2)+W_i/(2\sigma^4)$, which is \eqref{eq:NS_score}.

\medskip 
\noindent{\bf The score has mean zero.}
By part (a), $W_i=\|B_T'Y_i\|^2$ with $B_T'Y_i\sim N(0,\sigma^2I_{T-1})$,
whatever $\alpha_i$ is.  Hence $W_i/\sigma^2\sim\chi^2_{T-1}$, so
$\mathbb EW_i=(T-1)\sigma^2$ and
$\mathbb Es_{H,i}(\sigma^2)=0$ at the true value.

\medskip 
\noindent{\bf Comparison with the orthogonalised score.}
The second-order orthogonalised score in the Neyman--Scott example of
\citet{BonhommeJochmansWeidner2026} is
\[
    (2\sigma^2)^{-1}\{W_i/\sigma^2-(T-1)\}.
\]
This is the same function of the data and of $\sigma^2$ as
\eqref{eq:NS_score}.  The equality can also be verified without reference
to their example: model \eqref{eq:NS_model} is model \eqref{eq:orbit_model}
with $A=\mathbf 1_T$, $r=1$, $\mu_i\equiv0$ and spherical Gaussian errors,
so Theorem~\ref{thm:haar_orthogonal}(c), whose proof does not use the
present proposition, gives $u^*_{q,i}=s_{H,i}$ for every $q\ge2$.

\medskip
\noindent{\bf Part (c)}

\noindent{\bf The estimator.}
The equation $\sum_is_{H,i}(\sigma^2)=0$ is
$\sum_iW_i=N(T-1)\sigma^2$, whose solution is $\widehat\sigma_H^2$.

\medskip 
\noindent{\bf Its distribution.}
The variables $W_i/\sigma^2$ are independent across $i$, each with the
$\chi^2_{T-1}$ distribution.  Their sum,
$N(T-1)\widehat\sigma_H^2/\sigma^2$, is $\chi^2_{N(T-1)}$.

\medskip 
\noindent{\bf Mean and variance.}
A $\chi^2_k$ variable has mean $k$ and variance $2k$.  With $k=N(T-1)$,
$\mathbb E\widehat\sigma_H^2=\sigma^2$ and
$\operatorname{Var}(\widehat\sigma_H^2)
=\sigma^4\,2k/k^2=2\sigma^4/\{N(T-1)\}$.

\subsection{Proof of Proposition~\ref{prop:haar_translation}}
\label{app:proof_haar_translation}

Let $Q=A(A'A)^{-1/2}$.  Then $Q'Q=I_r$ and $B'Q=0$, so $(Q,B)$ is a
$T\times T$ orthogonal matrix, and
$\operatorname{col}(Q)=\operatorname{col}(A)$ is the null space of $B'$.

\medskip
\noindent{\bf Part (a)}

\noindent{\bf Invariant $B'Y_i$.}
Since $B'A=0$, $B'(Y_i+Aa)=B'Y_i$ for every $a\in\mathbb R^r$.

\medskip 
\noindent{\bf The invariant is maximal.}
If $B'Y_i=B'\widetilde{Y}_i$, then $Y_i-\widetilde{Y}_i$ is in the null space of $B'$,
which is $\operatorname{col}(A)$.  Hence $\widetilde{Y}_i=Y_i+Aa$ for some $a$,
and the two points lie on one orbit.

\medskip 
\noindent{\bf Its distribution is free of $\xi_i$.}
From \eqref{eq:orbit_model},
$B'Y_i=B'\mu_i(\beta)+B'A\xi_i+B'\varepsilon_i
=B'\mu_i(\beta)+B'\varepsilon_i$.  The distribution of the right-hand side
involves $\beta$ and $\eta$ only.

\medskip
\noindent{\bf Part (b)}

\noindent{\bf The density of the contrasts.}
The map $\varepsilon\mapsto(Q'\varepsilon,B'\varepsilon)$ is orthogonal, so
it has unit Jacobian and inverse $(c,w)\mapsto Qc+Bw$.  The joint density of
$(c,w)=(Q'\varepsilon_i,B'\varepsilon_i)$ is therefore $g(Qc+Bw;\eta)$, and
the marginal density of $B'\varepsilon_i$ is
$h(w;\eta)=\int_{\mathbb R^r}g(Qc+Bw;\eta)\,dc$.

\medskip 
\noindent{\bf A change of variable in the Haar integral.}
Write $z=Y_i-\mu_i(\beta)=Qc_z+Bw_z$, with $c_z=Q'z$ and $w_z=B'z$.  Since
$A=Q(A'A)^{1/2}$,
\[
    z-A\xi=Q\{c_z-(A'A)^{1/2}\xi\}+Bw_z .
\]
The substitution $c=c_z-(A'A)^{1/2}\xi$ has
$d\xi=|A'A|^{-1/2}dc$, so
\[
    \int_{\mathbb R^r}g(z-A\xi;\eta)\,d\xi
    =|A'A|^{-1/2}\int_{\mathbb R^r}g(Qc+Bw_z;\eta)\,dc
    =|A'A|^{-1/2}h(w_z;\eta) .
\]
Multiplying by the factor $|A'A|^{1/2}$ in \eqref{eq:haar_lik} gives
\eqref{eq:haar_equals_quotient}.

\medskip 
\noindent{\bf Finiteness.}
The function $h(\cdot;\eta)$ is a probability density on
$\mathbb R^{T-r}$, so it is finite almost everywhere.  The distribution of
$B'Y_i$ has a density, so $L_{H,i}(\theta)$ is finite for almost every
$Y_i$.

\medskip 
\noindent{\bf The choice of basis does not matter.}
Let $H$ be a nonsingular $r\times r$ matrix.  The substitution
$\zeta=H\xi$ gives
$\int g(z-AH\xi;\eta)\,d\xi=|\det H|^{-1}\int g(z-A\zeta;\eta)\,d\zeta$,
while $|(AH)'(AH)|^{1/2}=|\det H|\,|A'A|^{1/2}$.  The product is unchanged.

\medskip
\noindent{\bf Part (c)}

\noindent{\bf The Haar likelihood is a density for every $\theta$.}
By (a) and (b), $L_{H,i}(\theta)$ is the density
$p_\theta(w)=h\{w-B'\mu_i(\beta);\eta\}$ of $W_i=B'Y_i$, evaluated at
$W_i$, and this density does not involve $\xi_i$.  Hence
$\int p_\theta(w)\,dw=1$ for every $\theta$.

\medskip 
\noindent{\bf The score is centred.}
Under Assumption~\ref{ass:regular}, differentiating this identity with
respect to $\theta$ under the integral gives
$\int\nabla_\theta p_\theta(w)\,dw=0$.  Since
$\nabla_\theta p_\theta=(\nabla_\theta\log p_\theta)\,p_\theta$, this is
$\mathbb Es_{H,i}(\theta)=0$ at the true value, for every $\xi_i$, because
the expectation is an integral against $p_\theta$, which is free of $\xi_i$.

\medskip 
\noindent{\bf The information identity.}
Differentiating a second time gives $\int\nabla^2_\theta p_\theta(w)\,dw=0$.
Because
$\nabla^2_\theta\log p_\theta
=\nabla^2_\theta p_\theta/p_\theta-s_{H,i}s_{H,i}'$, taking expectations
gives
$-\mathbb E\nabla^2_\theta\log L_{H,i}(\theta)
=\mathbb E\{s_{H,i}(\theta)s_{H,i}(\theta)'\}$.

\medskip
\noindent{\bf Part (d)}

\noindent{\bf Functions of the invariant have a mean free of $\xi_i$.}
The first claim is immediate from (a). For the second, let $\psi$ be a
function with $\mathbb E|\psi(B'Y_i)|<\infty$ and for which the differentiated
products are integrable.  By (a), the expectation
\[
    \int\psi(B'y)\,g\{y-\mu_i(\beta)-A\xi;\eta\}\,dy
\]
does not depend on $\xi$.

\medskip 
\noindent{\bf Differentiation with respect to $\xi$.}
Differentiating $p$ times with respect to $\xi$ under the integral gives
$\int\psi(B'y)\,\nabla^{(p)}_\xi g\,dy=0$.  Since
$v_p=\nabla^{(p)}_\xi g/g$, this is $\mathbb E\{\psi(B'Y_i)v_p\}=0$.

\medskip 
\noindent{\bf Zero covariance.}
The same argument with $\psi\equiv1$ gives $\mathbb Ev_p=0$.  Hence
$\operatorname{Cov}\{\psi(B'Y_i),v_p\}=0$.

\subsection{Proof of Theorem~\ref{thm:haar_orthogonal}}
\label{app:proof_haar_orthogonal}

Write $C_i=Q'Y_i$ and $W_i=B'Y_i$, with $Q=A(A'A)^{-1/2}$.  Since $(Q,B)$ is
orthogonal, the map $Y_i\mapsto(C_i,W_i)$ has unit Jacobian, and the joint
density of $(C_i,W_i)$ at $(c,w)$ is
\[
    p(c,w;\theta,\xi_i)
    =g\big[Q\{c-Q'\mu_i(\beta)-(A'A)^{1/2}\xi_i\}
           +B\{w-B'\mu_i(\beta)\};\eta\big].
\]
By Proposition~\ref{prop:haar_translation}(b), the density of $W_i$ is
$L_{H,i}(\theta)=\int p(c,w;\theta,\xi_i)\,dc$ evaluated at $w=W_i$, and it
does not depend on $\xi_i$.

\medskip
\noindent{\bf Part (a)}

\noindent{\bf Differentiate the marginal density.}
Differentiating under the integral,
\[
    \frac{\partial L_{H,i}(\theta)}{\partial\theta}
    =\int\frac{\partial\log p(c,w;\theta,\xi_i)}{\partial\theta}\,
     p(c,w;\theta,\xi_i)\,dc .
\]

\medskip 
\noindent{\bf Divide by the marginal density.}
The ratio $p(c,w;\theta,\xi_i)/L_{H,i}(\theta)$ is the conditional density
of $C_i$ given $W_i=w$.  Dividing the display by $L_{H,i}(\theta)$ therefore
gives
$s_{H,i}=\mathbb E\{\partial\log p(C_i,W_i;\theta,\xi_i)/\partial\theta
\mid W_i\}$.

\medskip 
\noindent{\bf Identify the integrand with the original score.}
Because the Jacobian is one, $\log p(C_i,W_i;\theta,\xi_i)$ equals
$\log g\{Y_i-\mu_i(\beta)-A\xi_i;\eta\}$, so its derivative with respect to
$\theta$ is $u_i(\theta,\xi_i)$.  Hence
$s_{H,i}=\mathbb E(u_i\mid W_i)$.

\medskip
\noindent{\bf Part (b)}

\noindent{\bf The projection space has conditional mean zero.}
Let $v\in\mathcal V_q$ and let $\psi$ be a bounded measurable function.
Since $v$ is a linear combination of components of $v_1,\ldots,v_q$, which
are square integrable, the integrability conditions of
Proposition~\ref{prop:haar_translation}(d) hold for $\psi$, and that result
gives $\mathbb E\{\psi(W_i)v\}=0$.  Bounded functions of $W_i$ determine the
conditional expectation given $W_i$, so $\mathbb E(v\mid W_i)=0$.

\medskip 
\noindent{\bf The conditional mean of the orthogonalised score.}
The projection $\Pi(u_i\mid\mathcal V_q)$ is an element of $\mathcal V_q$,
with coefficients that depend on $(\theta,\xi_i)$ but not on the data.
Applying the previous step to it and using (a),
\[
    \mathbb E(u^*_{q,i}\mid W_i)
    =\mathbb E(u_i\mid W_i)-\mathbb E\{\Pi(u_i\mid\mathcal V_q)\mid W_i\}
    =s_{H,i}.
\]

\medskip 
\noindent{\bf The two consequences.}
For the first consequence, $s_{H,i}$ is a square-integrable function of
$W_i$ and, by the argument of the first step, $\mathbb E(v_p\mid W_i)=0$ and
$\mathbb Ev_p=0$.  Iterated expectations give
$\mathbb E(s_{H,i}v_p')=\mathbb E\{s_{H,i}\,\mathbb E(v_p'\mid W_i)\}=0$, so
$s_{H,i}$ and $v_p$ are uncorrelated.  The second is the law of
total variance: since $\mathbb E(u^*_{q,i}\mid W_i)=s_{H,i}$,
\[
    \operatorname{Var}(u^*_{q,i})
    =\operatorname{Var}(s_{H,i})
     +\mathbb E\operatorname{Var}(u^*_{q,i}\mid W_i)
    \ge\operatorname{Var}(s_{H,i}),
\]
in the ordering of positive semidefinite matrices.

\medskip
\noindent{\bf Part (c)}

\noindent{\bf The scores in rotated coordinates.}
Put $\widetilde{c}=\{Q'(Y_i-\mu_i(\beta))-(A'A)^{1/2}\xi_i\}/\sigma$ and
$\widetilde{w}=B'\{Y_i-\mu_i(\beta)\}/\sigma$.  Under the model these are
independent standard normal vectors of dimensions $r$ and $T-r$, and the log
density of $Y_i$ is
$-\tfrac T2\log(2\pi\sigma^2)-\tfrac12(\|\widetilde{c}\|^2+\|\widetilde{w}\|^2)$.
With $G_i=\partial\mu_i(\beta)/\partial\beta'$,
\[
    u_{\beta,i}=\frac{G_i'(Q\widetilde{c}+B\widetilde{w})}{\sigma},
    \qquad
    u_{\sigma^2,i}
    =-\frac T{2\sigma^2}+\frac{\|\widetilde{c}\|^2+\|\widetilde{w}\|^2}{2\sigma^2}.
\]
By \eqref{eq:haar_equals_quotient} the Haar log-likelihood is
$-\tfrac{T-r}2\log(2\pi\sigma^2)-\tfrac12\|\widetilde{w}\|^2$, so the Haar
scores are
\[
    s_{H\beta,i}=\frac{G_i'B\widetilde{w}}{\sigma},
    \qquad
    s_{H\sigma^2,i}
    =-\frac{T-r}{2\sigma^2}+\frac{\|\widetilde{w}\|^2}{2\sigma^2}.
\]
Subtracting,
\begin{equation}
    u_{\beta,i}-s_{H\beta,i}=\frac{G_i'Q\widetilde{c}}{\sigma},
    \qquad
    u_{\sigma^2,i}-s_{H\sigma^2,i}=\frac{\|\widetilde{c}\|^2-r}{2\sigma^2}.
\label{eq:app_score_split}
\end{equation}

\medskip 
\noindent{\bf The projection space.}
The density depends on $\xi_i$ only through $\widetilde{c}$, with
$\partial\widetilde{c}/\partial\xi_i'=-(A'A)^{1/2}/\sigma$, a nonsingular matrix.
Differentiating $\exp(-\|\widetilde{c}\|^2/2)$ $p$ times and dividing by it gives
Hermite polynomials of degree $p$ in $\widetilde{c}$.  The components of $v_p$
are therefore linear combinations of these polynomials, and, because the
matrix above is nonsingular, every Hermite polynomial of degree $p$ is a
linear combination of components of $v_p$.  Hence $\mathcal V_q$ is the
space of polynomials in $\widetilde{c}$ of degree at most $q$ with mean zero.

\medskip 
\noindent{\bf The projection.}
The first difference in \eqref{eq:app_score_split} is linear in $\widetilde{c}$
and belongs to $\mathcal V_1$.  The second is a quadratic with mean zero; it
belongs to $\mathcal V_2$, and it is orthogonal to $\mathcal V_1$ because
the third moments of a standard normal vector vanish.  The Haar scores are
functions of $\widetilde{w}$ with mean zero, so by independence they are
orthogonal to every $\mathcal V_q$.  Writing
$u_i=s_{H,i}+(u_i-s_{H,i})$, the projection of $u_i$ on $\mathcal V_q$ is
the difference \eqref{eq:app_score_split} when $q\ge2$.  When $q=1$ it is
that difference for $\beta$ and zero for $\sigma^2$.

\medskip 
\noindent{\bf Conclusion.}
Subtracting the projection from $u_i$ gives $u^*_{q,i}=s_{H,i}$ for every
$q\ge2$.  For $q=1$ the $\beta$ components agree, and the $\sigma^2$
component of $u^*_{1,i}$ is $u_{\sigma^2,i}$, which differs from
$s_{H\sigma^2,i}$ by the nondegenerate variable
$(\|\widetilde{c}\|^2-r)/(2\sigma^2)$ in \eqref{eq:app_score_split}.

\subsection{Groups other than translations}
\label{app:additional_groups}

Translations are the Euclidean case of a general construction.  For a locally
compact group acting properly on the sample space, the density of a maximal
invariant is an integral of the density over the group with respect to Haar
measure, weighted by a multiplier when the group is not unimodular
\citep{Wijsman1967,Eaton1989}.  The decomposition theorem of
\citet{Davis1954} states that every such Haar integral is built from three
operations: summation over a discrete part, Lebesgue integration over a
Euclidean part, and averaging over a compact part.  The three behave
differently in incidental-parameter problems.  Lebesgue integration over a
noncompact orbit, as in \eqref{eq:haar_lik}, eliminates the incidental
parameter but can produce an integral that is not finite.  Averaging a continuous likelihood over a
compact group is finite and can compress the nuisance parameter: in the instrumental
variables and dynamic panel models of \citet{Moreira2009}, orthogonal
transformations across units reduce an incidental vector $\xi\in\mathbb R^N$
to the scalar $\xi'\xi/N$.  The loading reduction in the likelihood approach
of \citet{Bai2024} can be read in the same way, since the cross-sectional
second moment of the loadings is the maximal invariant of the loading matrix
under orthogonal transformations of the units.  Finally, invariance reaches
further than Haar integration.  The monotone transformation model of
\citet{Moreira2009} has ranks as a maximal invariant, but the group of
increasing transformations is not locally compact and has no Haar measure.

\subsection{Proof of Proposition~\ref{prop:haar_reduction_principle}}
\label{app:proof_haar_reduction_principle}

The proposition is proved under the following conditions on the quotient
family.

\begin{assumption}[Quotient identification and regularity]
\label{ass:quotient}
The units are independent, and:
\begin{enumerate}
\item[(i)] $\sup_{\theta\in\Theta}|Q_N(\theta)-\mathbb EQ_N(\theta)|
\overset p\to0$;
\item[(ii)] for every $\epsilon>0$,
$\liminf_{N\to\infty}\inf_{\|\theta-\theta_0\|\ge\epsilon}
N^{-1}\sum_i\mathrm{KL}_i(\theta)>0$;
\item[(iii)] on a neighbourhood $\mathcal N_0$ of $\theta_0$, $Q_N$ is twice
continuously differentiable,
\[
    \sup_{\theta\in\mathcal N_0}
    \|\nabla^2Q_N(\theta)-\mathbb E\nabla^2Q_N(\theta)\|\overset p\to0,
\]
the matrices $\mathbb E\nabla^2Q_N(\theta)$ are continuous at $\theta_0$
uniformly in $N$, and the average $I_N$ of the matrices
$\mathbb E\{s_{H,i}(\theta_0)s_{H,i}(\theta_0)'\}$ converges to a positive
definite matrix $I_H$;
\item[(iv)] the array $\{s_{H,i}(\theta_0)\}$ satisfies Lindeberg's condition.
\end{enumerate}
\end{assumption}

\noindent{\bf Part (a)}

By Proposition~\ref{prop:haar_translation}(b),
$\log L_{H,i}(\theta_0)-\log L_{H,i}(\theta)$ is the logarithm of the ratio
of the density of $B'Y_i$ under $\theta_0$ to its density under $\theta$,
evaluated at $B'Y_i$.  Its expectation under $\theta_0$ is by definition the
divergence $\mathrm{KL}_i(\theta)$, which is nonnegative.  By
Proposition~\ref{prop:haar_translation}(a) the distribution of $B'Y_i$ is
free of $\xi_i$, so $\mathrm{KL}_i(\theta)$ is free of $\xi_i$ as well.
Averaging over $i$ gives
$\mathbb EQ_N(\theta_0)-\mathbb EQ_N(\theta)
=N^{-1}\sum_i\mathrm{KL}_i(\theta)$.

\medskip
\noindent{\bf Part (b)}

\noindent{\bf Separation.}
Fix $\epsilon>0$.  By Assumption~\ref{ass:quotient}(ii) there are
$\delta>0$ and $N_0$ such that, for $N\ge N_0$,
\[
    \inf_{\|\theta-\theta_0\|\ge\epsilon}
    N^{-1}\sum_i\mathrm{KL}_i(\theta)\ge2\delta .
\]

\medskip 
\noindent{\bf Comparison of criteria.}
On the event $\|\widehat\theta_H-\theta_0\|\ge\epsilon$, part (a) gives
$\mathbb EQ_N(\theta_0)-\mathbb EQ_N(\widehat\theta_H)\ge2\delta$, where
$\mathbb EQ_N(\cdot)$ is the nonrandom function $\theta\mapsto\mathbb
EQ_N(\theta)$ evaluated at the realised value of $\widehat\theta_H$.  Since
$\widehat\theta_H$ maximises $Q_N$, $Q_N(\widehat\theta_H)\ge
Q_N(\theta_0)$.  Adding and subtracting,
\[
    2\delta
    \le\{\mathbb EQ_N(\theta_0)-Q_N(\theta_0)\}
    +\{Q_N(\widehat\theta_H)-\mathbb EQ_N(\widehat\theta_H)\}
    \le2\sup_{\theta\in\Theta}|Q_N(\theta)-\mathbb EQ_N(\theta)|.
\]

\medskip 
\noindent{\bf Conclusion.}
Hence
$P(\|\widehat\theta_H-\theta_0\|\ge\epsilon)
\le P\{\sup_\theta|Q_N(\theta)-\mathbb EQ_N(\theta)|\ge\delta\}$ for
$N\ge N_0$, and the right-hand side tends to zero by
Assumption~\ref{ass:quotient}(i).

\medskip
\noindent{\bf Part (c)}

\noindent{\bf First-order condition and expansion.}
By (b), $\widehat\theta_H$ lies in the interior of $\Theta$ and in
$\mathcal N_0$ with probability tending to one, and then
$\nabla Q_N(\widehat\theta_H)=0$. Integrating the Hessian along the segment
from $\theta_0$ to $\widehat\theta_H$ gives
\[
\begin{aligned}
    0&=\sqrt N\,\nabla Q_N(\theta_0)
    +\widetilde H_N\sqrt N(\widehat\theta_H-\theta_0),
    \\
    \widetilde H_N&=\int_0^1\nabla^2Q_N\{\theta_0+t(\widehat\theta_H-\theta_0)\}\,dt.
\end{aligned}
\]

\medskip 
\noindent{\bf The Hessian.}
By Proposition~\ref{prop:haar_translation}(c),
$\mathbb E\nabla^2Q_N(\theta_0)=-I_N$. Write $\mathcal S_N$ for the segment
from $\theta_0$ to $\widehat\theta_H$, which lies in $\mathcal N_0$ with
probability tending to one. On that event,
\[
    \|\widetilde H_N+I_N\|
    \le\sup_{\theta\in\mathcal N_0}
       \|\nabla^2Q_N(\theta)-\mathbb E\nabla^2Q_N(\theta)\|
    +\sup_{\theta\in\mathcal S_N}
       \|\mathbb E\nabla^2Q_N(\theta)-\mathbb E\nabla^2Q_N(\theta_0)\| .
\]
The first term tends to zero in probability and the second does so because
the segment shrinks to $\theta_0$ in probability and the expected Hessians are continuous
at $\theta_0$ uniformly in $N$, both by Assumption~\ref{ass:quotient}(iii).
Since $I_N\to I_H$, $\widetilde H_N\overset p\to-I_H$.

\medskip 
\noindent{\bf The score.}
The scores $s_{H,i}(\theta_0)$ are independent with mean zero by
Proposition~\ref{prop:haar_translation}(c), and the average of their
covariance matrices is $I_N\to I_H$.  With Lindeberg's condition,
Assumption~\ref{ass:quotient}(iv), the Lindeberg--Feller theorem gives
$\sqrt N\,\nabla Q_N(\theta_0)\overset d\to N(0,I_H)$.

\medskip 
\noindent{\bf The limit.}
Since $I_H$ is positive definite, the expansion can be solved with
probability tending to one:
$\sqrt N(\widehat\theta_H-\theta_0)
=-\widetilde H_N^{-1}\sqrt N\,\nabla Q_N(\theta_0)$.  By
Slutsky's lemma the limit is $I_H^{-1}N(0,I_H)=N(0,I_H^{-1})$.

\subsection{The Nickell bias as an orbit drift}
\label{app:proof_dyn_drift}

In this and the following subsections on the dynamic panel model
\eqref{eq:dyn_model}, expectations are taken at $\theta_0$ and conditional
on $(y_{i0},X_i,\alpha_i)$.

\begin{proposition}[The Nickell bias is an orbit drift]
\label{prop:dyn_drift}
In model \eqref{eq:dyn_model}, for every $T\ge2$ and every value of
$(\alpha_i,y_{i0},X_i)$, the scores of \eqref{eq:dyn_naive} at $\theta_0$
satisfy
\[
    \mathbb E\,\frac{\partial\ell_{H,i}}{\partial\beta}=0,
    \quad
    \mathbb E\,\frac{\partial\ell_{H,i}}{\partial\sigma^2}=0,
    \quad
    \mathbb E\,\frac{\partial\ell_{H,i}}{\partial\rho}=-b_T(\rho_0),
\]
where $b_T$ is defined in \eqref{eq:dyn_bT}.  The drift $b_T(\rho_0)$ is strictly positive for $\rho_0>-1$, and it depends
on neither the unit effect, the initial observation, the covariates, nor the
error variance.
\end{proposition}

\noindent{\em Proof.}

\noindent{\bf The scores for $\beta$ and $\sigma^2$.}
At $\theta_0$, $e_i(\rho_0,\beta_0)=\alpha_i\mathbf 1_T+\varepsilon_i$ and
$M\mathbf 1_T=0$, so $Me_i=M\varepsilon_i$.  Differentiating
\eqref{eq:dyn_naive},
\[
    \frac{\partial\ell_{H,i}}{\partial\beta}
    =\frac{X_i'M\varepsilon_i}{\sigma_0^2},
    \qquad
    \frac{\partial\ell_{H,i}}{\partial\sigma^2}
    =-\frac{T-1}{2\sigma_0^2}
     +\frac{\varepsilon_i'M\varepsilon_i}{2\sigma_0^4},
    \qquad
    \frac{\partial\ell_{H,i}}{\partial\rho}
    =\frac{y_{i,-}'M\varepsilon_i}{\sigma_0^2}.
\]
The first has mean zero by strict exogeneity.  The second has mean zero
because
$\mathbb E\varepsilon_i'M\varepsilon_i=\sigma_0^2\operatorname{tr}(M)
=\sigma_0^2(T-1)$.

\medskip 
\noindent{\bf Covariances of lagged values and errors.}
The recursion gives, for $s=0,\ldots,T-1$,
\[
    y_{is}=\rho_0^{\,s}y_{i0}
    +\sum_{j=1}^s\rho_0^{\,s-j}(x_{ij}'\beta_0+\alpha_i)
    +\sum_{j=1}^s\rho_0^{\,s-j}\varepsilon_{ij},
\]
so that $\mathbb E(y_{is}\varepsilon_{it})=\sigma_0^2\rho_0^{\,s-t}$ for
$1\le t\le s$ and zero otherwise.

\medskip 
\noindent{\bf The score for $\rho$.}
The $t$-th component of $y_{i,-}$ is $y_{i,t-1}$, which is uncorrelated with
$\varepsilon_{it}$, so $\mathbb E(y_{i,-}'\varepsilon_i)=0$.  By the covariances just derived,
\[
    \mathbb E\{(\mathbf 1_T'y_{i,-})(\mathbf 1_T'\varepsilon_i)\}
    =\sigma_0^2\sum_{s=1}^{T-1}\sum_{t=1}^s\rho_0^{\,s-t}
    =\sigma_0^2\sum_{k=0}^{T-2}(T-1-k)\rho_0^{\,k}
    =\sigma_0^2\,T\,b_T(\rho_0),
\]
where the second equality counts the pairs $(s,t)$ with $s-t=k$.  Since
$M=I_T-T^{-1}\mathbf 1_T\mathbf 1_T'$,
\[
    \mathbb E(y_{i,-}'M\varepsilon_i)
    =\mathbb E(y_{i,-}'\varepsilon_i)
     -T^{-1}\mathbb E\{(\mathbf 1_T'y_{i,-})(\mathbf 1_T'\varepsilon_i)\}
    =-\sigma_0^2b_T(\rho_0),
\]
which gives the third claim.  None of these expectations involves
$(\alpha_i,y_{i0},X_i)$, and $\sigma_0^2$ cancels.

\medskip 
\noindent{\bf The sign of the drift.}
If $\rho_0\ge0$ every term of \eqref{eq:dyn_bT} is nonnegative and the first
equals $(T-1)/T>0$.  If $-1<\rho_0<0$ the sum is alternating, the absolute
values $(T-t)|\rho_0|^{t-1}$ of its terms decrease strictly in $t$, and the
first term is positive, so the sum is positive.

\subsection{Proof of Theorem~\ref{thm:dyn_ortho}}
\label{app:proof_dyn_ortho}

\noindent{\bf The lag identity.}
Fix $(\theta,\alpha)$, drop the unit index, and write
$\varepsilon=e(\rho,\beta)-\alpha\mathbf 1_T$,
$\overline{\varepsilon}=T^{-1}\mathbf 1_T'\varepsilon=\overline{e}(\theta)-\alpha$ and
$K=K(\rho)$.  Solving the recursion as in the proof of
Proposition~\ref{prop:dyn_drift} gives the identity
\begin{equation}
    y_{-}=m(\alpha)+K\varepsilon,
\label{eq:app_lag_identity}
\end{equation}
where the $t$-th component of $m(\alpha)$ is
$\rho^{\,t-1}y_0+\sum_{j<t}\rho^{\,t-1-j}(x_j'\beta+\alpha)$, so that
$m(\alpha)$ is linear in $\alpha$ with
$\partial m(\alpha)/\partial\alpha=K\mathbf 1_T$.  Under the distribution
indexed by $(\theta,\alpha)$, $\varepsilon\sim N(0,\sigma^2I_T)$, and
$\overline{\varepsilon}$ is independent of $M\varepsilon$.

\medskip 
\noindent{\bf The projection space.}
Since $\sum_t(e_t-\alpha)^2=e'Me+T(\overline{e}-\alpha)^2$, the Gaussian
likelihood depends on $\alpha$ only through the factor
$\exp\{-T(\overline{e}-\alpha)^2/(2\sigma^2)\}$.  The $p$-th Bhattacharyya
function is therefore proportional to the Hermite polynomial of degree $p$
in $\sqrt T\,\overline{\varepsilon}/\sigma$, and $\mathcal V_q$ is the space of
polynomials in $\overline{\varepsilon}$ of degree at most $q$ with mean zero,
because the Hermite polynomials of degrees $1,\ldots,q$ form a basis of that
space.

\medskip 
\noindent{\bf Projection as a conditional expectation.}
Let $u$ be a component of the score whose conditional mean given
$\overline{\varepsilon}$ is a polynomial of degree at most two.  That conditional
mean has mean zero, because the score has mean zero, and so lies in
$\mathcal V_2\subseteq\mathcal V_q$ by the previous step, and the
remainder $u-\mathbb E(u\mid\overline{\varepsilon})$ is orthogonal to every
function of $\overline{\varepsilon}$, in particular to $\mathcal V_q$.  The
projection of $u$ on $\mathcal V_q$ is therefore
$\mathbb E(u\mid\overline{\varepsilon})$, and for such a component and every
$q\ge2$,
\begin{equation}
    u^*_q=u-\mathbb E(u\mid\overline{\varepsilon}).
\label{eq:app_proj_cond}
\end{equation}
All three components of the score are quadratic in $\varepsilon$, and
$\varepsilon=\overline{\varepsilon}\mathbf 1_T+M\varepsilon$ with $M\varepsilon$
independent of $\overline{\varepsilon}$, so their conditional means are polynomials
of degree at most two and \eqref{eq:app_proj_cond} applies to each of them.

\medskip
\noindent{\bf Part (a)}

The scores for $\beta$ and $\sigma^2$ are $X'\varepsilon/\sigma^2$ and
$-T/(2\sigma^2)+\varepsilon'\varepsilon/(2\sigma^4)$.  Using
$\varepsilon=\overline{\varepsilon}\mathbf 1_T+M\varepsilon$,
$\varepsilon'\varepsilon=T\overline{\varepsilon}^2+\varepsilon'M\varepsilon$ and
$\mathbb E(\varepsilon'M\varepsilon)=(T-1)\sigma^2$,
\[
    \mathbb E(u_\beta\mid\overline{\varepsilon})
    =\frac{X'\mathbf 1_T\,\overline{\varepsilon}}{\sigma^2},
    \qquad
    \mathbb E(u_{\sigma^2}\mid\overline{\varepsilon})
    =-\frac1{2\sigma^2}+\frac{T\overline{\varepsilon}^2}{2\sigma^4}.
\]
Subtracting as in \eqref{eq:app_proj_cond} gives
$u^*_{\beta,q}=X'M\varepsilon/\sigma^2$ and
$u^*_{\sigma^2,q}=-(T-1)/(2\sigma^2)
+\varepsilon'M\varepsilon/(2\sigma^4)$.  Since $M\varepsilon=Me(\rho,\beta)$,
these do not involve $\alpha$ and are the derivatives of
\eqref{eq:dyn_naive}.

\medskip
\noindent{\bf Part (b)}

\noindent{\bf The score for $\rho$.}
By \eqref{eq:app_lag_identity} the score for $\rho$ is
$u_\rho=y_-'\varepsilon/\sigma^2
=\{m(\alpha)'\varepsilon+\varepsilon'K'\varepsilon\}/\sigma^2$.  Expanding
with $\varepsilon=\overline{\varepsilon}\mathbf 1_T+M\varepsilon$,
\[
    \varepsilon'K'\varepsilon
    =\overline{\varepsilon}^2\,\mathbf 1_T'K\mathbf 1_T
     +\overline{\varepsilon}\,\mathbf 1_T'(K+K')M\varepsilon
     +(M\varepsilon)'K'M\varepsilon .
\]

\medskip 
\noindent{\bf Its conditional mean.}
The matrix $K$ has zero diagonal, so $\operatorname{tr}(K)=0$, and
$\mathbf 1_T'K\mathbf 1_T=Tb_T(\rho)$.  Hence
$\operatorname{tr}(MK)=\operatorname{tr}(K)-T^{-1}\mathbf 1_T'K\mathbf 1_T
=-b_T(\rho)$ and
$\mathbb E\{(M\varepsilon)'K'M\varepsilon\}
=\sigma^2\operatorname{tr}(MK)=-\sigma^2b_T(\rho)$.  The middle term of the
expansion has conditional mean zero.  Therefore
\[
    \mathbb E(u_\rho\mid\overline{\varepsilon})
    =\frac{\overline{\varepsilon}\,\mathbf 1_T'm(\alpha)
           +Tb_T(\rho)\,\overline{\varepsilon}^2-\sigma^2b_T(\rho)}{\sigma^2}.
\]

\medskip 
\noindent{\bf The orthogonalised score.}
By \eqref{eq:app_proj_cond},
\begin{equation}
    u^*_{\rho,q}
    =\frac{m(\alpha)'M\varepsilon
           +\overline{\varepsilon}\,\mathbf 1_T'(K+K')M\varepsilon
           +(M\varepsilon)'K'M\varepsilon}{\sigma^2}
     +b_T(\rho).
\label{eq:app_ustar_rho}
\end{equation}

\medskip 
\noindent{\bf The adjusted score.}
By \eqref{eq:app_lag_identity},
$y_-'M\varepsilon
=m(\alpha)'M\varepsilon+\overline{\varepsilon}\,\mathbf 1_T'K'M\varepsilon
+(M\varepsilon)'K'M\varepsilon$.  Since $a_T'(\rho)=b_T(\rho)$ and
$Me=M\varepsilon$,
\[
    \frac{\partial\ell_A}{\partial\rho}
    =\frac{y_-'Me}{\sigma^2}+b_T(\rho)
    =\frac{m(\alpha)'M\varepsilon
           +\overline{\varepsilon}\,\mathbf 1_T'K'M\varepsilon
           +(M\varepsilon)'K'M\varepsilon}{\sigma^2}
     +b_T(\rho).
\]

\medskip 
\noindent{\bf The difference.}
Subtracting this from \eqref{eq:app_ustar_rho} leaves
$\overline{\varepsilon}\,\mathbf 1_T'KM\varepsilon/\sigma^2
=\{\overline{e}(\theta)-\alpha\}\,g(\rho)'Me/\sigma^2$, with
$g(\rho)=K(\rho)'\mathbf 1_T$.  This is \eqref{eq:dyn_ortho}.

\medskip
\noindent{\bf Part (c)}

Setting $\alpha=\overline{e}(\theta)$ in \eqref{eq:dyn_ortho} removes the second
term, so the $\rho$ component is $\partial\ell_A/\partial\rho$.  By (a) the
other components are the derivatives of $\ell_H$, and $\ell_A$ and $\ell_H$
have the same derivatives with respect to $\beta$ and $\sigma^2$ because
$a_T$ depends on $\rho$ only.

\subsection{Identification and information with two periods}
\label{subsec:dyn_T2}

Centred scores do not settle requirements H4 and H5, and in this model both
need attention.  The case $T=2$ without covariates can be solved in closed
form and shows what is involved.  Let
$\Delta_{i1}=y_{i1}-y_{i0}$, $\Delta_{i2}=y_{i2}-y_{i1}$,
$S_{jk}=\sum_i\Delta_{ij}\Delta_{ik}$ and $\widehat r=S_{12}/S_{11}$, which is
the within-group estimator.  The initial deviation of unit $i$ from its
steady-state path is measured by
\[
    w_i=\alpha_i-(1-\rho_0)y_{i0},
\]
so that $\Delta_{i1}=w_i+\varepsilon_{i1}$.

\begin{proposition}[Two periods]
\label{prop:dyn_T2}
Consider model \eqref{eq:dyn_model} with $T=2$ and no covariates.  Suppose the
errors are independent across $i$ and $t$ with mean zero, variance
$\sigma_0^2$, third moment $\kappa_3$ and fourth moment $\mu_4<\infty$, and
that $N^{-1}\sum_iw_i\to\omega_1$, $N^{-1}\sum_iw_i^2\to\omega^2$ and
$\sup_NN^{-1}\sum_iw_i^4<\infty$.
\begin{enumerate}
\item[(a)] After $\sigma^2$ is concentrated out, the stationary points of
$\sum_i\ell_{A,i}$ are the roots of the quadratic equation
$\sum_i\psi_i(\rho)=0$, where
\begin{equation}
    \psi_i(\rho)
    =(\Delta_{i2}-\rho\Delta_{i1})\{\Delta_{i2}+(2-\rho)\Delta_{i1}\},
\label{eq:dyn_psi}
\end{equation}
and $\mathbb E\psi_i(\rho_0)=0$ for every $w_i$.
\item[(b)] The within-group estimator satisfies
$\widehat r\overset p\to\rho_0-\sigma_0^2/(\omega^2+\sigma_0^2)$.
\item[(c)] The limit of $N^{-1}\sum_i\psi_i(\rho)$ has the two roots
$\rho_0$ and $\rho_0+2\omega^2/(\omega^2+\sigma_0^2)$.  If $\omega^2>0$, the
first is a local maximiser and the second a local minimiser of the limiting
concentrated criterion, and
\[
    \widehat\rho_A
    =1+\widehat r-\sqrt{1+\widehat r^{\,2}-S_{22}/S_{11}}
\]
is well defined with probability tending to one and satisfies
$\widehat\rho_A\overset p\to\rho_0$.
\item[(d)] If $\omega^2>0$,
\[
    \sqrt N(\widehat\rho_A-\rho_0)\overset d\to
    N\!\left(0,\;
    \frac{2(\mu_4-\sigma_0^4)+8\omega^2\sigma_0^2+8\omega_1\kappa_3}
         {4\omega^4}\right),
\]
which equals $\sigma_0^2(\sigma_0^2+2\omega^2)/\omega^4$ for Gaussian errors.
\end{enumerate}
\end{proposition}

\noindent{\em Proof.}
Write $\sigma^2$ for $\sigma_0^2$ and $m=\omega^2+\sigma^2$.  From the model,
$\Delta_{i1}=w_i+\varepsilon_{i1}$ and
$\Delta_{i2}=\rho_0\Delta_{i1}+\varepsilon_{i2}-\varepsilon_{i1}$.

\medskip
\noindent{\bf Part (a)}

\noindent{\bf The adjusted criterion for two periods.}
For $T=2$,
$e_i'Me_i=\tfrac12(e_{i2}-e_{i1})^2
=\tfrac12(\Delta_{i2}-\rho\Delta_{i1})^2$ and $a_2(\rho)=\rho/2$, so
\[
    \ell_{A,i}(\rho,\sigma^2)
    =-\tfrac12\log(2\pi\sigma^2)
     -\frac{(\Delta_{i2}-\rho\Delta_{i1})^2}{4\sigma^2}+\frac\rho2.
\]

\medskip 
\noindent{\bf Concentrating out the variance.}
Let $q_N(\rho)=N^{-1}\sum_i(\Delta_{i2}-\rho\Delta_{i1})^2$.  Maximising
over $\sigma^2$ gives $\widehat\sigma^2(\rho)=q_N(\rho)/2$ and, up to a
constant, the concentrated criterion
$-\tfrac N2\log q_N(\rho)+\tfrac N2\rho$.

\medskip 
\noindent{\bf The stationary points.}
Its derivative is
\[
\frac N2\cdot
    \frac{2\sum_i\Delta_{i1}(\Delta_{i2}-\rho\Delta_{i1})
          +\sum_i(\Delta_{i2}-\rho\Delta_{i1})^2}
         {\sum_i(\Delta_{i2}-\rho\Delta_{i1})^2}
    =\frac N2\cdot\frac{\sum_i\psi_i(\rho)}
                       {\sum_i(\Delta_{i2}-\rho\Delta_{i1})^2},
\]
because
$(\Delta_{i2}-\rho\Delta_{i1})\{2\Delta_{i1}+\Delta_{i2}-\rho\Delta_{i1}\}
=\psi_i(\rho)$.  The stationary points are the roots of
$\sum_i\psi_i(\rho)=0$.

\medskip 
\noindent{\bf The estimating function at the truth.}
At $\rho_0$,
$\Delta_{i2}-\rho_0\Delta_{i1}=\varepsilon_{i2}-\varepsilon_{i1}$ and
$\Delta_{i2}+(2-\rho_0)\Delta_{i1}
=\varepsilon_{i2}+\varepsilon_{i1}+2w_i$, so
\begin{equation}
    \psi_i(\rho_0)
    =\varepsilon_{i2}^2-\varepsilon_{i1}^2
     +2w_i(\varepsilon_{i2}-\varepsilon_{i1}),
\label{eq:app_psi0}
\end{equation}
which has mean zero for every $w_i$.

\medskip
\noindent{\bf Part (b)}

The summands of $S_{jk}$ are independent with variances bounded on average,
by the moment conditions, so Chebyshev's inequality gives a law of large
numbers for each $S_{jk}/N$.  Using
$\mathbb E\Delta_{i1}^2=w_i^2+\sigma^2$ and
$\mathbb E\{\Delta_{i1}(\varepsilon_{i2}-\varepsilon_{i1})\}=-\sigma^2$,
\[
    \frac{S_{11}}N\overset p\to m,
    \qquad
    \frac{S_{12}}N\overset p\to\rho_0m-\sigma^2,
    \qquad
    \frac{S_{22}}N\overset p\to\rho_0^2m-2\rho_0\sigma^2+2\sigma^2 .
\]
Hence $\widehat r=S_{12}/S_{11}\overset p\to\rho_0-\sigma^2/m$.

\medskip
\noindent{\bf Part (c)}

\noindent{\bf The limiting equation and its roots.}
Put $\delta=\rho_0-\rho$.  Then
$\Delta_{i2}-\rho\Delta_{i1}
=\delta\Delta_{i1}+\varepsilon_{i2}-\varepsilon_{i1}$, and by (b) the limit
of $q_N(\rho)$ is
$q(\rho)=\delta^2m-2\delta\sigma^2+2\sigma^2$, with
$q'(\rho)=2\sigma^2-2\delta m$ and $q''(\rho)=2m$.  By the derivative
computed in part (a),
$N^{-1}\sum_i\psi_i(\rho)=q_N(\rho)-q_N'(\rho)$, whose limit is
\[
    q(\rho)-q'(\rho)=\delta^2m+2\delta(m-\sigma^2)
    =\delta\,(\delta m+2\omega^2).
\]
The roots are $\delta=0$ and $\delta=-2\omega^2/m$, that is, $\rho=\rho_0$
and $\rho=\rho_0+2\omega^2/m$.

\medskip 
\noindent{\bf Which root is a maximiser.}
The limiting concentrated criterion per unit is
$f(\rho)=-\tfrac12\log q(\rho)+\tfrac12\rho$, with $f'=(q-q')/(2q)$.  At a
root, where $q=q'$, the second derivative is $f''=(q'-q'')/(2q)
=(q-q'')/(2q)$.  At $\rho_0$, $q=2\sigma^2$ and
$f''=-\omega^2/(2\sigma^2)<0$.  At the second root,
$q=2(2\omega^2+\sigma^2)$ and
$f''=\omega^2/\{2(2\omega^2+\sigma^2)\}>0$.

\medskip 
\noindent{\bf The sample roots.}
Expanding \eqref{eq:dyn_psi}, the equation $\sum_i\psi_i(\rho)=0$ is
$S_{11}\rho^2-2(S_{11}+S_{12})\rho+S_{22}+2S_{12}=0$, with roots
$1+\widehat r\pm\{1+\widehat r^{\,2}-S_{22}/S_{11}\}^{1/2}$.  By (b),
\[
    1+\widehat r^{\,2}-\frac{S_{22}}{S_{11}}
    \overset p\to\Big(1-\frac{\sigma^2}m\Big)^2
    =\frac{\omega^4}{m^2}>0,
\]
so the roots are real with probability tending to one, and the smaller root
$\widehat\rho_A$ converges to
$1+\rho_0-\sigma^2/m-(1-\sigma^2/m)=\rho_0$.

\medskip 
\noindent{\bf The smaller root is the interior local maximiser.}
The quadratic has a positive leading coefficient, so it is positive to the
left of the smaller root and negative between the roots.  The derivative in
part (a) has the same sign, so the concentrated criterion increases up to
the smaller root and decreases after it.

\medskip
\noindent{\bf Part (d)}

\noindent{\bf Expansion.}
Let $\Psi_N(\rho)=N^{-1}\sum_i\psi_i(\rho)$, a quadratic in $\rho$ with
second derivative $2S_{11}/N$.  On the event that the roots are real, the
exact Taylor expansion at $\rho_0$ is
\[
    0=\Psi_N(\rho_0)+\Psi_N'(\rho_0)(\widehat\rho_A-\rho_0)
     +\frac{S_{11}}N(\widehat\rho_A-\rho_0)^2.
\]
Differentiating \eqref{eq:dyn_psi} and using (b),
\[
    \Psi_N'(\rho_0)
    =-\frac2N\sum_i\Delta_{i1}\{\Delta_{i2}+(1-\rho_0)\Delta_{i1}\}
    \overset p\to-2(\rho_0m-\sigma^2)-2(1-\rho_0)m
    =-2\omega^2.
\]

\medskip 
\noindent{\bf The estimating function at the truth.}
By \eqref{eq:app_psi0}, the variables $\psi_i(\rho_0)$ are independent with
mean zero and variance
$2(\mu_4-\sigma^4)+8w_i^2\sigma^2+8w_i\kappa_3$, since
$\operatorname{Var}(\varepsilon_{i2}^2-\varepsilon_{i1}^2)
=2(\mu_4-\sigma^4)$,
$\operatorname{Var}(\varepsilon_{i2}-\varepsilon_{i1})=2\sigma^2$ and
$\operatorname{Cov}(\varepsilon_{i2}^2-\varepsilon_{i1}^2,
\varepsilon_{i2}-\varepsilon_{i1})=2\kappa_3$.  Their average variance
converges to
$\Omega=2(\mu_4-\sigma^4)+8\omega^2\sigma^2+8\omega_1\kappa_3$.  The bound on
$N^{-1}\sum_iw_i^4$ implies $\max_i|w_i|=O(N^{1/4})=o(N^{1/2})$, which
together with the finite fourth moment of the errors gives Lindeberg's
condition.  Hence $\sqrt N\,\Psi_N(\rho_0)\overset d\to N(0,\Omega)$.

\medskip 
\noindent{\bf The limit.}
Since $\widehat\rho_A-\rho_0=o_p(1)$ and $S_{11}/N=O_p(1)$, the quadratic
term is of smaller order than the linear term, and
\[
    \sqrt N(\widehat\rho_A-\rho_0)
    =-\frac{\sqrt N\,\Psi_N(\rho_0)}{\Psi_N'(\rho_0)}\{1+o_p(1)\}
    \overset d\to N\Big(0,\frac{\Omega}{4\omega^4}\Big).
\]
For Gaussian errors $\mu_4=3\sigma^4$ and $\kappa_3=0$, so
$\Omega/(4\omega^4)=\sigma^2(\sigma^2+2\omega^2)/\omega^4$.

Three points deserve comment.  First, part (c) is a multiple-root problem
related to the distinction between local and global identification discussed
in Section~\ref{subsec:information}.  The recentred equation has a second
root, and the adjusted criterion is unbounded as $\rho\to\infty$, so the
estimator must be defined as the interior local maximiser and not as a
global one.  The alias belongs to the estimating equation and not to the
experiment.

Second, part (d) shows where the information comes from.  With two periods
the variance is of order $\omega^{-4}$: the autoregressive coefficient is
identified by this estimating equation only through the cross-sectional mean
square of the initial deviations $w_i$.  If every unit starts on its
steady-state path, so that $\omega^2=0$, the two roots coincide, the derivative of the estimating
equation vanishes, and H5 fails.  Under a stationary start
$\omega^2=\sigma_0^2(1-\rho_0)/(1+\rho_0)$, which tends to zero as
$\rho_0\to1$, so that persistent panels are weakly identified by
construction.

Third, at the true value
$\psi_i(\rho_0)=\varepsilon_{i2}^2-\varepsilon_{i1}^2
+2w_i(\varepsilon_{i2}-\varepsilon_{i1})$.  This statistic changes when
$w_i$ changes, but its mean is free of $w_i$.  It is a moment of the larger
class in \eqref{eq:inv_subset_robust}, and its variance grows with $w_i^2$.
Unlike in Theorem~\ref{thm:minimax}, however, greater initial-deviation variation also
strengthens the sensitivity of the estimating equation in this model.  It
also relies on equal error variances in the two periods, which is the price
of not using instruments.

For $T>2$ and with covariates the adjusted criterion is maximised
numerically, with the same selection of an interior local maximiser; the
analysis of its roots is given by \citet{DhaeneJochmans2016}.

\subsection{Simulation evidence for the dynamic panel model}
\label{subsec:dyn_numerical}

Table~\ref{tab:dyn_sim} reports a small experiment for model
\eqref{eq:dyn_model} without covariates.  The effects and the errors are
independent standard normal, the initial observations are drawn from the
stationary distribution, and each entry is based on 1{,}000 replications.  The
within-group estimator is the maximiser of \eqref{eq:dyn_naive}.  The
recentred estimator is the interior local maximiser of the concentrated
version of \eqref{eq:dyn_adjusted}, searched over $\rho\in[-0.99,1.6]$.

\begin{table}[ht]
\centering
\caption{Within-group and recentred estimators of $\rho_0$ in the dynamic
panel model.  ``Share'' is the proportion of replications in which the
adjusted criterion has an interior local maximum; the median and standard
deviation of the recentred estimator are computed over those replications.}
\label{tab:dyn_sim}
\begin{tabular}{rrrcrcrrr}
\hline
 & & && Within-group && \multicolumn{3}{c}{Recentred}\\
\cline{5-5}\cline{7-9}
$T$ & $\rho_0$ & $N$ && Mean && Share & Median & s.d.\\
\hline
2 & 0.5 & 500 && $-0.250$ && 0.776 & 0.455 & 0.115\\
2 & 0.5 & 5000 && $-0.250$ && 0.987 & 0.501 & 0.058\\
2 & 0.9 & 500 && $-0.049$ && 0.525 & 0.710 & 0.107\\
2 & 0.9 & 5000 && $-0.051$ && 0.543 & 0.810 & 0.058\\
4 & 0.5 & 500 && 0.089 && 0.999 & 0.498 & 0.054\\
4 & 0.5 & 5000 && 0.089 && 1.000 & 0.500 & 0.016\\
4 & 0.9 & 500 && 0.339 && 0.538 & 0.832 & 0.057\\
4 & 0.9 & 5000 && 0.340 && 0.663 & 0.877 & 0.030\\
8 & 0.5 & 500 && 0.295 && 1.000 & 0.499 & 0.021\\
8 & 0.5 & 5000 && 0.295 && 1.000 & 0.500 & 0.007\\
8 & 0.9 & 500 && 0.598 && 0.711 & 0.881 & 0.030\\
8 & 0.9 & 5000 && 0.598 && 0.953 & 0.898 & 0.017\\
\hline
\end{tabular}
\end{table}

The experiment confirms the two halves of the argument.  The within-group
estimator converges to the wrong value, and for $T=2$ its mean agrees with the
limit $\rho_0-(1+\rho_0)/2$ implied by Proposition~\ref{prop:dyn_T2}(b).
Increasing $N$ tenfold does not move it.  For $\rho_0=0.5$, the recentred
estimator approaches the truth as $N$ increases, and for $T=2$ and $N=5000$ its standard
deviation is close to the value $0.055$ given by
Proposition~\ref{prop:dyn_T2}(d).  The experiment also shows the limit of the
repair.  When $\rho_0=0.9$ the initial deviations are small, the two roots of
the estimating equation are close, and in a large share of samples they merge
so that no interior maximum exists; conditional on existence the estimator is
biased downward.  With $T=2$ the share of such samples is almost the same at
$N=500$ and at $N=5000$.  Recentring removes the incidental-parameter bias,
but it cannot supply information that a persistent short panel does not contain.

\subsection{Dynamic panel simulations with non-Gaussian errors}
\label{subsec:dyn_nongaussian}

The drift in Proposition~\ref{prop:dyn_drift} and the centring of the
recentred equation use only the first two moments of the errors, although
Theorem~\ref{thm:dyn_ortho} is stated for the Gaussian likelihood.
Table~\ref{tab:dyn_nongaussian} repeats the experiment of
Table~\ref{tab:dyn_sim} with two non-Gaussian error laws of unit variance: a
centred chi-square law with one degree of freedom, which is strongly skewed,
and a Student law with five degrees of freedom, which has heavy tails.  The
effects are centred exponential variables, and the process is started 50
periods before the sample.

\begin{table}[ht]
\centering
\caption{Within-group and recentred estimators of $\rho_0$ with non-Gaussian
errors; 1{,}000 replications.  Columns are as in Table~\ref{tab:dyn_sim}.}
\label{tab:dyn_nongaussian}
\setlength{\tabcolsep}{5pt}
\begin{tabular}{rrrcrcrrrcrcrrr}
\hline
 & & && \multicolumn{5}{c}{Skewed errors} && \multicolumn{5}{c}{Heavy-tailed errors}\\
\cline{5-9}\cline{11-15}
 & & && Within && \multicolumn{3}{c}{Recentred} && Within && \multicolumn{3}{c}{Recentred}\\
\cline{5-5}\cline{7-9}\cline{11-11}\cline{13-15}
$T$ & $\rho_0$ & $N$ && Mean && Share & Median & s.d. && Mean && Share & Median & s.d.\\
\hline
2 & 0.5 & 500 && $-0.249$ && 0.627 & 0.366 & 0.156 && $-0.250$ && 0.691 & 0.425 & 0.147\\
2 & 0.5 & 5000 && $-0.250$ && 0.884 & 0.488 & 0.093 && $-0.249$ && 0.921 & 0.490 & 0.079\\
2 & 0.9 & 500 && $-0.047$ && 0.504 & 0.573 & 0.154 && $-0.047$ && 0.499 & 0.647 & 0.134\\
2 & 0.9 & 5000 && $-0.050$ && 0.521 & 0.725 & 0.092 && $-0.050$ && 0.542 & 0.761 & 0.081\\
4 & 0.5 & 500 && 0.087 && 0.969 & 0.494 & 0.067 && 0.088 && 0.993 & 0.496 & 0.057\\
4 & 0.5 & 5000 && 0.089 && 1.000 & 0.499 & 0.023 && 0.088 && 1.000 & 0.500 & 0.019\\
4 & 0.9 & 500 && 0.340 && 0.541 & 0.791 & 0.077 && 0.338 && 0.560 & 0.798 & 0.066\\
4 & 0.9 & 5000 && 0.340 && 0.590 & 0.858 & 0.042 && 0.340 && 0.601 & 0.867 & 0.038\\
8 & 0.5 & 500 && 0.293 && 1.000 & 0.498 & 0.023 && 0.295 && 1.000 & 0.500 & 0.021\\
8 & 0.5 & 5000 && 0.295 && 1.000 & 0.500 & 0.007 && 0.295 && 1.000 & 0.500 & 0.007\\
8 & 0.9 & 500 && 0.598 && 0.663 & 0.872 & 0.035 && 0.598 && 0.683 & 0.877 & 0.034\\
8 & 0.9 & 5000 && 0.598 && 0.885 & 0.897 & 0.019 && 0.598 && 0.912 & 0.897 & 0.018\\
\hline
\end{tabular}
\end{table}

The within-group estimator converges to the same wrong values as with
Gaussian errors, as Proposition~\ref{prop:dyn_drift} implies.  The recentred
estimator remains centred on the truth wherever it was centred before: for
$\rho_0=0.5$ its median is within $0.01$ of the true value for $T\ge4$, and
within $0.012$ for $T=2$ and $N=5000$.  What changes is its precision.  With
two periods the standard deviation at $N=5000$ rises from $0.058$ under
Gaussian errors to $0.093$ under the skewed law and $0.079$ under the
heavy-tailed law, in line with the fourth-moment term in the variance of
Proposition~\ref{prop:dyn_T2}(d), and an interior maximum is found less
often.  For $\rho_0=0.9$ the weak identification near the unit root is the
same under all three laws.

\subsection{Alternative groups for the dynamic panel model}
\label{subsec:dyn_groups}

Recentring repairs the score of an integrated criterion that is not the
likelihood of a quotient free of the parameter.  The alternative is to look
for a group that does satisfy H1.  The model admits two, and they correspond
to elimination and compression, the two alternatives in requirement H3 of
Section~\ref{subsec:principle}.

\medskip 
\noindent{\bf Translations of the whole path.}
Suppose there are no covariates and $|\rho|<1$, and write
$\alpha_i=(1-\rho)m_i$, so that
$y_{it}-m_i=\rho(y_{i,t-1}-m_i)+\varepsilon_{it}$.  The translation
\[
    (y_{i0},y_{i1},\ldots,y_{iT})\longmapsto
    (y_{i0}+a,y_{i1}+a,\ldots,y_{iT}+a),
    \qquad m_i\longmapsto m_i+a,
\]
acts on the full path of $T+1$ observations and does not involve any
parameter.  It preserves the model provided the distribution of the initial
deviation $y_{i0}-m_i$ does not depend on $m_i$, for instance under a
stationary start.  Then
Proposition~\ref{prop:haar_translation} applies with $A=\mathbf 1_{T+1}$: the
maximal invariant can be represented by the vector of first differences
$(\Delta_{i1},\ldots,\Delta_{iT})$.  Its density is proportional to the
normalised Haar likelihood by a parameter-independent Jacobian.  Its scores
are centred, and Proposition~\ref{prop:haar_reduction_principle} gives
consistency and asymptotic normality under its identification and regularity
conditions.  This is the route of the
stationary example in \citet{Moreira2009} and of the transformed likelihood
of \citet{HsiaoPesaranTahmiscioglu2002}.  It is an exact Haar solution.  Its
cost is that H1 has been bought with an assumption about the initial
condition, which was not needed in Section~\ref{sec:dynamic}.  If the
initial deviations are instead left unrestricted, they form a second set of
incidental parameters that the translation does not remove.

\medskip 
\noindent{\bf Orthogonal transformations across units.}
\citet{Moreira2009} conditions on the initial observations and lets the
orthogonal group act on the cross-section, $Y\mapsto gY$ for
$g\in O(N)$. Here $Y$ is the $N\times T$ matrix with entries
$y_{it}-y_{i0}$, so the transformed initial observations are zero and the
transformed effects are $\alpha_i-(1-\rho)y_{i0}$.  With Gaussian errors
this action is known and preserves $(\rho,\sigma^2)$.  The maximal invariant
is $Y'Y$, and the effects enter its distribution only through
$\sum_i\{\alpha_i-(1-\rho)y_{i0}\}^2/(N\sigma^2)$.  This is compression and
not elimination: one common nuisance parameter remains in the invariant likelihood, and that
likelihood does not factor across units.  In return the reduction keeps the
between-unit variation that translations discard, and the resulting estimator
is consistent whenever $NT\to\infty$.

\medskip 
\noindent{\bf No group at all.}
Instrumental-variable estimation of the differenced equation
\citep{ArellanoBond1991} requires neither a group nor a likelihood.  It is
valid for fixed $T$ under sequential moment conditions, which are weaker than
the homoskedasticity used in Section~\ref{sec:dynamic}, at the cost of
weak instruments when $\rho_0$ is close to one.


\section{Proofs and additional results for Section~\ref{sec:entropy}}
\label{app:proofs_entropy}

Throughout, expectations are taken at the true parameter values and
conditional on the covariates.

\subsection{Proof of Theorem~\ref{thm:entropy}}
\label{app:proof_entropy}

\noindent{\bf Part (a)}

\noindent{\bf The contrasts at a candidate orbit.}
At $\theta_0$, $Y_i-X_i\beta_0=Q_0a_i+\varepsilon_i$, so
$B_U'(Y_i-X_i\beta_0)=V_i+\nu_i$ with $V_i=B_U'\varepsilon_i$ and
$\nu_i=B_U'Q_0a_i$.  By definition $V_i$ has density
$h_U=h_U(\cdot;\eta_0)$, and $\nu_i$ is not random.

\medskip 
\noindent{\bf The expected Haar criterion.}
The Haar log-likelihood at $(\theta_0,U)$ is $\log h_U(V_i+\nu_i)$, so
\[
    \mathbb E\,\ell_{H,i}(\theta_0,U)=\int h_U(v)\log h_U(v+\nu_i)\,dv .
\]
The integral is well defined with values in $[-\infty,\infty)$: by the next step it
equals $-\mathcal E(U;\eta_0)$, which is finite by assumption, minus a
divergence, which is nonnegative and may be infinite.

\medskip 
\noindent{\bf Adding the entropy.}
Adding $\mathcal E(U;\eta_0)=-\int h_U(v)\log h_U(v)\,dv$ gives
\[
    \mathbb E\,\ell_{E,i}(\theta_0,U)
    =-\int h_U(v)\log\frac{h_U(v)}{h_U(v+\nu_i)}\,dv,
\]
which is $-\operatorname{KL}\{h_U\,\|\,h_U(\cdot+\nu_i)\}$, with
$h_U$ as the reference density in the expectation.

\medskip 
\noindent{\bf When the divergence vanishes.}
The divergence is nonnegative and is zero only if the two densities agree
almost everywhere.  A probability density cannot agree with a nontrivial
translate of itself, since repeated translation would then give it the same
mass on infinitely many disjoint sets.  Equality therefore requires
$\nu_i=0$.  The expression does not depend on the choice of $B_U$, because a
rotation of the basis rotates $V_i$ and $\nu_i$ together.

\medskip
\noindent{\bf Part (b)}

\noindent{\bf The average vanishes only if every shift is zero.}
By (a) each term of the average is nonpositive, so the average is zero if
and only if $\nu_i=B_U'Q_0a_i=0$ for every $i$.

\medskip 
\noindent{\bf From the shifts to the factor space.}
Since $N^{-1}\sum_i\nu_i\nu_i'=B_U'Q_0\mathcal A_NQ_0'B_U$ and the trace of
the left-hand side is $N^{-1}\sum_i\|\nu_i\|^2$, all shifts vanish if and
only if $B_U'Q_0\mathcal A_NQ_0'B_U=0$.  When $\mathcal A_N$ is positive
definite this gives $B_U'Q_0\mathcal A_N^{1/2}=0$ and hence $B_U'Q_0=0$.

\medskip 
\noindent{\bf The two spaces coincide.}
The equation $B_U'Q_0=0$ says that $\operatorname{span}(Q_0)$ is orthogonal
to $\operatorname{span}(B_U)$, that is, contained in
$\operatorname{span}(U)$.  The two spaces have dimension $r$, so they
coincide.

\medskip
\noindent{\bf Part (c)}

\noindent{\bf A maximum at the true orbit.}
By (a), for every loading the function
$U\mapsto\mathbb E\,\ell_{E,i}(\theta_0,U)$ is nonpositive on
$\operatorname{Gr}(r,T)$ and equals zero at $\operatorname{span}(Q_0)$.  It
therefore attains its maximum there.

\medskip 
\noindent{\bf The first-order condition.}
This function is the sum of $\mathbb E\,\ell_{H,i}(\theta_0,U)$ and
$\mathcal E(U;\eta_0)$, both differentiable at $Q_0$ by assumption.  The
Grassmann manifold has no boundary, so the derivative at a maximiser is
zero.  Differentiating the first term under the expectation gives
\[
    \mathbb E\,D_U\ell_{H,i}(\theta_0,Q_0)+D_U\mathcal E(Q_0;\eta_0)=0 .
\]
This is the first statement.  The second follows because
$D_U\ell_{E,i}=D_U\ell_{H,i}+D_U\mathcal E$ and the entropy is not random.

\medskip 
\noindent{\bf The drift formula.}
Comparison with \eqref{eq:orbit_drift} gives
$D_U\mathcal E(Q_0;\eta_0)[B_0H]=\mathbb E\{s_w(W_i)'HC_i\}$.

\medskip
\noindent{\bf Part (d)}

\noindent{\bf The Haar term and the entropy term.}
With $z=Y_i-X_i\beta-m_\eta$, the contrasts $B_U'z$ are
$N(0,B_U'\Sigma_\eta B_U)$ under the candidate values, so
\[
    \ell_{H,i}(\theta,U)
    =-\tfrac12\log|2\pi B_U'\Sigma_\eta B_U|
     -\tfrac12z'B_U(B_U'\Sigma_\eta B_U)^{-1}B_U'z .
\]
The entropy of that Gaussian law is
$\tfrac12\log|2\pi B_U'\Sigma_\eta B_U|+\tfrac12(T-r)$.  The determinants
cancel in the sum, which leaves
$\ell_{E,i}(\theta,U)=\tfrac12(T-r)
-\tfrac12z'B_U(B_U'\Sigma_\eta B_U)^{-1}B_U'z$.

\medskip 
\noindent{\bf The quadratic form.}
The identity
\[
    B_U(B_U'\Sigma_\eta B_U)^{-1}B_U'
    =\Sigma_\eta^{-1}
     -\Sigma_\eta^{-1}U(U'\Sigma_\eta^{-1}U)^{-1}U'\Sigma_\eta^{-1}
\]
holds because both sides map every column of $U$ to zero and both map
$\Sigma_\eta B_U$ to $B_U$, using $B_U'U=0$, and the columns of $U$ and of
$\Sigma_\eta B_U$ together span $\mathbb R^T$: if
$Uc+\Sigma_\eta B_Ud=0$, then premultiplying by $B_U'$ gives
$B_U'\Sigma_\eta B_Ud=0$, so $d=0$ and then $c=0$.

\medskip 
\noindent{\bf Generalised least squares.}
The right-hand side is the matrix of the residual quadratic form of the
generalised least-squares regression of $z$ on $U$:
$z'\{\Sigma_\eta^{-1}-\Sigma_\eta^{-1}U(U'\Sigma_\eta^{-1}U)^{-1}
U'\Sigma_\eta^{-1}\}z=\min_c(z-Uc)'\Sigma_\eta^{-1}(z-Uc)$.  This gives the
stated minimum.

\subsection{Latent-index models}
\label{app:additional_latent}

In the binary model
$Y_{it}=\mathbf 1\{X_{it}'\beta+\alpha_i+\varepsilon_{it}>0\}$ with standard
normal errors, the translation acts on the latent index and not on the
observed data, whose sample space $\{0,1\}^T$ admits no translation.
Integrating the likelihood over $\alpha_i$ with Lebesgue measure is then only
a flat-prior integrated likelihood.  It violates H2: for a unit with
$Y_{it}=1$ in every period the integrand
$\prod_t\Phi(X_{it}'\beta+a)$ tends to one as $a\to\infty$ and the integral
diverges.  For the remaining units the integral is finite but is not the
density of a maximal invariant, and Proposition~\ref{prop:haar_translation}(c)
gives no centring.  This is consistent with the result of
\citet{Chamberlain2010} that in binary panels with fixed $T$ the slope can
fail to be point identified outside the logistic case, where elimination
rests on a sufficient statistic and not on invariance.

\subsection{Proof of Theorem~\ref{thm:entropy_limit}}
\label{app:proof_entropy_limit}

Part (c) of the theorem uses the following regularity conditions.

\begin{assumption}[Regular estimating equations]
\label{ass:entropy}
Assumption~\ref{ass:regular} holds at the true orbit, and so does the
differentiability condition of Theorem~\ref{thm:entropy}(c).
On a neighbourhood of $\varphi_0$, $G_N$ is continuously differentiable,
$\partial G_N/\partial\varphi'$ converges in probability, uniformly, to a
matrix function that is continuous at $\varphi_0$ with nonsingular value
$\Gamma$ there.  The array $\{g_i(\varphi_0)\}$ satisfies Lindeberg's
condition, and the average of the covariance matrices of the
$g_i(\varphi_0)$ converges to a matrix $\Omega$.
\end{assumption}

\noindent{\bf Part (a)}

\noindent{\bf The structural block of the estimating function.}
At the true orbit, model \eqref{eq:IFE_model} is model
\eqref{eq:orbit_model} with the known matrix $A=F$, and
$\ell_{H,i}(\theta,Q_0)$ is its Haar log-likelihood.  The first block of
$g_i(\varphi_0)$ therefore has mean zero by
Proposition~\ref{prop:haar_translation}(c).

\medskip 
\noindent{\bf The orbit block.}
The path $U(H)$ has derivative $B_0H$ at $H=0$, so the second block is
$D_U\ell_{E,i}(\theta_0,Q_0)$ applied to the directions $B_0H$.  It has mean
zero by Theorem~\ref{thm:entropy}(c).

\medskip 
\noindent{\bf The gap at the truth.}
The first bracket in \eqref{eq:equilibrium_gap} at $(\theta_0,Q_0)$ is zero
because
\[
    \overline{Q}_N(\theta_0,Q_0)-\overline{Q}_N(\theta,Q_0)
    =N^{-1}\sum_i\mathrm{KL}_i(\theta)\ge0
\]
by Proposition~\ref{prop:haar_reduction_principle}(a), so that
$\theta_0$ attains the supremum.  The second bracket is zero because, by
Theorem~\ref{thm:entropy}(a), each $\mathbb E\,\ell_{E,i}(\theta_0,U)$ is
nonpositive and vanishes at $U=Q_0$, so that $Q_0$ attains the supremum.
Hence $\Delta_N(\theta_0,Q_0)=0$.

\medskip
\noindent{\bf Part (b)}

\noindent{\bf Sample and population gaps.}
Let $\widehat\Delta_N$ be defined as in \eqref{eq:equilibrium_gap} with
$Q_N$ and $L_N$ in place of $\overline{Q}_N$ and $\overline{L}_N$, and let
$e_N=\sup|Q_N-\overline{Q}_N|+\sup|L_N-\overline{L}_N|$, which tends to zero in
probability by Assumption~\ref{ass:entropy_consistency}(i).  Each supremum in
\eqref{eq:equilibrium_gap} changes by at most the corresponding uniform
error when the sample criteria replace the population ones, and so does
each criterion value.  Therefore
$|\widehat\Delta_N(\theta,U)-\Delta_N(\theta,U)|\le2e_N$ uniformly in
$(\theta,U)$.

\medskip 
\noindent{\bf The gap at the estimator.}
The two conditions that define the estimator give
$\widehat\Delta_N(\widehat\theta_E,\widehat U_E)=o_p(1)$, because each
bracket of $\widehat\Delta_N$ is nonnegative and, by those conditions, at
most $o_p(1)$ there.  By the uniform bound just established,
\[
    \Delta_N(\widehat\theta_E,\widehat U_E)
    \le\widehat\Delta_N(\widehat\theta_E,\widehat U_E)+2e_N=o_p(1).
\]

\medskip 
\noindent{\bf Separation.}
Given $\epsilon>0$, Assumption~\ref{ass:entropy_consistency}(ii) supplies
$\delta>0$ and $N_0$ such that $\Delta_N\ge\delta$ whenever
$d\{(\theta,U),(\theta_0,Q_0)\}\ge\epsilon$ and $N\ge N_0$.  Therefore, for
$N\ge N_0$,
\[
    P\big[d\{(\widehat\theta_E,\widehat U_E),(\theta_0,Q_0)\}\ge\epsilon\big]
    \le P\{\Delta_N(\widehat\theta_E,\widehat U_E)\ge\delta\}\to0.
\]

\medskip
\noindent{\bf Part (c)}

\noindent{\bf Local coordinates.}
If $Q_0'U$ is nonsingular, then
$U(Q_0'U)^{-1}=(Q_0Q_0'+B_0B_0')U(Q_0'U)^{-1}=Q_0+B_0H$ with
$H=B_0'U(Q_0'U)^{-1}$.  Every subspace close enough to
$\operatorname{span}(Q_0)$ is therefore the span of $Q_0+B_0H$ for exactly
one $H$, and this $H$ is continuous in $UU'$ and zero at $Q_0$.  By (b),
$\widehat U_E$ spans the same space as $U(\widehat H)$ with
$\widehat H\overset p\to0$, so $\widehat\varphi\overset p\to\varphi_0$.

\medskip 
\noindent{\bf First-order conditions.}
Because $\theta_0$ is interior and $\widehat\theta_E$ maximises
$Q_N(\cdot,\widehat U_E)$ over $\Theta$, the first block of $G_N$ vanishes at
$\widehat\varphi$ with probability tending to one.  Because the Grassmann
manifold has no boundary and $\widehat U_E$ maximises
$L_N(\widehat\theta_E,\cdot)$, the second block vanishes as well.  Hence
$G_N(\widehat\varphi)=0$ with probability tending to one, and
$\widehat\varphi$ lies in the neighbourhood of
Assumption~\ref{ass:entropy}.

\medskip 
\noindent{\bf Expansion.}
Integrating the Jacobian along the segment from $\varphi_0$ to
$\widehat\varphi$ gives
\[
    0=\sqrt N\,G_N(\varphi_0)
     +\widetilde\Gamma_N\sqrt N(\widehat\varphi-\varphi_0),
    \qquad
    \widetilde\Gamma_N=\int_0^1
    \frac{\partial G_N\{\varphi_0+t(\widehat\varphi-\varphi_0)\}}
         {\partial\varphi'}\,dt.
\]
The segment shrinks to $\varphi_0$ in probability.  By the uniform
convergence and the continuity of the limit in
Assumption~\ref{ass:entropy}, $\widetilde\Gamma_N$ converges in probability
to $\Gamma$, which is nonsingular.

\medskip 
\noindent{\bf The limit.}
The summands of $G_N(\varphi_0)$ are independent with mean zero by (a), so
by the Lindeberg--Feller theorem $\sqrt N\,G_N(\varphi_0)$ converges in
distribution to $N(0,\Omega)$.  Solving the expansion for
$\sqrt N(\widehat\varphi-\varphi_0)$ and applying Slutsky's lemma gives a
normal limit with mean zero and covariance matrix
$\Gamma^{-1}\Omega(\Gamma^{-1})'$.

\medskip
\noindent{\bf Part (d)}

\noindent{\bf The expected criterion along the path.}
By Theorem~\ref{thm:entropy}(a),
\[
    \frac1N\sum_{i=1}^N\mathbb E\,\ell_{E,i}\{\theta_0,U(H)\}
    =-\frac1N\sum_{i=1}^N
      \operatorname{KL}\{h_{U(H)}\,\|\,h_{U(H)}(\cdot+\nu_i(H))\},
\]
where $\nu_i(H)=B_{U(H)}'Q_0a_i$.

\medskip 
\noindent{\bf The retained shift.}
The matrix $B_0-Q_0H'$ is orthogonal to $Q_0+B_0H$, since
$(B_0-Q_0H')'(Q_0+B_0H)=H-H=0$, and
$(B_0-Q_0H')'(B_0-Q_0H')=I+HH'$.  An orthonormal basis of the complement of
$\operatorname{span}\{U(H)\}$ is therefore $(B_0-Q_0H')(I+HH')^{-1/2}$,
which gives
$\nu_i(H)=-(I+HH')^{-1/2}Ha_i=-Ha_i+O(\|H\|^3)$.

\medskip 
\noindent{\bf Second-order expansion.}
For a small shift the divergence between a density and its translate is
$\operatorname{KL}=\tfrac12\nu'J_W\nu+o(\|\nu\|^2)$, where $J_W$ is the
Fisher information for location: expanding $\log h(w+\nu)$ to the second
order in $\nu$ under the integral, the linear term vanishes because
$\int\nabla h(w)\,dw=0$, and
$-\int h(w)\nabla^2\log h(w)\,dw=J_W$.  At $H=0$ the contrast density is
that of $W_i=B_0'\varepsilon_i$, and its dependence on $H$ affects only
terms of higher order.  With the retained shift above,
\[
    \frac1N\sum_{i=1}^N\mathbb E\,\ell_{E,i}\{\theta_0,U(H)\}
    =-\frac1{2N}\sum_{i=1}^Na_i'H'J_WHa_i+o(\|H\|^2)
    =-\tfrac12\operatorname{tr}(H'J_WH\mathcal A_N)+o(\|H\|^2),
\]
and $\operatorname{tr}(H'J_WH\mathcal A_N)
=\operatorname{vec}(H)'(\mathcal A_N\otimes J_W)\operatorname{vec}(H)$.

\medskip 
\noindent{\bf Hessian.}
The orbit block of $\Gamma$ is the limit of the Hessian of the left-hand
side at $H=0$.  When the expansion may be differentiated twice, this is
$-\lim_N(\mathcal A_N\otimes J_W)$.

\subsection{Proof of Theorem~\ref{thm:separable}}
\label{app:proof_separable}

\noindent{\bf Part (a)}

\noindent{\bf Separability is sufficient.}
By \eqref{eq:entropy_adjusted},
$L_N(\theta,U)=Q_N(\theta,U)+\mathcal E(U;\eta)$.  If
$\mathcal E(U;\eta)=a(U)+b(\eta)$, then $P_N=Q_N+a$ satisfies
\[
    \frac{\partial P_N}{\partial\theta}=\frac{\partial Q_N}{\partial\theta},
    \qquad
    D_UP_N=D_UQ_N+D_Ua(U)=D_UQ_N+D_U\mathcal E(U;\eta)=D_UL_N,
\]
because $b(\eta)$ does not depend on $U$.  The estimating functions are
therefore the gradient of $P_N$.

\medskip
\noindent{\bf Separability is necessary.}
Suppose that there is a continuously differentiable function $\Psi$ on
$\Theta\times\operatorname{Gr}(r,T)$ with
$\partial\Psi/\partial\theta=\partial Q_N/\partial\theta$ and
$D_U\Psi=D_UL_N$, and let $c=\Psi-Q_N$.  Then
\[
    \frac{\partial c}{\partial\theta}=0,
    \qquad
    D_Uc(\theta,U)=D_UL_N-D_UQ_N=D_U\mathcal E(U;\eta).
\]
For each $U$, the first equation and the chain rule along the paths in
$\Theta$ make $c$ constant in $\theta$. Write it as $c(U)$. The second equation gives
$D_U\{\mathcal E(U;\eta)-c(U)\}=0$ for every $\eta$ and every $U$.  The
Grassmann manifold is connected, so for each $\eta$ the function
$U\mapsto\mathcal E(U;\eta)-c(U)$ is constant.  Calling the constant
$b(\eta)$ gives \eqref{eq:separable} with $a=c$, and
$\Psi=Q_N+c=P_N$. The same path argument on the product domain shows that
two potentials with the same gradient differ by a constant.

\medskip
\noindent{\bf Part (b)}

\noindent{\bf One criterion for both steps.}
The function $a(U)$ does not depend on $\theta$, so the structural step,
which maximises $Q_N(\cdot,U)$, maximises $P_N(\cdot,U)$.  The orbit step
maximises $L_N(\theta,\cdot)=P_N(\theta,\cdot)+b(\eta)$, and $b(\eta)$ does
not depend on $U$, so it maximises $P_N(\theta,\cdot)$.  Neither step can
decrease $P_N$.

\medskip
\noindent{\bf The expected criterion.}
Under the true parameter values,
$B_U'(Y_i-X_i\beta)=B_U'\varepsilon_i+s_i(\beta,U)$, which has density
$h_U(\cdot-s_i;\eta_0)$ with $s_i=s_i(\beta,U)$.  Hence
\[
    \mathbb E\,\ell_{H,i}(\theta,U)
    =\int h_U(w-s_i;\eta_0)\log h_U(w;\eta)\,dw
    =-\mathcal E(U;\eta_0)
     -\operatorname{KL}\big\{h_U(\cdot-s_i;\eta_0)\,\big\|\,
       h_U(\cdot;\eta)\big\},
\]
because translation does not change entropy.  By \eqref{eq:separable},
$a(U)-\mathcal E(U;\eta_0)=-b(\eta_0)$.  Adding $a(U)$ and averaging over
$i$ gives the equality in \eqref{eq:potential_KL}.  At $(\theta_0,Q_0)$,
$s_i(\beta_0,Q_0)=B_0'F\lambda_i=0$ and $\eta=\eta_0$, so every divergence
vanishes and $\overline{P}_N(\theta_0,Q_0)=-b(\eta_0)$.  Divergences are
nonnegative, which gives the inequality. The identity also holds when a
divergence is infinite, since the contrast entropy at $\eta_0$ is finite.

\medskip
\noindent{\bf Part (c)}

\noindent{\bf Continuity of the contrast densities.}
Write $d=T-r$ and use the compact set of orthonormal $T\times d$ frames
$B$, with $U$ denoting the orthogonal complement of $\operatorname{span}(B)$.
Let $P_{B,\eta}$ be the law of $B'\varepsilon_i$ under $\eta$ and
$P_{B,\eta_0}^{s}$ the law of $B'\varepsilon_i+s$ under $\eta_0$. Put
\[
    D(B,s,\eta)=\|P_{B,\eta_0}^{s}-P_{B,\eta}\|_{\mathrm{TV}},
    \qquad \|P-Q\|_{\mathrm{TV}}=\sup_A|P(A)-Q(A)|.
\]
Rotating the frame and the coordinates of the shift together preserves
$D(B,s,\eta)$ and $\|s\|$. If $B_n\to B$, complete the frames to orthogonal
matrices with a common limit. Such completions can be obtained by
projecting a fixed complement of $B$ onto the complement of $B_n$ and
orthonormalising it. This gives $B_n=R_nB$ with $R_n\to I_T$. The law of
$B_n'\varepsilon_i$ is then the image under $B'$ of the law of
$R_n'\varepsilon_i$. Rotations and translations are continuous in $L^1$,
as is $\eta\mapsto g(\cdot;\eta)$ by Assumption~\ref{ass:design}(ii).
Total variation contracts under $B'$. It follows that
$P_{B_n,\eta_n}\to P_{B,\eta}$ and $P_{B_n,\eta_0}^{s_n}\to P_{B,\eta_0}^s$
in total variation if $(B_n,s_n,\eta_n)\to(B,s,\eta)$. Thus $D$ is continuous.

\medskip
\noindent{\bf A uniform lower bound on total variation.}
For $\delta>0$ and $\epsilon'>0$, let $\tau(\delta,\epsilon')$ be the
infimum of $D(B,s,\eta)$ over all frames and all $(s,\eta)$ with
$\|s\|\ge\delta$ or $\|\eta-\eta_0\|\ge\epsilon'$. The error laws form a
compact set in total variation by Assumption~\ref{ass:design}(ii).
For any prescribed accuracy one can cover this set by a finite net in
total variation and choose a common radius controlling the tails of the
net laws. Thus $p(R)=\sup_\eta P_\eta(\|\varepsilon_i\|\ge R)\to0$.
Choose $R$ with $p(R)<1/4$, and let $A_R=\{w:\|w\|<R\}$. Then
$P_{B,\eta}(A_R)\ge 1-p(R)$. If $\|s\|\ge 2R$ and $B'\varepsilon_i+s\in A_R$,
the triangle inequality gives $\|\varepsilon_i\|\ge\|B'\varepsilon_i\|>R$,
so $P_{B,\eta_0}^{s}(A_R)\le p(R)$. Consequently
$D(B,s,\eta)\ge 1-2p(R)>1/2$ for all such shifts. For $\|s\|\le 2R$,
the separation restriction defines a closed subset of the product of
the frame space, the closed ball and the compact error-parameter set.
On this subset $D$ is continuous and positive by
Assumption~\ref{ass:design}(ii), so it has a positive minimum whenever
the subset is nonempty. Combining the bounds gives $\tau(\delta,\epsilon')>0$.

\medskip
\noindent{\bf Separation.}
Given $\epsilon>0$, take $\delta$ and $\pi\le1$ from
Assumption~\ref{ass:design}(i) for $\epsilon/2$, and let
$\tau=\tau(\delta,\epsilon/2)$.  Let $(\theta,U)$ be at distance $\epsilon$
or more from $(\theta_0,Q_0)$.  Since
$\|\theta-\theta_0\|\le\|\beta-\beta_0\|+\|\eta-\eta_0\|$, one of two cases
holds.  If $\|\beta-\beta_0\|+\|UU'-Q_0Q_0'\|\ge\epsilon/2$, at least
$\pi N$ units have $\|s_i(\beta,U)\|\ge\delta$, and for each of them the
divergence in \eqref{eq:potential_KL} is at least $2\tau^2$ by Pinsker's
inequality \citep[Chapter~2]{Tsybakov2009}.  Otherwise
$\|\eta-\eta_0\|\ge\epsilon/2$, and the divergence is at least $2\tau^2$
for every unit.  In both cases, by part (b), for all large $N$,
\[
    \overline{P}_N(\theta,U)\le\overline{P}_N(\theta_0,Q_0)-2\pi\tau^2 .
\]

\medskip
\noindent{\bf Consistency.}
Since $P_N-\overline{P}_N=Q_N-\overline{Q}_N$,
$e_N=\sup|P_N-\overline{P}_N|\overset p\to0$ by
Assumption~\ref{ass:entropy_consistency}(i).  Then
\[
    \overline{P}_N(\widehat\theta,\widehat U)
    \ge P_N(\widehat\theta,\widehat U)-e_N
    \ge P_N(\theta_0,Q_0)-e_N-o_p(1)
    \ge\overline{P}_N(\theta_0,Q_0)-2e_N-o_p(1).
\]
The right-hand side exceeds $\overline{P}_N(\theta_0,Q_0)-2\pi\tau^2$ with
probability tending to one, so by the separation bound the probability that
$(\widehat\theta,\widehat U)$ is at distance $\epsilon$ or more from
$(\theta_0,Q_0)$ tends to zero.

\medskip
\noindent{\bf Part (d)}

\noindent{\bf A known error law.}
With $\eta_0$ known the entropy is a function of $U$ alone, so
\eqref{eq:separable} holds with $b=0$.  The set of values of $\eta$ is a
single point, and $h_U(\cdot-s;\eta_0)=h_U(\cdot;\eta_0)$ implies $s=0$,
because no probability density is invariant under a nonzero translation.

\medskip
\noindent{\bf A scale family.}
If $\varepsilon_i=\sigma\varepsilon_i^\circ$, then
$B_U'\varepsilon_i=\sigma B_U'\varepsilon_i^\circ$ and
$h_U(w;\sigma)=\sigma^{-(T-r)}h_U^\circ(w/\sigma)$, where $h_U^\circ$ is the
density of $B_U'\varepsilon_i^\circ$. The change of variable
$v=w/\sigma$ gives
\[
    \mathcal E(U;\sigma)=a(U)+(T-r)\log\sigma,
\]
where $a(U)$ is the entropy of $B_U'\varepsilon_i^\circ$.
Rescaling is continuous in $L^1$ and the
interval $[\underline\sigma,\overline\sigma]$ is compact.  If
$h_U(\cdot-s;\sigma_0)=h_U(\cdot;\sigma)$, the two densities have the same
entropy, and translation does not change entropy, so
$(T-r)\log\sigma_0=(T-r)\log\sigma$ and $\sigma=\sigma_0$.  The density
$h_U(\cdot;\sigma_0)$ is then invariant under translation by $s$, which
requires $s=0$.

\medskip
\noindent{\bf An elliptical family.}
Let $d=T-r\ge1$ and $C=B_U'\Sigma_0^{1/2}$, a $d\times T$ matrix of full
row rank. Write $C=(CC')^{1/2}O$, where $O=(CC')^{-1/2}C$ has orthonormal
rows. Complete $O$ to an orthogonal $T\times T$ matrix. Spherical symmetry
then gives $O\zeta_i$ the same law as the first $d$ coordinates of
$\zeta_i$. Their density $k(\cdot;\eta)$ is spherically symmetric on
$\mathbb R^d$ and independent of $U$. Hence
$B_U'\varepsilon_i=(B_U'\Sigma_0B_U)^{1/2}O\zeta_i$ has density
\[
    h_U(w;\eta)
    =\det(B_U'\Sigma_0B_U)^{-1/2}\,
     k\{(B_U'\Sigma_0B_U)^{-1/2}w;\eta\},
\]
and
$\mathcal E(U;\eta)=\tfrac12\log\det(B_U'\Sigma_0B_U)+b(\eta)$, where
$b(\eta)$ is the entropy of $k(\cdot;\eta)$. The determinant is invariant
under orthogonal changes of basis and is smooth in local coordinates on
the Grassmann manifold. The fixed invertible map $\Sigma_0^{1/2}$ preserves
$L^1$ distances between the full densities, so continuity in $\eta$
carries over to $g(\cdot;\eta)$. To verify the identification requirement
of Assumption~\ref{ass:design}(ii), suppose that
$h_U(\cdot-s;\eta_0)=h_U(\cdot;\eta)$.  The density on the right is symmetric
about the origin and the density on the left is symmetric about $s$.  A
density that is symmetric about both points is invariant under translation
by $2s$, so $s=0$. Undoing the invertible map $(CC')^{1/2}$ gives
$k(\cdot;\eta)=k(\cdot;\eta_0)$ almost everywhere. In particular the
first coordinates have the same law. If $\varphi_\eta$ is the
characteristic function of $\zeta_i$ under $\eta$ and $e_1$ is the first
coordinate vector, spherical symmetry gives
$\varphi_\eta(t)=\varphi_\eta(\|t\|e_1)$ for every $t\in\mathbb R^T$.
The marginal equality gives $\varphi_\eta(\|t\|e_1)=
\varphi_{\eta_0}(\|t\|e_1)$. Equality of the full characteristic functions
therefore gives equality of the full laws, so the identified family
has $\eta=\eta_0$.

\subsection{A counterexample: unrestricted Gaussian covariance}
\label{app:gaussian_gap}

\begin{proposition}[Unrestricted Gaussian covariance]
\label{prop:gaussian_gap}
Consider model \eqref{eq:IFE_model} with $T>r$ fixed, no regressors and
errors that are $N(0,\Sigma)$ with $\eta=\Sigma$ unrestricted.  Then
$\Delta_N(\Sigma^\dagger,U)=0$ for every $U$, where
$\Sigma^\dagger=\Sigma_0+N^{-1}\sum_iF\lambda_i\lambda_i'F'$, and
Assumption~\ref{ass:entropy_consistency}(ii) fails.
\end{proposition}

\noindent{\em Proof.}

\noindent{\bf The structural step.}
The contrast distributions under the true and candidate models are,
respectively,
\[
    N(B_U'F\lambda_i,B_U'\Sigma_0B_U)
    \quad\text{and}\quad N(0,B_U'\Sigma B_U).
\]
Hence
\[
    \overline{Q}_N(\Sigma,U)
    =-\tfrac12\log\det(2\pi B_U'\Sigma B_U)
     -\tfrac12\operatorname{tr}\{(B_U'\Sigma B_U)^{-1}B_U'\Sigma^\dagger B_U\}.
\]
For a positive definite matrix $S$, the function
$V\mapsto-\log\det V-\operatorname{tr}(V^{-1}S)$ is maximised at $V=S$
only.  The criterion is therefore maximised over $\Sigma$ exactly when
$B_U'\Sigma B_U=B_U'\Sigma^\dagger B_U$, in particular at
$\Sigma=\Sigma^\dagger$ for every $U$, with maximal value
$-\tfrac12\log\det(2\pi e\,B_U'\Sigma^\dagger B_U)$.

\medskip 
\noindent{\bf The orbit step.}
The entropy of $N(0,B_U'\Sigma^\dagger B_U)$ is the negative of that value,
so $\overline{L}_N(\Sigma^\dagger,U)=0$ for every $U$.

\medskip 
\noindent{\bf The gap vanishes at every orbit.}
Both brackets of \eqref{eq:equilibrium_gap} vanish at $(\Sigma^\dagger,U)$:
$\Sigma^\dagger$ maximises $\overline{Q}_N(\cdot,U)$, and
$\overline{L}_N(\Sigma^\dagger,\cdot)$ is constant.  Since
$\|UU'-Q_0Q_0'\|$ can be made as large as the manifold allows, the infimum
in Assumption~\ref{ass:entropy_consistency}(ii) is zero.

\subsection{A numerical check of equilibrium identification}
\label{app:gap_check}

The frontier family is not covered by Theorem~\ref{thm:separable}, because
its entropy is not separable in the ratio $\lambda$ or in period-specific
noise variances.  This subsection checks
Assumption~\ref{ass:entropy_consistency}(ii) numerically in the designs in
which the family is used.  For a candidate orbit $U$, let $\theta^*(U)$
maximise $\overline{Q}_N(\cdot,U)$, and define the profile gap
\begin{equation}
    g(U)=\Delta_N\{\theta^*(U),U\}
    =\sup_{U'}\overline{L}_N\{\theta^*(U),U'\}-\overline{L}_N\{\theta^*(U),U\}.
\label{eq:profile_gap}
\end{equation}
The first term of \eqref{eq:equilibrium_gap} vanishes at $\theta^*(U)$, so,
when the maximiser is unique, the population system has an equilibrium at
the orbit $U$ if and only if $g(U)=0$.  A profile gap that is positive away
from the true orbit therefore means that the system has no second
equilibrium, which is the global requirement left open in
Section~\ref{subsec:entropy_limit}.

\medskip
\noindent{\bf Designs.}
The first two designs are the homoskedastic and heteroskedastic designs of
Table~\ref{tab:ife_sim}, with $T=3$; the structural parameter consists of
the noise standard deviations and $\sigma_u$.  The third is calibrated to
the application, with $T=5$: the factor direction, the time effects, the
five noise standard deviations and $\sigma_u$ are the entropy-adjusted
estimates of Table~\ref{tab:rice}, and the loadings are the generalised
least-squares estimates for the 800 households.  In this design the slope
coefficients are held at their true values, so the structural parameter
consists of the time effects and the six scale parameters.  Expectations
are computed from simulated units with common random numbers, 30{,}000 in
the first two designs and 4{,}800 in the third, in which each household's
loading is used six times.  Candidate orbits are placed at eleven angles
from the true orbit, between $0.05$ and $\pi/2$ radians: eight equally
spaced directions at each angle when $T=3$, and five random directions at
each of the ten angles from $0.1$ when $T=5$.  The maximiser $\theta^*(U)$
is computed from two starting values, the true value and a random
perturbation of it, and the supremum over $U'$ from three, namely the true
orbit, the candidate and equal weights. At a fixed structural estimate,
the orbit search gives a lower bound on the gain from optimising the orbit
criterion, conditional on the numerical approximation used for that
criterion.

\medskip
\noindent{\bf Time effects.}
At a given unit direction $U$, the Haar criterion depends on the wave
effects only through $B_U'\gamma$. Let $\gamma_\perp=B_UB_U'\gamma$ be
the orthogonal representative of these contrasts. When $U_1\ne0$, the
representative used in the application is
\[
    \gamma^{(1)}=\gamma_\perp-\frac{\gamma_{\perp,1}}{U_1}U,
    \qquad \gamma^{(1)}_1=0.
\]
This representation becomes ill conditioned as $U_1$ approaches zero.
In particular the coefficient of $U$ diverges if $\gamma_{\perp,1}$ stays
away from zero. At the entropy-adjusted fit in Table~\ref{tab:rice}, the
first factor weight is $0.246$ (standard error $0.038$), so the
$\gamma_1=0$ normalisation is locally regular. The gap check uses
$\gamma_\perp$ in the structural step
and holds it fixed in the comparison across orbits. The profile gap can
depend on this choice because it compares criterion values at other orbits.

At an interior structural solution with $U_1\ne0$, the two normalisations
give the same first-order conditions. The wave-effect score sets the
average contrast location score $\overline{s}_w$ to zero because
$d\gamma\mapsto B_U'd\gamma$ maps the admissible wave-effect directions
onto the contrast space under either normalisation. For a tangent
direction $\dot U$, replacing $\gamma$ by $\gamma+cU$ changes the orbit
derivative at $U$ by $c\,\overline{s}_w'B_U'\dot U$. This follows from
$D_UB_U'[\dot U]U=-B_U'\dot U$, obtained by differentiating $B_U'U=0$.
The derivative change vanishes at the structural solution.

\begin{table}[h]
\centering
\caption{Profile gap \eqref{eq:profile_gap} in the frontier family: minimum
and median over the candidate orbits at each angle from the true orbit
(eight orbits per angle in the two designs with $T=3$, of which four are
distinct at the angle $\pi/2$, and five in the design calibrated to the
application).}
\label{tab:gap_check}
\begin{tabular}{lcccccc}
\hline
 & \multicolumn{2}{c}{Homoskedastic, $T=3$} & \multicolumn{2}{c}{Heteroskedastic, $T=3$}
 & \multicolumn{2}{c}{Rice design, $T=5$}\\
Angle & Minimum & Median & Minimum & Median & Minimum & Median\\
\hline
0.05 & 0.001 & 0.002 & 0.001 & 0.002 & -- & --\\
0.10 & 0.007 & 0.011 & 0.005 & 0.008 & 0.008 & 0.010\\
0.20 & 0.022 & 0.040 & 0.023 & 0.034 & 0.027 & 0.033\\
0.30 & 0.047 & 0.078 & 0.059 & 0.098 & 0.042 & 0.065\\
0.45 & 0.041 & 0.108 & 0.035 & 0.121 & 0.095 & 0.138\\
0.60 & 0.092 & 0.163 & 0.042 & 0.129 & 0.166 & 0.210\\
0.80 & 0.173 & 0.235 & 0.085 & 0.213 & 0.208 & 0.274\\
1.00 & 0.255 & 0.297 & 0.066 & 0.246 & 0.320 & 0.450\\
1.20 & 0.314 & 0.342 & 0.057 & 0.270 & 0.141 & 0.525\\
1.40 & 0.348 & 0.368 & 0.139 & 0.303 & 0.472 & 0.497\\
$\pi/2$ & 0.360 & 0.375 & 0.163 & 0.297 & 0.377 & 0.523\\
\hline
\end{tabular}
\end{table}

\medskip
\noindent{\bf Results.}
Table~\ref{tab:gap_check} reports the profile gap.  It is positive at every
candidate orbit; there are 218 distinct ones, since opposite directions at
the angle $\pi/2$ give the same orbit when $T=3$. The small-angle values
are compatible with quadratic growth. For angles of $0.1$ or more
the ratio of the gap to the squared sine of the angle is at least $0.21$,
$0.06$ and $0.16$ in the three designs.  The row for the angle $0.05$ is at
the limit of the numerical resolution, because the equilibrium of a
simulated population is itself displaced from the true orbit by about
$0.01$.  The heteroskedastic design has the smallest gaps away from the
truth, with a minimum of $0.035$ for angles of $0.3$ or more.

The integral over the loading in the Haar likelihood is computed by a
20-node Gauss--Hermite rule, which loses accuracy when a noise standard
deviation is close to zero.  At 50 of the candidate orbits the maximiser
$\theta^*(U)$ had a scale parameter more than $e^{1.5}$ times larger or
smaller than its true value, and at these the criterion was recomputed with
200 nodes.  The two values differed by more than $5\times10^{-4}$ at twelve
orbits, eleven of them in the heteroskedastic design, and for these the
maximiser and the gap were recomputed with 60 nodes and three starting
values; the table reports the recomputed gaps.  At three of the twelve the
criterion with 60 nodes still differs from that with 200 nodes, by at most
$0.015$, at orbits whose gaps are $0.066$ or more.

As a second check, the alternating algorithm was applied to the simulated
populations themselves, started from the candidate orbits with the smallest
gaps among those at an angle of $0.3$ or more, and in the rice design also
from equal weights.  The orbit step was started from the current orbit only,
and the first structural step from a perturbed parameter value.
In the rice design the algorithm reached the equilibrium of the simulated
population from all three starting orbits, in ten or eleven iterations.  In
the heteroskedastic design it did so from two of four starting orbits, and
from the other two it stopped at orbits with errors $0.24$ and $0.41$.
Neither point is an equilibrium, because at neither is the structural
parameter a maximiser of $\overline{Q}_N(\cdot,U)$.  At the first the structural
step had converged to a local maximiser with $\sigma_u$ close to zero, a
boundary solution of the frontier likelihood at which the criterion is
$0.056$ below its maximum.  At the second it had converged to a point with a
noise standard deviation of $0.011$, where the 20-node rule overstates the
criterion: with 200 nodes the criterion there is $0.015$ below its value at
the regular maximiser.  When the structural step was restarted from its
maximiser, the algorithm reached the equilibrium from both orbits.  The
practical lesson is that the structural step should be started from several
values, and that the quadrature should be checked when an estimated noise
standard deviation is close to zero.

The gap checks cover a finite set of orbits in three designs and hold
the slope coefficients fixed in the rice design. The global separation
in Assumption~\ref{ass:entropy_consistency}(ii) remains an assumption.

\subsection{A longer panel and two factors}
\label{app:ife_rk}

Table~\ref{tab:ife_rk} extends the heteroskedastic experiment of
Section~\ref{subsec:entropy_numerical} in two directions, with $N=1000$,
$\sigma_u=1$ and the three estimators defined there.  In the first design
$T=8$ and $r=1$, with $F=(1,0.6,1.4,0.8,1.2,1,0.7,1.3)'$, loadings
$N(0.5,0.5^2)$ and noise standard deviations\\
$(0.3,0.5,0.9,0.3,0.5,0.9,0.3,0.9)$.  In the second $T=5$ and $r=2$: the
factors are a constant and the linear trend $(-1,-0.5,0,0.5,1)'$, with
independent loadings $N(0.5,0.5^2)$ and $N(0,0.7^2)$ and noise standard
deviations $(0.3,0.4,0.9,0.5,0.7)$.  The entropy is computed from $4N$
simulated error vectors in the first design and $2N$ in the second, and the orbit error is the sine of the
largest principal angle between the estimated and true factor spaces.  For
one factor the integral over the loading is computed by the piecewise rule
of Appendix~\ref{app:quadrature}.  For
two factors the integral over the loadings is computed by a nested adaptive
Gauss--Hermite rule with $16^2$ nodes, all derivatives are analytic, and the
entropy-adjusted estimator is obtained by solving its estimating equations
from the true values, with alternating maximisation as a fallback, which was
used in 17 per cent of the replications; where both converge they give the same
estimates.

\begin{table}[h]
\centering
\caption{Estimators with period-specific noise variances, $N=1000$: mean
and, in parentheses, standard deviation over 100 replications ($T=8$, $r=1$)
and 86 replications ($T=5$, $r=2$).  True values are $\sigma_{v1}=0.3$ and
$\sigma_u=1$.}
\label{tab:ife_rk}
\begin{tabular}{llccc}
\hline
Design & Estimator & $\widehat\sigma_{v1}$ & $\widehat\sigma_u$ & Orbit error\\
\hline
$T=8$, $r=1$ & Raw Haar         & 0.342 (0.032) & 0.965 (0.028) & 0.095 (0.013)\\
             & Second moment    & 0.338 (0.033) & 0.968 (0.028) & 0.090 (0.015)\\
             & Entropy adjusted & 0.291 (0.039) & 1.004 (0.025) & 0.036 (0.010)\\[1mm]
$T=5$, $r=2$ & Raw Haar         & 0.554 (0.115) & 0.593 (0.239) & 0.245 (0.069)\\
             & Second moment    & 0.541 (0.120) & 0.626 (0.246) & 0.234 (0.072)\\
             & Entropy adjusted & 0.323 (0.132) & 0.954 (0.124) & 0.085 (0.033)\\
\hline
\end{tabular}
\end{table}

With eight periods the raw and second-moment estimators are less distorted
than with three, but they retain appreciable orbit error.  Their mean orbit
error is about $0.09$ and $\sigma_u$ is underestimated, whereas the
entropy-adjusted estimator has smaller structural bias and mean orbit error
$0.036$.  With two factors every estimator is noisier, since the second factor is weaker than the first.  The raw and
second-moment estimators are bimodal: in 52 and 46 of the 86 replications
they settle on an orbit that roughly halves the estimate of $\sigma_u$.  The
entropy-adjusted estimator did so in two replications.  It has the smallest
orbit error in 82 of the 86 and is close to unbiased for $\sigma_u$,
although the noise standard deviation of the first period is imprecisely
estimated at this sample size by all three.

The rule used for the integral matters little for the first design and
for one column of the second.  With eight periods the adaptive
Gauss--Hermite rule and the piecewise rule give the same estimates to three
decimal places in all 100 replications, and no estimated noise standard
deviation is below $0.16$.  With two factors the piecewise rule is applied
to one coordinate of the loading at each node of the other.  It is accurate
but slow when an estimate approaches a boundary, and the table therefore
uses the adaptive rule.  To check this, 86 further panels were drawn from
the same design with a different seed, and each was fitted with the
adaptive rule and then refitted with the piecewise rule, starting from the
adaptive-rule estimates.  The raw and second-moment estimates are
essentially unchanged.  For the entropy-adjusted estimator the means of
$\widehat\sigma_u$ and of the orbit error move from $0.972$ to $0.962$ and
from $0.078$ to $0.081$, while the mean of $\widehat\sigma_{v1}$ moves from
$0.339$ to $0.322$ and its standard deviation rises from $0.121$ to
$0.179$.  With the piecewise rule, $\widehat\sigma_{v1}$ is at its lower
bound in 15 of the 86 panels, against none with the adaptive rule.  The
estimating equations cannot be solved exactly at such a boundary, and the
root-finder stopped short in 25 of the piecewise refits and 19 of the
adaptive-rule fits.  In the 45 panels where it converged under both rules,
the means agree to within $0.006$ and the orbit error changes by at most
$0.03$.  The column for $\widehat\sigma_{v1}$ in the two-factor design is
thus subject to the boundary effect documented in
Appendix~\ref{app:quadrature}, while the columns for $\widehat\sigma_u$ and
the orbit error are not materially affected.

\subsection{Accuracy of the quadrature}
\label{app:quadrature}

The integral over the loading in the Haar likelihood is one-dimensional.
For unit $i$ it has the form
\[
    \int\phi(z)\prod_{t=1}^T\Phi(a_{it}+b_tz)\,dz,
\]
where $a_{it}$ depends on the data and $b_t$ is proportional to
$\sigma_u/\sigma_{vt}$.  A Gauss--Hermite rule in $z$ is accurate when every
slope $b_t$ is moderate.  When a noise standard deviation is small relative
to $\sigma_u$, the corresponding factor approaches a step function, and a
fixed set of nodes does not resolve it.  Appendix~\ref{app:gap_check} met
this difficulty at candidate orbits.  It also arises at estimates.

\medskip
\noindent{\bf A piecewise rule.}
The simulations of Table~\ref{tab:ife_sim} use the following rule.  For
each unit the line is cut at the points $z_{it}=-a_{it}/b_t$, at
$z_{it}\pm3/|b_t|$ and $z_{it}\pm7/|b_t|$, and at fifteen fixed points
between $-12$ and $12$.  A six-point Gauss--Legendre rule is applied on each
piece.  Every factor is then either resolved or flat on every piece, and
the pieces move continuously with the parameters.  Against adaptive
integration the error of this rule in a unit log density was below
$10^{-7}$ at the parameter values of Table~\ref{tab:quadrature_error}.

Table~\ref{tab:quadrature_error} reports the error of each rule in the mean
log density of $20{,}000$ units drawn from the heteroskedastic design of
Table~\ref{tab:ife_sim}.  The density is evaluated at the true orbit and
the true values of the other parameters, while $\sigma_{v1}$ is lowered from
its true value $0.3$ to $0.0067$, the lower bound used in the numerical
search.  The reference is the piecewise rule with sixteen points on each
piece.

\begin{table}[h]
\centering
\caption{Error in the mean log density of $20{,}000$ units, multiplied by
$10^3$, in the heteroskedastic design of Table~\ref{tab:ife_sim} when
$\sigma_{v1}$ is set to the value shown.  A negative entry means that the
rule understates the likelihood.}
\label{tab:quadrature_error}
\begin{tabular}{lrrrrr}
\hline
 & \multicolumn{4}{c}{Gauss--Hermite nodes} & \\
\cline{2-5}
$\sigma_{v1}$ & 16 & 20 & 60 & 200 & Piecewise\\
\hline
0.3    & $0.00$   & $0.00$   & $0.00$  & $0.00$  & $0.00$\\
0.1    & $-2.1$   & $-1.3$   & $-0.03$ & $0.00$  & $0.00$\\
0.05   & $-7.1$   & $-4.9$   & $-0.64$ & $-0.03$ & $0.00$\\
0.02   & $-12$    & $-9.1$   & $-2.4$  & $-0.56$ & $0.00$\\
0.0067 & $-14$    & $-11$    & $-3.5$  & $-1.2$  & $0.00$\\
\hline
\end{tabular}
\end{table}

Every Gauss--Hermite rule understates the likelihood when $\sigma_{v1}$ is
small, and the error at the bound is still $1.2\times10^{-3}$ with 200
nodes.  Errors of this size exceed the differences in the criterion that
separate an interior solution from a boundary solution.  A coarse rule
therefore keeps an estimate of a small noise standard deviation away from
zero, and a finer Gauss--Hermite rule changes the estimates without making
them accurate.

\medskip
\noindent{\bf Effect on the estimates.}
Table~\ref{tab:quadrature} compares the estimates from the 16-node
Gauss--Hermite rule with those from the piecewise rule, on the same panels
and from the same starting values.  An estimate of $\sigma_{v1}$ below
$0.01$ is counted as a boundary solution.

\begin{table}[ht]
\centering
\caption{Heteroskedastic design of Table~\ref{tab:ife_sim} with a 16-node
Gauss--Hermite rule and with the piecewise rule: mean and, in parentheses,
standard deviation, and the number of samples in which
$\widehat\sigma_{v1}<0.01$.  There are 200 replications at $N=500$ and 100
at $N=2000$.}
\label{tab:quadrature}
\small
\setlength{\tabcolsep}{4pt}
\begin{tabular}{@{}rlcrccrc@{}}
\hline
 & & \multicolumn{3}{c}{16 nodes} & \multicolumn{3}{c}{Piecewise rule}\\
$N$ & Estimator & $\widehat\sigma_{v1}$ & Bound & Orbit error & $\widehat\sigma_{v1}$ & Bound & Orbit error\\
\hline
500 & Raw Haar & 0.492 (0.119) & 3 & 0.281 (0.052) & 0.476 (0.143) & 13 & 0.277 (0.043)\\
 & Second moment & 0.485 (0.104) & 1 & 0.263 (0.047) & 0.468 (0.135) & 12 & 0.262 (0.046)\\
 & Entropy adjusted & 0.303 (0.156) & 6 & 0.088 (0.059) & 0.281 (0.184) & 41 & 0.091 (0.063)\\[1mm]
2000 & Raw Haar & 0.509 (0.048) & 0 & 0.279 (0.030) & 0.502 (0.063) & 1 & 0.277 (0.024)\\
 & Second moment & 0.499 (0.041) & 0 & 0.264 (0.027) & 0.499 (0.041) & 0 & 0.264 (0.027)\\
 & Entropy adjusted & 0.293 (0.090) & 0 & 0.040 (0.021) & 0.285 (0.109) & 6 & 0.041 (0.022)\\
\hline
\end{tabular}
\end{table}

The rule matters for the smallest noise standard deviation and for little
else.  With the piecewise rule the entropy-adjusted estimate of
$\sigma_{v1}$ is at the bound in 41 of the 200 samples with $N=500$ and in
6 of the 100 with $N=2000$, against 6 and none with 16 nodes; its median is
$0.303$ and $0.305$.  The means of $\widehat\sigma_u$ and of the orbit error
change by at most $0.004$, and the ranking of the three estimators is the
same.  In the homoskedastic design no estimate is at the bound, and the two
rules give the same means to within $0.001$.

The boundary solutions are a property of the sample criterion.  In eight of
the entropy-adjusted boundary fits at $N=500$, the structural step was
restarted from four interior values of $\sigma_{v1}$.  In six it returned to
the bound.  In the other two it stopped at $0.06$ and $0.10$ with the same
value of the criterion to five decimal places, so the criterion is nearly
flat in $\sigma_{v1}$ below these values.  With three periods and
$\sigma_u/\sigma_{v1}$ above three, a sample of 500 units often cannot
distinguish a small noise standard deviation in one period from none, which
is the fixed-effects counterpart of the boundary solutions of
Appendix~\ref{app:sfm_estimators}.

\medskip
\noindent{\bf The application.}
At the reported entropy-adjusted estimates for the 800 rice households
in Table~\ref{tab:rice}, the smallest estimated noise standard deviation
is $0.083$ and $\widehat\sigma_u=0.313$, so the slopes are moderate.  The
empirical log-likelihood is $338.40$ with the 20-node rule used in
estimation, and $338.47$ both with 200 nodes and with the piecewise rule.
With 200 nodes the largest element of the average estimating function is
$1.6\times10^{-3}$.  Appendix~\ref{app:rice_quadrature} reports a
calibration using simulated residuals at the reported rice scale and
direction estimates.  The gap check of Appendix~\ref{app:gap_check}, the
two-factor design of Appendix~\ref{app:ife_rk}, the experiments of
Appendix~\ref{app:misspec} and those of
Appendix~\ref{app:proofs_application} use Gauss--Hermite rules.

\subsection{Misspecified error laws}
\label{app:misspec}

The entropy-adjusted estimator uses the error law, so the question arises of
what it estimates when that law is misspecified.  Let the errors have a
density whose contrasts $B_U'\varepsilon_i$ have density $h_U^0$ and entropy
$\mathcal E^0(U)$, while the fitted family is $h_U(\cdot;\eta)$ with entropy
$\mathcal E(U;\eta)$.  The argument for Theorem~\ref{thm:separable}(b) gives,
for every $(\theta,U)$,
\[
    \mathbb E\,\ell_{E,i}(\theta,U)
    =-\operatorname{KL}\big\{h_U^0(\cdot-s_i(\beta,U))\,\big\|\,
       h_U(\cdot;\eta)\big\}
     -\big\{\mathcal E^0(U)-\mathcal E(U;\eta)\big\}.
\]
Under correct specification the second term vanishes at $\eta_0$. Under
misspecification the population orbit criterion balances the divergence
against the difference between the true and the fitted entropy. Recovery
of the true orbit requires the adjusted population criterion to attain its
maximum there at the structural parameter selected by the population
system.

Table~\ref{tab:misspec} reports the three estimators of
Section~\ref{subsec:entropy_numerical}, computed under the normal and
half-normal specification, when the data of the two designs of
Table~\ref{tab:ife_sim} come from another law.  In the first case the
inefficiency is exponential, with the standard deviation of the half-normal
distribution that it replaces.  In the second the noise is Student with five
degrees of freedom, rescaled to the same variances.  The sample size is
$N=1000$, and the entropy is computed from $8N$ simulated error vectors.  The
last two columns give the estimates from a single sample of $100{,}000$
units, as an approximation to the pseudo-true values.

\begin{table}[h]
\centering
\caption{Estimators under a misspecified error law, $T=3$ and $N=1000$: mean
and, in parentheses, standard deviation over 80 replications, and the
estimates from one sample of $100{,}000$ units.  The value of $\sigma_u$
that matches the standard deviation of the inefficiency is 1.}
\label{tab:misspec}
\small
\begin{tabular}{@{}lllcccc@{}}
\hline
 & & & \multicolumn{2}{c}{$N=1000$} & \multicolumn{2}{c}{$N=100{,}000$}\\
Design & True law & Estimator & $\widehat\sigma_u$ & Orbit error & $\widehat\sigma_u$ & Orbit error\\
\hline
Homosk. & Exponential $u$ & Raw Haar & 1.076 (0.069) & 0.061 (0.021) & 1.087 & 0.062\\
 &  & Second moment & 1.098 (0.068) & 0.076 (0.023) & 1.112 & 0.079\\
 &  & Entropy adjusted & 1.107 (0.066) & 0.081 (0.022) & 1.108 & 0.076\\
 & Student $v$ & Raw Haar & 0.989 (0.054) & 0.035 (0.021) & 1.000 & 0.009\\
 &  & Second moment & 0.996 (0.056) & 0.035 (0.020) & 0.998 & 0.003\\
 &  & Entropy adjusted & 1.017 (0.057) & 0.039 (0.021) & 1.004 & 0.003\\[1mm]
Heterosk. & Exponential $u$ & Raw Haar & 0.633 (0.143) & 0.235 (0.053) & 0.584 & 0.251\\
 &  & Second moment & 0.691 (0.185) & 0.205 (0.069) & 0.621 & 0.228\\
 &  & Entropy adjusted & 1.068 (0.076) & 0.074 (0.027) & 1.083 & 0.069\\
 & Student $v$ & Raw Haar & 0.733 (0.057) & 0.282 (0.047) & 0.742 & 0.276\\
 &  & Second moment & 0.750 (0.053) & 0.263 (0.038) & 0.752 & 0.264\\
 &  & Entropy adjusted & 1.000 (0.068) & 0.057 (0.043) & 1.010 & 0.001\\
\hline
\end{tabular}
\end{table}

With heavy-tailed noise the entropy-adjusted estimator is little affected:
its pseudo-true orbit error is below $0.01$ in both designs.  With
exponential inefficiency it is not consistent for the orbit.  Its
pseudo-true orbit error is about $0.07$ in both designs, and in the
homoskedastic design the three estimators are displaced by similar amounts,
between $0.06$ and $0.08$.  In the heteroskedastic design the raw and
second-moment estimators have pseudo-true orbit errors of $0.23$ to $0.28$
under both laws, as large as under correct specification in
Table~\ref{tab:ife_sim}, and they underestimate $\sigma_u$ by a quarter or
more, while the entropy-adjusted estimator stays within $0.07$ of the true
orbit.  In these experiments, misspecification of the shape of the error
law costs the entropy-adjusted estimator less than period-specific noise
variances cost the other two, but it does not leave the estimator
consistent, and a specification check of the error law remains advisable in
applications.


\section{Proofs and additional results for Section~\ref{sec:limits}}
\label{app:proofs_limits}

Throughout, expectations are taken at the true parameter values and
conditional on the covariates.

\subsection{Proof of Proposition~\ref{prop:two_period}}
\label{app:proof_two_period}

All three parts are about one unit at a time.  To keep the formulas short,
the subscript $i$ is left out, so that $Y_t$, $Z_t$, $\alpha$, $A$, $S$ and
$K$ mean $Y_{it}$, $Z_{it}$, $\alpha_i$, $A_i$, $S_i$ and $K_i$.

\medskip
\noindent{\bf Part (a)}

Under the reflected law the errors are $-Z_1$ and $-Z_2$, so the difference
is $(-Z_1)-(-Z_2)=Z_2-Z_1$.  Because $Z_1$ and $Z_2$ are independent with
the same distribution, $Z_2-Z_1$ has the same distribution as $Z_1-Z_2$.
The difference $Y_1-Y_2=Z_1-Z_2$ therefore has the same law under $F_+$ and
$F_-$.

\medskip
\noindent{\bf Part (b)}

\noindent{\bf The mean of $A$.}
With $D=Z_1-Z_2$, $A=D^2$ and $\mathbb EA=\operatorname{Var}(Z_1)+
\operatorname{Var}(Z_2)=2v$.

\medskip 
\noindent{\bf The covariance of $A$ and $U$.}
Let $U=Z_1+Z_2$, so that $S=2\alpha+U$ and $\mathbb EU=0$.  Expanding,
\[
    AU=(Z_1-Z_2)^2(Z_1+Z_2)=Z_1^3-Z_1^2Z_2-Z_1Z_2^2+Z_2^3 .
\]
The errors are independent and centred, so
$\mathbb E(Z_1^2Z_2)=\mathbb E(Z_1Z_2^2)=0$ and $\mathbb E(AU)=2\kappa$.

\medskip 
\noindent{\bf The moment.}
Hence
$\mathbb E\{(A-2v)S\}=2\alpha\,\mathbb E(A-2v)+\mathbb E(AU)-2v\,\mathbb EU
=2\kappa$, whatever the value of $\alpha$.

\medskip
\noindent{\bf Part (c)}

\noindent{\bf The deviations are free of the effect.}
$Y_t-\overline{Y}=Z_t-\overline{Z}$ does not involve $\alpha$.

\medskip 
\noindent{\bf The third moment of one deviation.}
For $T=3$, $Z_1-\overline{Z}=(2Z_1-Z_2-Z_3)/3$.  The third moment of a sum of
independent centred variables is the sum of their third moments, so
$\mathbb E(Z_1-\overline{Z})^3=(2^3-1-1)\kappa/27=2\kappa/9$.

\medskip 
\noindent{\bf The statistic.}
The same value holds for $t=2$ and $t=3$.  Summing over the three periods
and multiplying by $3/2$ gives $\mathbb EK=\kappa$.

\subsection{Proof of Proposition~\ref{prop:geometric}}
\label{app:proof_geometric}

\noindent{\bf Part (a)}

If $X_ih=Aa_i$ for every $i$, then
$X_i(\beta+h)+A(\xi_i-a_i)=X_i\beta+A\xi_i$.  The parameter values
$(\beta+h,\eta,\{\xi_i-a_i\})$ and $(\beta,\eta,\{\xi_i\})$ therefore give
the same systematic component for every unit. With the error law held
fixed, they induce the same distribution of $(Y_1,\ldots,Y_N)$.

\medskip
\noindent{\bf Part (b)}

\noindent{\bf The score and the information.}
By \eqref{eq:haar_equals_quotient},
$\log L_{H,i}=\log h\{B'(Y_i-X_i\beta);\eta\}$.  The score for $\beta$ at
the truth is $-X_i'Bs_w(W_i)$ with $W_i=B'\varepsilon_i$, and with
$J_W=\mathbb E\{s_w(W_i)s_w(W_i)'\}$ the information for $\beta$ is
$\sum_iX_i'BJ_WB'X_i$.

\medskip 
\noindent{\bf When it is singular.}
For $h\neq0$,
\[
    h'\Big\{\sum_iX_i'BJ_WB'X_i\Big\}h=\sum_i\|J_W^{1/2}B'X_ih\|^2 .
\]
Since $J_W$ is positive definite, this is zero if and only if $B'X_ih=0$
for every $i$.  The null space of $B'$ is $\operatorname{col}(A)$, so the
condition is $X_ih\in\operatorname{col}(A)$ for every $i$.

\subsection{Proof of Theorem~\ref{thm:minimax}}
\label{app:proof_minimax}

Write $U_i=Z_{i1}+Z_{i2}$, so that $S_i=2\alpha_i+U_i$, while
$A_i=(Z_{i1}-Z_{i2})^2$ does not involve $\alpha_i$.  The pairs $(A_i,U_i)$
are independent and identically distributed with $\mathbb EU_i=0$, and
$\operatorname{Cov}(A_i,U_i)=\mathbb E(A_iU_i)$ equals twice the third moment
of the error law, by the proof of Proposition~\ref{prop:two_period}(b).

\medskip
\noindent{\bf Part (a)}

\noindent{\bf Decomposition of the statistic.}
Substituting $S_i-\overline{S}=2(\alpha_i-\overline{\alpha}_N)+(U_i-\overline{U})$ in
\eqref{eq:kappa_hat_boundary} and using
$\sum_i(\alpha_i-\overline{\alpha}_N)=0$,
\[
    \widehat\kappa_N
    =\underbrace{\frac{1}{N-1}\sum_{i=1}^NA_i(\alpha_i-\overline{\alpha}_N)}_{T_{1N}}
    +\underbrace{\frac{1}{2(N-1)}\sum_{i=1}^N(A_i-\overline{A})(U_i-\overline{U})}_{T_{2N}}.
\]

\medskip 
\noindent{\bf The term that carries the effects.}
The first term has mean $2v\sum_i(\alpha_i-\overline{\alpha}_N)/(N-1)=0$.  The
$A_i$ are independent, so for $\alpha\in\mathcal A_N(H_N)$ and $N\ge2$,
\[
    \operatorname{Var}(T_{1N})
    =\frac{\operatorname{Var}(A_1)}{(N-1)^2}
     \sum_{i=1}^N(\alpha_i-\overline{\alpha}_N)^2
    \le\frac{N}{(N-1)^2}\operatorname{Var}(A_1)H_N
    \le\frac{4\operatorname{Var}(A_1)H_N}{N}.
\]
The last step uses $N\le2(N-1)$.

\medskip 
\noindent{\bf The term that is free of the effects.}
The sample covariance of the pairs $(A_i,U_i)$ is the $U$-statistic $C_N$
with kernel $k(i,j)=\tfrac12(A_i-A_j)(U_i-U_j)$, and $T_{2N}=C_N/2$.  Hence
$T_{2N}$ is unbiased for $\tfrac12\operatorname{Cov}(A_1,U_1)$, the third
moment of the error law.  By the variance bound for $U$-statistics of order
two \citep{Hoeffding1948},
$\operatorname{Var}(C_N)\le2\operatorname{Var}\{k(1,2)\}/N$, so
$\operatorname{Var}(T_{2N})\le\operatorname{Var}\{k(1,2)\}/(2N)$.  The
variance of the kernel is finite because the error law has a finite sixth
moment.

\medskip 
\noindent{\bf Conclusion.}
By the two preceding steps, $\widehat\kappa_N$ is unbiased.  Since
$\operatorname{Var}(X+Y)\le2\operatorname{Var}(X)+2\operatorname{Var}(Y)$,
\[
    \operatorname{Var}(\widehat\kappa_N)
    \le\frac{8\operatorname{Var}(A_1)H_N}{N}
       +\frac{\operatorname{Var}\{k(1,2)\}}{N}
    \le\frac{c_F(1+H_N)}{N},
\]
with $c_F=\max[8\operatorname{Var}(A_1),\operatorname{Var}\{k(1,2)\}]$.
These moments are the same under $F_+$ and $F_-$.

\medskip
\noindent{\bf Part (b)}

Under $F_+$ the mean of $\widehat\kappa_N$ is $\kappa>0$.  The event
$\widehat\kappa_N\le0$ implies $|\widehat\kappa_N-\kappa|\ge\kappa$, so by
Chebyshev's inequality
$P(\widehat\kappa_N\le0)\le\operatorname{Var}(\widehat\kappa_N)/\kappa^2$.
Under $F_-$ the mean is $-\kappa$ and the same bound holds for
$P(\widehat\kappa_N>0)$.  Part (a) bounds the variance uniformly over
$\mathcal A_N(H_N)$ by $c_F(1+H_N)/N$, which tends to zero when
$H_N/N\to0$.

\medskip
\noindent{\bf Part (c)}

\noindent{\bf Reduction to a mixture.}
The test $\phi\equiv1/2$ has both errors equal to $1/2$, so
$R_N^*\le1/2$.  For the reverse inequality, note that for any test $\phi$
and any priors on the effects under the two hypotheses, the maximum of the
two worst-case errors is at least the average of the two prior-weighted
errors.  With $\overline{P}_0$ and $\overline{P}_1$ the two mixtures, this gives
\begin{equation}
    R_N^*\ge\tfrac12\{1-\operatorname{TV}(\overline{P}_0,\overline{P}_1)\}
\label{eq:TV_test_bound}
\end{equation}
\citep[Chapter~2]{Tsybakov2009}.  Take the
same prior under both hypotheses, with $\alpha_1,\ldots,\alpha_N$
independent $N(0,\tau^2)$.

\medskip 
\noindent{\bf One unit.}
Consider one unit and drop its index.  The data are equivalent to
$(D,S)$ with $D=Z_1-Z_2$ and $S=U+2\alpha$.  Let $p_+(d,u)$ and $p_-(d,u)$
be the joint densities of $(D,U)$ under $F_+$ and $F_-$, and let
$\varphi_\sigma$ be the $N(0,\sigma^2)$ density.  Under the prior,
$2\alpha\sim N(0,4\tau^2)$ is independent of $(D,U)$, so the joint density
of $(D,S)$ is
$m_\pm^\tau(d,x)=\int p_\pm(d,u)\varphi_{2\tau}(x-u)\,du$.  By
Proposition~\ref{prop:two_period}(a), $D$ has the same marginal density under
both laws, so $q_d(u)=p_+(d,u)-p_-(d,u)$ satisfies $\int q_d(u)\,du=0$ for
almost every $d$. Therefore
\[
    \int|m_+^\tau(d,x)-m_-^\tau(d,x)|\,dx
    =\int\Big|\int q_d(u)\{\varphi_{2\tau}(x-u)-\varphi_{2\tau}(x)\}\,du
     \Big|\,dx
    \le\int|q_d(u)|\,\rho_\tau(u)\,du,
\]
where
$\rho_\tau(u)=\int|\varphi_{2\tau}(x-u)-\varphi_{2\tau}(x)|\,dx
=2\{2\Phi(|u|/(4\tau))-1\}$, the $L^1$ distance between two normal densities
with common standard deviation $2\tau$ whose means differ by $u$.  The
inequality moves the absolute value inside the integral over $u$ and then
integrates over $x$ first.

\medskip 
\noindent{\bf The one-unit distance vanishes as the prior spreads.}
For every $u$, $\rho_\tau(u)\le2$ and $\rho_\tau(u)\to0$ as
$\tau\to\infty$, and $\int\!\!\int|q_d(u)|\,du\,dd\le2$.  By dominated
convergence the total-variation distance between the two one-unit mixtures,
\[
    \operatorname{TV}_1(\tau)
    =\tfrac12\int\!\!\int|m_+^\tau(d,x)-m_-^\tau(d,x)|\,dx\,dd
    \le\tfrac12\int\!\!\int|q_d(u)|\,\rho_\tau(u)\,du\,dd,
\]
tends to zero as $\tau\to\infty$.

\medskip 
\noindent{\bf $N$ units.}
The units are independent under the prior, so each $N$-unit mixture is a
product measure, and the total-variation distance between two product
measures is at most the sum of the distances between their factors, here
$N\operatorname{TV}_1(\tau)$.  Hence
$R_N^*\ge\tfrac12\{1-N\operatorname{TV}_1(\tau)\}$ for every $\tau$, and
letting $\tau\to\infty$ with $N$ fixed gives $R_N^*\ge1/2$.

\subsection{Sharpness of the rate in Theorem~\ref{thm:minimax}}
\label{app:minimax_sharp}

Theorem~\ref{thm:minimax}(b) shows that the skewness sign is recovered
uniformly when $H_N=o(N)$.  This subsection shows that the rate cannot be
improved under a tail condition.  Let $D=Z_1-Z_2$ and $U=Z_1+Z_2$ for two
independent draws from $F_+$, let $U^\dagger$ be a draw from the conditional
law of $U$ given $D$ that is conditionally independent of $U$, and define
$\kappa_F(s)=\mathbb E\exp(|UU^\dagger|/s^2)$ for $s>0$.  This function is
finite for large $s$ whenever $F_+$ has bounded support, and more generally
whenever the conditional law of $U$ given $D$ has uniformly sub-Gaussian
tails.

\begin{proposition}[Sharpness of the recovery rate]
\label{prop:minimax_sharp}
In the setting of Theorem~\ref{thm:minimax}, suppose that
$\kappa_F(s_0)<\infty$ for some $s_0>0$.  If $H_N\ge cN$ for some $c>0$ and
all large $N$, then
\[
    \liminf_{N\to\infty}R_N^*
    \ge\tfrac12\Big[1-\big\{1-e^{-4s_0^2\{\kappa_F(s_0)-1\}/c}\big\}^{1/2}\Big]
    >0,
\]
and the bound tends to $1/2$ as $c$ grows.  Consequently $R_N^*\to0$ if and
only if $H_N/N\to0$.
\end{proposition}

\noindent{\em Proof.}
The argument for Theorem~\ref{thm:minimax}(c) in
Appendix~\ref{app:proof_minimax} is refined in two ways: the prior is
restricted to
the class $\mathcal A_N(H_N)$, and the distance between the mixtures is
bounded at a rate that gives a positive lower bound on the error when
$H_N$ is of order $N$.

\medskip 
\noindent{\bf A chi-square bound for one unit.}
Keep the notation of that argument, let $s=2\tau$ be the standard deviation of
$2\alpha$, and let $f_D$ be the density of $D$, which is the same under
$F_+$ and $F_-$ and is symmetric.  Write $G_d^\pm$ for the conditional law
of $U$ given $D=d$, so that
$m_\pm^\tau(d,x)=f_D(d)\int\varphi_s(x-u)\,G_d^\pm(du)$, and let
$R_s(d,x)=f_D(d)\varphi_s(x)$.  The Gaussian identity
$\varphi_s(x-u)\varphi_s(x-u')/\varphi_s(x)
=\varphi_s(x-u-u')\exp(uu'/s^2)$, integrated over $x$, gives
\[
    \chi^2(m_+^\tau\,\|\,R_s)
    =\int\!\!\int\frac{m_+^\tau(d,x)^2}{R_s(d,x)}\,dx\,dd-1
    =\mathbb E\exp(UU^\dagger/s^2)-1,
\]
with $U,U^\dagger$ as in the definition of $\kappa_F$ above.

\medskip 
\noindent{\bf The bound is of order $s^{-2}$.}
For $s\ge s_0$ the expectation is finite and may be expanded as a power
series, because $\kappa_F(s)\le\kappa_F(s_0)<\infty$.  Since $U$ and
$U^\dagger$ are conditionally independent and identically distributed given
$D$, $\mathbb E\{(UU^\dagger)^k\}
=\mathbb E[\{\mathbb E(U^k\mid D)\}^2]\ge0$, and
\[
    \chi^2(m_+^\tau\,\|\,R_s)
    =\sum_{k\ge1}\frac{\mathbb E\big[\{\mathbb E(U^k\mid D)\}^2\big]}{k!\,s^{2k}}.
\]
Every term is nonnegative and the $k$th term is proportional to $s^{-2k}$,
so for $s\ge s_0$ each term is at most $(s_0/s)^2$ times its value at
$s_0$, and
\[
    \chi^2(m_+^\tau\,\|\,R_s)
    \le\frac{s_0^2}{s^2}\sum_{k\ge1}
    \frac{\mathbb E\big[\{\mathbb E(U^k\mid D)\}^2\big]}{k!\,s_0^{2k}}
    \le\frac{s_0^2}{s^2}\{\kappa_F(s_0)-1\}
    =:\chi_s^2.
\]
Under $F_-$ the pair $(D,U)$ has the law of $(-D,-U)$ under $F_+$, so
$m_-^\tau(d,x)=m_+^\tau(-d,-x)$.  Since $R_s(-d,-x)=R_s(d,x)$, the same
bound holds for $\chi^2(m_-^\tau\,\|\,R_s)$.

\medskip 
\noindent{\bf From chi-square to affinity.}
Let $\operatorname{H}^2(P,Q)=\int(\sqrt p-\sqrt q)^2$ be the squared
Hellinger distance.  Since
$(\sqrt p-\sqrt r)^2=(p-r)^2/(\sqrt p+\sqrt r)^2\le(p-r)^2/r$,
$\operatorname{H}^2(P,R)\le\chi^2(P\,\|\,R)$.  This inequality and the
triangle inequality
$\operatorname{H}(P,Q)\le\operatorname{H}(P,R)+\operatorname{H}(Q,R)$ give
$\operatorname{H}(m_+^\tau,m_-^\tau)\le2\chi_s$, that is,
$\operatorname{H}^2(m_+^\tau,m_-^\tau)\le4\chi_s^2$.  The affinity
$\rho_1=\int\!\!\int(m_+^\tau m_-^\tau)^{1/2}=1-\tfrac12\operatorname{H}^2$
therefore satisfies $\rho_1\ge1-2\chi_s^2$.

\medskip 
\noindent{\bf $N$ units.}
Under the prior the units are independent and identically distributed, so
the affinity of the two $N$-unit mixtures $\overline{P}_+$ and $\overline{P}_-$ is
$\rho_1^N$.  By Le~Cam's inequality \citep[Chapter~2]{Tsybakov2009},
$\operatorname{TV}(\overline{P}_+,\overline{P}_-)\le(1-\rho_1^{2N})^{1/2}$.  Since
$\log(1-x)\ge-2x$ for $0\le x\le\tfrac12$, taking $x=2\chi_s^2$ gives
$\rho_1^{2N}\ge\exp(-8N\chi_s^2)$ whenever $\chi_s^2\le\tfrac14$.

\medskip 
\noindent{\bf Restriction of the prior to the class.}
Take $\tau^2=H_N/2$, so that $s^2=2H_N$, and let $E_N$ be the event that
$N^{-1}\sum_i(\alpha_i-\overline{\alpha}_N)^2\le H_N$.  Under the prior this
statistic is distributed as $\tau^2\chi^2_{N-1}/N$, so
$\pi(E_N^c)=P(\chi^2_{N-1}>2N)$.  The $\chi^2_{N-1}$ law has mean $N-1$ and
variance $2(N-1)$, and Chebyshev's inequality gives
$\pi(E_N^c)\le2(N-1)/(N+1)^2\to0$.  Let $\widetilde{\pi}$ be the prior conditioned
on $E_N$, which is supported in $\mathcal A_N(H_N)$, and let
$\widetilde{P}_\pm$ be the corresponding mixtures.  Mixing the same kernel
against two priors does not increase total variation, and
$\operatorname{TV}(\widetilde{\pi},\pi)=\pi(E_N^c)$, so by the triangle inequality
$\operatorname{TV}(\widetilde{P}_+,\widetilde{P}_-)
\le\operatorname{TV}(\overline{P}_+,\overline{P}_-)+2\pi(E_N^c)$.

\medskip 
\noindent{\bf The lower bound.}
Applying \eqref{eq:TV_test_bound} to $\widetilde{\pi}$ and using the two preceding steps,
\[
    R_N^*\ge\tfrac12\Big[1-\{1-\exp(-8N\chi_s^2)\}^{1/2}\Big]-\pi(E_N^c) .
\]
If $H_N\ge cN$ then $s\to\infty$, so eventually $s\ge s_0$ and
$\chi_s^2\le\frac14$, and
$8N\chi_s^2=4s_0^2\{\kappa_F(s_0)-1\}N/H_N\le4s_0^2\{\kappa_F(s_0)-1\}/c$.
Letting $N\to\infty$ gives the bound in the proposition.

\medskip 
\noindent{\bf The equivalence.}
If $H_N/N\to0$ then $R_N^*\to0$ by Theorem~\ref{thm:minimax}(b). Otherwise there are $c>0$ and a
subsequence along which $H_N\ge cN$, and along it $R_N^*$ is bounded away
from zero by the lower bound just obtained.  Hence $R_N^*\to0$ if and only if $H_N/N\to0$.

\subsection{Proof of Proposition~\ref{prop:sfm_orientation}}
\label{app:proof_sfm_orientation}

The proof is for one unit, so the subscript $i$ is omitted, and
$\widetilde{e}=Me(\beta)$.  Part (a) shows that, with two periods, reversing the
sign $\varsigma$ does not change the Haar likelihood.  Part (b) shows that a
third period reveals the sign through the skewness of the errors.

\medskip
\noindent{\bf Part (a)}

\noindent{\bf The expectation under a cost frontier.}
For $T=2$, $\widetilde{e}_1+\widetilde{e}_2=0$, so $\widetilde{e}_2=-\widetilde{e}_1$.  Put
$c=\widetilde{e}_1/\sigma$.  Under $\varsigma=-1$ the expectation in
\eqref{eq:sfm_haar} is
\[
    \mathbb E_Z\big[\Phi\{\lambda(c+Z/\sqrt2)\}\,
                    \Phi\{\lambda(-c+Z/\sqrt2)\}\big].
\]

\medskip 
\noindent{\bf Reflection of $Z$.}
The variable $-Z$ has the same distribution as $Z$.  Replacing $Z$ by $-Z$
turns the expectation into
\[
    \mathbb E_Z\big[\Phi\{-\lambda(-c+Z/\sqrt2)\}\,
                    \Phi\{-\lambda(c+Z/\sqrt2)\}\big],
\]
which is the expectation in \eqref{eq:sfm_haar} under $\varsigma=1$.

\medskip 
\noindent{\bf The remaining factors.}
The factors of \eqref{eq:sfm_haar} outside the expectation do not involve
$\varsigma$.  The Haar likelihood therefore takes the same value for
$\varsigma=1$ and $\varsigma=-1$.

\medskip
\noindent{\bf Part (b)}

\noindent{\bf Reduction to a third central moment.}
The statistic $K_i$ depends on the errors only through deviations from
their mean, so Proposition~\ref{prop:two_period}(c) applies with $\kappa$
equal to the third central moment of
$\varepsilon_{it}=v_{it}-\varsigma u_{it}$.  The noise $v_{it}$ is Gaussian
and independent of $u_{it}$, and third central moments of independent
variables add, so $\kappa$ is $-\varsigma$ times the third central moment of
$u_{it}$.

\medskip 
\noindent{\bf The half-normal distribution.}
For the half-normal distribution,
$\mathbb Eu=\sigma_u\sqrt{2/\pi}$, $\mathbb Eu^2=\sigma_u^2$ and
$\mathbb Eu^3=2\sigma_u^3\sqrt{2/\pi}$.  The third central moment is
\[
    \mathbb Eu^3-3\,\mathbb Eu\,\mathbb Eu^2+2(\mathbb Eu)^3
    =\sigma_u^3\sqrt{2/\pi}\,\{2-3+4/\pi\}
    =\sigma_u^3\sqrt{2/\pi}\,(4/\pi-1)>0 .
\]
Hence $\mathbb EK_i=-\varsigma\,\sigma_u^3\sqrt{2/\pi}\,(4/\pi-1)$.

\subsection{Orthogonalised scores with sample splitting}
\label{app:sfm_split}

Table~\ref{tab:sfm_scores} evaluates the orthogonalised scores at the
unit-level estimate of the effect computed from the same data.  The
procedure of \citet{BonhommeJochmansWeidner2026} uses preliminary estimates
that are independent of the data entering the score, and
Table~\ref{tab:sfm_split} reports the resulting estimators.  The periods of
every unit are divided into two halves, the odd and the even periods.  The
effect is estimated by maximum likelihood on one half, as a function of
$(\sigma,\lambda)$; the orthogonalised score of order $q$ is evaluated on
the other half at that estimate; the roles are exchanged and the two
estimating functions are added.  The projection coefficients are computed by
simulation as functions of $(\sigma,\lambda)$, and the two equations are
solved from the true values; a solution was found in every replication.  The
design is that of Table~\ref{tab:sfm_scores}, with standard normal effects
and $N=1000$.

\begin{table}[h]
\centering
\caption{Haar estimator and orthogonalised-score estimators with sample
splitting in the frontier model: mean and, in parentheses, standard
deviation over 200 replications, $N=1000$.  True values are $\sigma=1.118$
and $\lambda=2$.}
\label{tab:sfm_split}
\begin{tabular}{rlcc}
\hline
$T$ & Estimator & $\widehat\sigma$ & $\widehat\lambda$\\
\hline
4  & Haar                 & 1.119 (0.038) & 2.025 (0.268)\\
   & Orthogonalised, $q=1$ & 1.203 (0.033) & 1.166 (0.098)\\
   & Orthogonalised, $q=2$ & 0.974 (0.022) & 1.267 (0.101)\\[1mm]
10 & Haar                 & 1.120 (0.016) & 2.016 (0.095)\\
   & Orthogonalised, $q=1$ & 1.164 (0.017) & 1.617 (0.068)\\
   & Orthogonalised, $q=2$ & 1.084 (0.015) & 1.823 (0.081)\\[1mm]
20 & Haar                 & 1.118 (0.011) & 2.003 (0.061)\\
   & Orthogonalised, $q=1$ & 1.141 (0.012) & 1.792 (0.051)\\
   & Orthogonalised, $q=2$ & 1.106 (0.011) & 1.952 (0.061)\\
\hline
\end{tabular}
\end{table}

The Haar estimator has little bias at the reported panel lengths.  The
orthogonalised estimators approach the true values as $T$ grows and the second order
improves on the first, as the theory of
\citet{BonhommeJochmansWeidner2026} predicts, but with $N=1000$ the bias of
the second-order estimator of $\lambda$ exceeds seven standard deviations at
$T=4$, is about two at $T=10$ and remains close to one at $T=20$.
Splitting halves the number of periods available to each step, which is why these biases exceed
those implied by Table~\ref{tab:sfm_scores}.  At $T=20$ the second-order
estimator is as precise as the Haar estimator, so the information that the
Haar reduction gives up is no longer visible; what remains of the comparison
is the centring.

\subsection{Estimating the effects: a numerical comparison}
\label{app:sfm_estimators}

Table~\ref{tab:sfm_estimators} compares the maximum likelihood estimator that
treats the effects as parameters with the maximiser of \eqref{eq:sfm_haar},
in model \eqref{eq:sfm_model} without covariates, with $\sigma_v=0.5$,
$\sigma_u=1$, standard normal effects, $N=500$ and 500 replications.  Both
criteria are maximised numerically from the true values.

\begin{table}[h]
\centering
\caption{Estimators of $(\sigma_v,\sigma_u)=(0.5,1)$ in the frontier model
with fixed effects: mean and standard deviation over 500 replications,
$N=500$.  ``Effects estimated'' is the maximum likelihood estimator with one
parameter per unit; for $T=10$ it is the interior local maximiser nearest the
true values.}
\label{tab:sfm_estimators}
\begin{tabular}{rrrcrrcrrcrr}
\hline
 & \multicolumn{5}{c}{Effects estimated} && \multicolumn{5}{c}{Haar}\\
\cline{2-6}\cline{8-12}
 & \multicolumn{2}{c}{$\widehat\sigma_v$} && \multicolumn{2}{c}{$\widehat\sigma_u$}
 && \multicolumn{2}{c}{$\widehat\sigma_v$} && \multicolumn{2}{c}{$\widehat\sigma_u$}\\
\cline{2-3}\cline{5-6}\cline{8-9}\cline{11-12}
$T$ & Mean & S.d. && Mean & S.d. && Mean & S.d. && Mean & S.d.\\
\hline
3 & 0.001 & 0.000 && 0.974 & 0.021 && 0.498 & 0.071 && 0.991 & 0.105\\
5 & 0.002 & 0.022 && 1.160 & 0.022 && 0.499 & 0.034 && 0.998 & 0.057\\
10 & 0.420 & 0.025 && 1.029 & 0.037 && 0.501 & 0.020 && 0.996 & 0.033\\
\hline
\end{tabular}
\end{table}

The Haar estimator has little bias in these simulations at each panel
length.  The estimator that keeps the effects in the likelihood fails in a
particular way.
With $T=3$ and $T=5$ its criterion increases as $\sigma_v\to0$: each unit
effect can be placed at the largest observation of the unit, which makes the
noise component unnecessary, and the fitted model attributes all within-unit
variation to inefficiency.  The value $0.001$ in the table is the lower bound
imposed in the numerical search.  With $T=10$ an interior local maximum
exists, and the table reports it, but $\sigma_v$ is underestimated by about
16 per cent, and the boundary solution attained a higher value of the
criterion in each of ten samples examined separately.  The
incidental-parameter problem does not appear here as a small bias in a
variance parameter.  It removes one of the two error components from the
fitted model.


\section{Additional results for Section~\ref{sec:application}}
\label{app:proofs_application}

\subsection{Standard errors and the Wald test in a calibrated design}
\label{app:rice_coverage}

The standard errors of Table~\ref{tab:rice} come from the limit distribution
of Theorem~\ref{thm:entropy_limit}(c) with the simulated entropy treated as
exact.  Table~\ref{tab:rice_coverage} checks them in two designs calibrated
to the application.  Both keep the regressors of the 800 households and
five waves.  In the first the parameters are the entropy-adjusted estimates
of Table~\ref{tab:rice}; in the second they are the equal-weights estimates,
so that the hypothesis of equal weights is true.  The household loadings are
fixed at their generalised least-squares estimates, and new noise and
inefficiency are drawn in each of 100 replications.  The model is
re-estimated as in the application, with the entropy computed from $8N$
simulated error vectors, by solving the estimating equations from the true
values.

\begin{table}[h]
\centering
\caption{Calibrated designs, $N=800$, $T=5$, 100 replications: mean of the
estimates, their standard deviation, the mean reported standard error, and
the coverage of nominal 95 per cent intervals.}
\label{tab:rice_coverage}
\begin{tabular}{lrrccc}
\hline
 & True & Mean & s.d. & Mean s.e. & Coverage\\
\hline
\multicolumn{6}{l}{\emph{Design at the entropy-adjusted estimates}}\\
~~~Direction, 2008 & 0.246 & 0.240 & 0.026 & 0.025 & 0.93\\
~~~Direction, 2010 & 0.247 & 0.245 & 0.024 & 0.026 & 0.97\\
~~~Direction, 2012 & 0.519 & 0.521 & 0.020 & 0.021 & 0.96\\
~~~Direction, 2014 & 0.638 & 0.638 & 0.025 & 0.023 & 0.91\\
~~~Direction, 2016 & 0.448 & 0.448 & 0.022 & 0.022 & 0.96\\
~~~Returns to scale & 0.955 & 0.958 & 0.013 & 0.014 & 0.96\\
~~~$\log\sigma_u$ & $-1.162$ & $-1.160$ & 0.041 & 0.032 & 0.85\\[1mm]
\multicolumn{6}{l}{\emph{Design at the equal-weights estimates}}\\
~~~Direction, 2008 & 0.447 & 0.445 & 0.028 & 0.026 & 0.93\\
~~~Direction, 2010 & 0.447 & 0.447 & 0.028 & 0.026 & 0.91\\
~~~Direction, 2012 & 0.447 & 0.449 & 0.020 & 0.024 & 0.97\\
~~~Direction, 2014 & 0.447 & 0.442 & 0.027 & 0.026 & 0.95\\
~~~Direction, 2016 & 0.447 & 0.450 & 0.026 & 0.024 & 0.94\\
~~~Returns to scale & 0.998 & 0.999 & 0.017 & 0.016 & 0.93\\
~~~$\log\sigma_u$ & $-1.281$ & $-1.282$ & 0.047 & 0.048 & 0.95\\
\hline
\end{tabular}
\end{table}

The estimator has little bias in both calibrated designs, and the reported
standard errors of the factor direction and of the returns to scale agree with the standard
deviations of the estimates, with coverage between $0.91$ and $0.97$.  With
100 replications the Monte Carlo standard error of a coverage of $0.95$ is
$0.02$. In the first design the mean reported standard error for
$\log\sigma_u$ is $0.032$, about a fifth below the simulation standard
deviation of $0.041$. The Wald test of equal weights rejects in $9$ of the
100 replications of the second design at the nominal 5 per cent level and in
one at the 1 per cent level. The empirical 95th percentile of the simulated statistics
is $10.4$, compared with the chi-square critical value of $9.5$. One of
the 100 statistics exceeds the application value of $13.4$. The test
rejects in every replication of the first design. The fitted loadings
used in that design include estimation noise, which can increase their
dispersion and raise power in the calibrated design.

\subsection{Measurement error in the early waves and farm size}
\label{app:rice_contaminated}

Section~\ref{sec:application} asks how fragile the agreement between the
three estimated directions in Table~\ref{tab:rice} is.  In
Table~\ref{tab:rice_contaminated}, independent normal errors with standard
deviation $0.25$ are added to log output in the first two waves, as a
stylised model of additional recall error in the early survey rounds.  If
the original model is correctly specified, the contaminated model remains within the same family because it has a separate Gaussian noise
variance for each wave, and the true factor direction,
production function and inefficiency scale are unchanged; only the noise
standard deviations of the two waves rise, to about three times those of the
later waves.  The entropy-adjusted factor direction and returns to scale
remain close to their original estimates: in each of three draws the direction stays within $0.06$ of the original estimate in every
component, and a restart from the second-moment solution returns the same
point.  The second-moment and raw estimators now place almost all the weight
of the household effect on the contaminated waves, and both report returns
to scale above one and a smaller inefficiency scale.  In these three draws,
the added heteroskedastic measurement error changes the estimated returns to scale from below one to above one under the raw
and second-moment estimators, while the entropy-adjusted estimate remains
below one.  This is a targeted sensitivity exercise, rather than evidence of
robustness to arbitrary measurement error.

\begin{table}[h]
\centering
\caption{Rice frontier with measurement error added to the first two waves.
Independent $N(0,0.25^2)$ errors are added to log output in 2008 and 2010,
and the model of Table~\ref{tab:rice} is re-estimated.  Entries are means
over three draws of the added errors; the first column repeats the
entropy-adjusted estimates from the original data.}
\label{tab:rice_contaminated}
\begin{tabular}{lcccc}
\hline
 & Original data & \multicolumn{3}{c}{With added measurement error}\\
\cline{3-5} & Entropy adjusted & Second moment & Raw Haar & Entropy adjusted\\
\hline
Direction, 2008 & 0.246 & 0.853 & 0.941 & 0.219\\
Direction, 2010 & 0.247 & 0.460 & 0.199 & 0.231\\
Direction, 2012 & 0.519 & 0.189 & 0.161 & 0.513\\
Direction, 2014 & 0.638 & 0.086 & 0.112 & 0.658\\
Direction, 2016 & 0.448 & 0.106 & 0.089 & 0.448\\
Returns to scale       & 0.955 & 1.034 & 1.026 & 0.958\\
$\sigma_u$             & 0.313 & 0.237 & 0.237 & 0.290\\
\hline
\end{tabular}
\end{table}

Evidence against equal weights is stronger among the larger farms.  Among
the 400 households with rice area above the median the estimated direction
has components $0.19$, $0.27$, $0.50$, $0.68$ and $0.44$, with standard
errors near $0.045$; among
the smaller farms the loadings are less dispersed and the direction is
poorly determined.  The Wald statistics for equal weights are $9.4$ and
$8.1$ in the two halves ($p$-values $0.05$ and $0.09$).

\subsection{Quadrature accuracy at the rice estimates}
\label{app:rice_quadrature}

For each column of Table~\ref{tab:rice}, 40{,}000 residual vectors were
generated under the reported Gaussian noise and half-normal inefficiency
scales, with common standard normal draws across the four calibrations and
each reported factor direction normalised to unit length. The largest
absolute difference between the 60- and 200-node log-density evaluations
across all simulated residuals was $2.87\times10^{-7}$. At the
entropy-adjusted calibration, the corresponding 20- versus 200-node
difference was $0.0029$. Appendix~\ref{app:quadrature}
reports the empirical log-likelihood check at the entropy-adjusted fit
for the 800 households.

\end{document}